\documentclass[aps,prr,reprint,groupedaddress]{revtex4-2}

\usepackage{amsmath}
\usepackage{mathtools}
\usepackage{physics}
\usepackage{amssymb}
\usepackage{amsfonts}
\usepackage{dsfont}
\usepackage{graphicx}
\usepackage{bm}
\usepackage{color}
\usepackage{appendix}
\usepackage{epsfig}
\usepackage{dcolumn}
\usepackage{bbm}
\usepackage{enumitem}
\usepackage{cancel}

\def\Tr{\mathrm{Tr}}

\def\pdagger{{\phantom{\dagger}}}

\newcommand{\hop}{J}

\newcommand{\DOS}{D}

\newcommand{\sign}{\mathrm{sgn}}
\newcommand{\identity}{1}

\newcommand{\TCL}{{STCL}}
\newcommand{\convolutionless}{steady-state  time-convolutionless }
\newcommand{\Convolutionless}{Steady-state time-convolutionless  }

\newcommand{\rs}{\rm \scriptscriptstyle}

\newcommand{\rsR}{{\rm \scriptscriptstyle R}}

\newcommand{\sys}{\mathrm{sys}}
\newcommand{\env}{\mathrm{env}}
\newcommand{\reg}{\mathrm{loc}}

\newcommand{\CG}{\mathcal{G}}

\newcommand{\CA}{\mathcal{A}}
\newcommand{\CB}{\mathcal{B}}
\newcommand{\CZ}{\mathcal{Z}}

\newcommand{\CU}{\mathcal{U}}
\newcommand{\CS}{\mathcal{S}}
\newcommand{\CR}{\mathcal{R}}
\newcommand{\CGL}{\dot{\CG}}
\newcommand{\CGH}{\CG '}

\newcommand{\kernel}{\mathcal{K}}
\newcommand{\projector}{\mathcal{P}}

\newcommand{\rhop}{\rho_\mathrm{p}}
\newcommand{\rhopb}{\bar{\rho}_\mathrm{p}}
\newcommand{\rhopt}{\rho_\mathrm{\tilde{p}}}

\newcommand{\liouvillian}{\mathcal{L}}
\newcommand{\CUP}{\CU_\projector}

\newcommand{\timeOrder}{\mathcal{T}}
\newcommand{\pt}{\partial_t}
\newcommand{\rhoptb}{\bar{\rho}_{\mathrm{\tilde{p}}}}
\newcommand{\spectral}{C}
\newcommand{\Qr}{Q_\lambda}
\newcommand{\Nr}{N_\lambda}
\newcommand{\Ir}{I_\lambda}

\newcommand{\bigO}{\mathcal{O}}

\usepackage{tikz}
\usepackage{etoolbox}

\BeforeBeginEnvironment{tabular}{\tikzpicture\node[rounded corners=2mm,draw=black,line width=1pt,inner sep=0pt]\bgroup}

\AfterEndEnvironment{tabular}{\egroup;\endtikzpicture}

\begin{document}
\title{Open quantum systems beyond Fermi's golden rule:\\ Diagrammatic expansion of the \convolutionless master equation} 
\author{Michael Sven Ferguson} 
\affiliation{Institute for theoretical physics, ETH Zurich, 8093 Zurich, Switzerland} 
\author{Oded Zilberberg} 
\affiliation{Institute for theoretical physics, ETH Zurich, 8093 Zurich, Switzerland} 
\author{Gianni Blatter} 
\affiliation{Institute for theoretical physics, ETH Zurich, 8093 Zurich, Switzerland} 
\date{\today} 
\begin{abstract}
	Steady-state observables, such as occupation numbers and currents, are crucial experimental signatures in open quantum systems.
	The time-convolutionless (TCL) master equation, which is both exact and time-local, is an ideal candidate for the perturbative computation of such observables.
	We develop a diagrammatic approach to evaluate the steady-state TCL generator based on operators rather than superoperators.
	We obtain the steady-state occupation numbers, extend our formulation to the calculation of currents, and provide a simple physical interpretation of the diagrams.
	We further benchmark our method on a single non-interacting level coupled to Fermi reservoirs, where we recover the exact expansion to next-to-leading order.
	The low number of diagrams appearing in our formulation makes the extension to higher orders accessible.
	Combined, these properties make the steady-state time-convolutionless master equation an effective tool for the calculation of steady-state properties in open quantum systems.
\end{abstract}

\maketitle 

\section{Introduction}
\label{sec:introduction}  %
Open quantum systems constitute a wide research area that permeates both fundamental and applied physics~\cite{Breuer2002TheoryOpenQuantum,Rivas2012OpenQuantumSystems}.
Specific examples include transport phenomena in semiconductor devices~\cite{Beenakker1991TheoryCoulombblockadeOscillations,Hanson2007SpinsFewelectronQuantum}, quantum simulations based on cold atom experiments~\cite{Diehl2008QuantumStatesPhases,Bloch2008ManybodyPhysicsUltracold,Gross2017QuantumSimulationsUltracold}, as well as quantum information processing with trapped ions~\cite{Wineland1998ExperimentalIssuesCoherent,Bruzewicz2019TrappedIonQuantumComputing} or with superconducting circuits~\cite{Devoret2004SuperconductingQubitsShort,Kjaergaard2020SuperconductingQubitsCurrent}.
Given the recent developments in quantum technologies, such systems promise great advances in computing~\cite{Steane1997IonTrapQuantum,Gyongyosi2019SurveyQuantumComputing}, simulation~\cite{Georgescu2014QuantumSimulation}, and sensing~\cite{Neumann1955MathematicalFoundationsQuantum,Gurvitz1997MeasurementsNoninvasiveDetector,Degen2017QuantumSensing}.
Regardless of the specific realisation, open systems can all be broadly described as containing a small system of central interest that is coupled to a large environment. 
The presence of the environment can fundamentally change the dynamics of the system~\cite{Sieberer2016KeldyshFieldTheory,Ferguson2020QuantumMeasurementInduces}, while leaving it sufficiently coherent for quantum effects to be crucial in explaining its behaviour. 
In electronic mesoscopic transport several phenomena, such as cotunneling~\cite{Bruus2004ManyBodyQuantumTheory,Ihn2009SemiconductorNanostructuresQuantum,Ferguson2017LongrangeSpinCoherence} or the Kondo effect~\cite{Ihn2009SemiconductorNanostructuresQuantum,Bruus2004ManyBodyQuantumTheory,Kondo1964ResistanceMinimumDilute,Ferguson2017LongrangeSpinCoherence}, fall into this category.
The standard way to cope with open systems is to construct the \textit{effective} dynamics of the system by integrating out the environmental degrees of freedom.
A common \textit{phenomenological} approach to do so is based on the Lindblad master equation~\cite{Breuer2002TheoryOpenQuantum}, where the most general and physically-allowed evolution of the system is parametrised and then constrained by physical assumptions and experimental data.
Conversely, bottom-up methods start from a \textit{microscopic} description of the entire setup including the system, the environment, and their coupling~\cite{ Bethe1931TheoryMetalsEigenvalues, Nakajima1958QuantumTheoryTransport, Zwanzig1960EnsembleMethodTheory, Bruus2004ManyBodyQuantumTheory, Bulla2008NumericalRenormalizationGroup}.
This latter approach offers greater predictive power by reducing the number of (or even eliminating the need for) fitting parameters~\cite{ Bruus2004ManyBodyQuantumTheory}.
Furthermore, a microscopic description can be readily extended to include higher-order effects.
In setting up such a microscopic formalism, several assumptions have to be made about the environment's state and its coupling to the system.
Commonly, we assume an environment that is equilibrated and decoupled from the system in the far distant past~\cite{Bruus2004ManyBodyQuantumTheory}.
The coupling between the system and the environment then involves a slow switch-on.
Technically, this is done with the introduction of a switch-on rate~\(\eta/\hbar\) that defines the time scale over which the system and environment are coupled.
Such a slow switch-on appears, explicitly or implicitly, in a range of methods, from the functional renormalisation group~\cite{Berges2002NonperturbativeRenormalizationFlow} to simpler perturbative master equations~\cite{Breuer2002TheoryOpenQuantum,Bruus2004ManyBodyQuantumTheory,Timm2008TunnelingMoleculesQuantum}.
The latter can be broadly categorised into three families:
(i) Formally exact \emph{time-non-local} methods, such as the equivalent real-time-diagrammatic (RT) method~\cite{Schoeller1994MesoscopicQuantumTransport,Schoeller1994ResonantTunnelingCharge,Konig1995ResonantTunnelingCoulomb,Konig1996ResonantTunnelingUltrasmall,Konig1995ResonantTunnelingCharging,Konig1996ZeroBiasAnomaliesBosonAssisted,Konig1997CotunnelingResonanceSingleElectron,Konig1999QuantumFluctuationsSingleElectron}, Nakajima-Zwanzig (NZ) master equation~\cite{Nakajima1958QuantumTheoryTransport,Zwanzig1960EnsembleMethodTheory},  and Bloch-Redfield (BR) master equation~\cite{Wangsness1953DynamicalTheoryNuclear,Bloch1957GeneralizedTheoryRelaxation,Redfield1965TheoryRelaxationProcesses}. (ii) Formally exact \emph{time-local} master approaches such as the time-convolutionless master equation (TCL)~\cite{Tokuyama1975StatisticalMechanicalApproachRandom, Tokuyama1976StatisticalMechanicalTheoryRandom, Breuer2001TimeConvolutionlessProjectionOperator}.
(iii) Approximate methods, that include approximations on top of (i) and (ii), in particular Fermi's golden rule~\cite{Bruus2004ManyBodyQuantumTheory} and the T-matrix master equation~\cite{ Averin1994PeriodicConductanceOscillations,Turek2002CotunnelingThermopowerSingle,Koch2004ThermopowerSinglemoleculeDevices}.
The T-matrix approach is often used to generalise Fermi's golden rule~\cite{Bruus2004ManyBodyQuantumTheory}, however, it predicts unphysical divergences in the switch-on rate~\(\hbar/\eta\) due to the time non-local nature of the rates~\cite{Timm2008TunnelingMoleculesQuantum}.
Deep in the perturbative regime, it has been shown that a physically motivated regularisation scheme for the T-matrix~\cite{Averin1994PeriodicConductanceOscillations,Turek2002CotunnelingThermopowerSingle,Koch2004ThermopowerSinglemoleculeDevices} becomes an acceptable approximation when computing currents, but not occupation probabilities~\cite{Koller2010DensityoperatorApproachesTransport, FergusonThesis}.
On the other hand, the RT or NZ master equation is a \textit{time non-local} method, that naturally avoids divergences in~\(\eta\)~\cite{ Schoeller1994MesoscopicQuantumTransport, Schoeller1994ResonantTunnelingCharge}.
The TCL master equation provides a further formally exact description, naturally free of divergences, and produces a conceptually simpler {\it time-local} master equation~\cite{ Tokuyama1975StatisticalMechanicalApproachRandom, Tokuyama1976StatisticalMechanicalTheoryRandom, Breuer2001TimeConvolutionlessProjectionOperator, Nestmann2020HowQuantumEvolution}.
Furthermore, the TCL has recently been combined with the slow switch-on approximation~\cite{Timm2011TimeconvolutionlessMasterEquation, Nestmann2019TimeconvolutionlessMasterEquation, Nestmann2020HowQuantumEvolution} such that it can directly be used to compute a perturbative expansion of the steady-state; a development which we call the steady-state time convolutionless {\TCL} master equation.
In this work, we focus on the {\TCL} master equation approach and demonstrate that it serves as a practical tool to compute the steady-state of open quantum systems.
We provide a brief overview of current approaches to open systems dynamics and highlight the merits of using the {\TCL}.
We then develop a diagrammatic approach to compute the {\TCL} generator and perform the expansion explicitly up to fourth-order for quadratic environments.
For practical applications, we extend the {\TCL} to the calculation of currents and again perform the expansion explicitly to fourth-order for quadratic environments.
We demonstrate the implementation of our formalism on a non-interacting setup that serves as a test bed.
%

%
The paper is structured as follows: in Section~\ref{sec:results}, we briefly highlight and discuss the main results of this work.
In Section~\ref{sec:background}, we review the state of the art in the field.
We introduce the T-matrix approach in both the usual operator formalism and in terms of superoperators.
The real-time-diagrammatic and \convolutionless master equations then are directly formulated in the superoperator language.
En route, we show that the {\TCL}, which relies on a switch-on process in the distant past, is suitable to compute the steady-state, order by order, whereas it cannot directly be used to compute dynamics without further assumptions.
In Section~\ref{sec:formal}, we develop a diagrammatic formulation of the {\TCL} generator~\(\CS\).
We use both the operator and superoperator formalisms to minimise the complexity of the diagrams.
In Section~\ref{sec:quadratic}, we apply the {\TCL} master equation to setups with quadratic environments and take advantage of Wick's theorem~\cite{Wick1950EvaluationCollisionMatrix}.
We then show how the {\TCL} recovers exact results for the occupation numbers in a non-interacting setup.
In Section~\ref{sec:currents}, we extend the {\TCL} to compute currents flowing through the system in steady-state, and again show that we recover exact results for the currents in a non-interacting setup.
Finally, in Section~\ref{sec:conclusion}, we summarise the results of our work and give an outlook on future applications for the {\TCL} master equation.

\section{Main results} \label{sec:results} 
We start with a short tour through our main results and discuss their implications.
The goal of the present work is to develop a practical, though exact at every order, method to calculate steady-state probability distributions and transport currents in driven open quantum systems, see e.g., Fig.~\ref{Fig:openSystem}, where we illustrate the electronic mesoscopic setup serving as our physical motivation.
We achieve this goal with the help of the \convolutionless master equation~\cite{ Tokuyama1975StatisticalMechanicalApproachRandom, Tokuyama1976StatisticalMechanicalTheoryRandom, Timm2011TimeconvolutionlessMasterEquation, Nestmann2019TimeconvolutionlessMasterEquation}
\begin{align}
\label{eq:TCLMEv0}
\pt{\rhop}(t) = -\frac{i}{\hbar}\liouvillian_0\rhop(t)
+\CS(t,\eta)\rhop(t),
\end{align}
for the (projected) density matrix $\rhop(t)$, with $\liouvillian_0$ the Liouvillian of the uncoupled system--environment setup, $\CS$ the {\TCL} generator and~\(\eta\to 0\) a slow switch on rate, see the discussion around Eq.~\eqref{eq:TCLME}.
Specifically, we perform three steps: 

\begin{figure}
	\includegraphics[width=8.6cm]{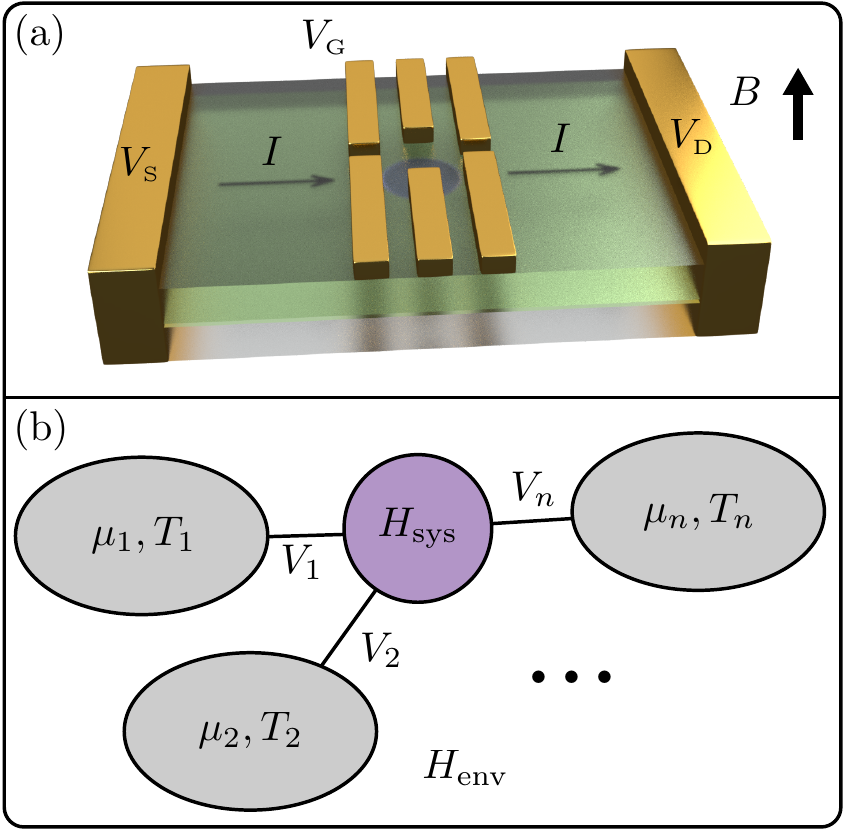}
	\caption{
		\label{Fig:openSystem}
		Open quantum system, device schematic and model.
		(a) Sketch of a gate-defined quantum dot device.
		The semiconductor heterostructure defines a two-dimensional electron gas (2DEG) at the boundary (green) between two semiconductor layers (transparent)~\cite{Ihn2009SemiconductorNanostructuresQuantum}.
		The gold top-gates (centre, biased with a set of gate voltages~\(V_{{\rs G}}\)) deplete the 2DEG and form a quantum dot (purple), which is coupled to metallic leads on each side.
		These are, in turn, contacted with wires (gold, sides) that impose a bias voltage (source~\(V_{{\rs S}}\) and drain~\(V_{{\rs D}}\)) across the device.
		The latter drives the measurable current~\(I\) through the dot.
		A magnetic field~\(B\) can be applied to the entire setup making the setup spinless for large fields.
		(b) Generic setup for an open quantum system out of equilibrium, see Eqs.~\eqref{eq:H_0} and~\eqref{eq:V}.
		A system (purple disc) is coupled (via~\(V_r\)) to the~\(n\) different reservoirs (grey ovals) constituting the environment.
		Each individual reservoir~\(r\) is at equilibrium with a temperature~\(T_r\) and a chemical potential~\(\mu_r\).
		The environment is usually assumed to be out-of-equilibrium, e.g.,~\(\mu_r \neq \mu_{r'}\) and/or~\(T_r \neq T_{r'}\).
	}
\end{figure}
(i) We construct a set of diagrammatic rules to compute the {\TCL} generator~\(\CS\) for arbitrary system--environment setups, order by order in the system--environment coupling~\(V\).
The results are found in Eqs.~\eqref{eq:simpleSuperS}--\eqref{eq:superGdot} describing the recursive expansion of the {\TCL} generator~\(\CS\) in terms of the propagator~\(\CG\) presented in Eq.~\eqref{eq:superMatrixElements}.
The latter involves the expansion of the evolution operator in Eq.~\eqref{eq:recursiveU}.
The corresponding diagrammatic representation is depicted in Figs.~\ref{Fig:genericDiagram},~\ref{Fig:compositionRule}, and~\ref{Fig:genericSDiagram}.
(ii) We apply the diagrammatic expansion to setups with quadratic environments, where Wick's theorem greatly simplifies the calculations, and use the corresponding diagrammatic formulation from Figs.~\ref{Fig:specificDiagram} and~\ref{Fig:WickTheorem} to obtain explicit fourth-order rates for the \TCL.
These rates are reported in Eqs.~\eqref{eq:S22a},~\eqref{eq:S22b}--\eqref{eq:S22c}, and~\eqref{eq:S31a}--\eqref{eq:S31c}.
In Fig.~\ref{Fig:nonInteracting}, we highlight the result of computing these rates for a non-interacting setup and then calculating the steady state occupation $P_1$ of a single-level in a non-interacting quantum-dot at equilibrium.
We compare this analysis (up to next-to-leading order) to the expansion of the exact result and find perfect agreement.
(iii) We extend the {\TCL} to compute steady-state currents for setups with quadratic environments.
We modify the diagrammatic formulation to include so-called current rates.
This provides us with the results in Eq.~\eqref{eq:currentTCLdef} for the currents as expressed through the current generator~\eqref{eq:currentS}, and the charge transfer propagator~\eqref{eq:superGlambda}, see also Fig.~\ref{Fig:currentWick}.
In Fig.~\ref{Fig:nonInteractingCurrent}, we again show up to fourth-order, that these results recover the exact results for a non-interacting single-level setup.
The diagrammatic expansion presented here can be systematically expanded beyond fourth-order, e.g., we predict that only three times more diagrams appear at sixth order as compared with what we have computed at fourth-order.
There are several effects, such as measurement backaction~\cite{Zilberberg2014MeasuringCotunnelingIts} and the Kondo effect~\cite{Bruus2004ManyBodyQuantumTheory}, that manifest at sixth order, motivating further development of our diagrammatic scheme.
Lastly, we highlight that the {\TCL} lends itself to resummation schemes, similar in spirit to those that have already been developed for the RT method~\cite{ Schoeller1994MesoscopicQuantumTransport, Schoeller1994ResonantTunnelingCharge,Konig1995ResonantTunnelingCoulomb, Konig1997CotunnelingResonanceSingleElectron}, see Ref.~\cite{FergusonThesis} for a brief discussion.
\section{Background} \label{sec:background} 
We consider a setup composed of a system and an environment as shown in Fig.~\ref{Fig:openSystem}, and compute its steady-state properties.
To this end, we focus on the time evolution of the system after a slow switch-on using a master equation.
Expanding the rates that govern this evolution order-by-order in the system--environment coupling~\(V\), we obtain a power series for the steady-state properties of interest.
In this section, we provide an introduction to the T-matrix~\cite{Bruus2004ManyBodyQuantumTheory} and \convolutionless~\cite{ Tokuyama1975StatisticalMechanicalApproachRandom, Tokuyama1976StatisticalMechanicalTheoryRandom, Breuer2001TimeConvolutionlessProjectionOperator, Nestmann2019TimeconvolutionlessMasterEquation} master equations.
Concurrently, we comment on the relationship between these methods and the real-time diagrammatic approach~\cite{ Nakajima1958QuantumTheoryTransport, Zwanzig1960EnsembleMethodTheory, Breuer2002TheoryOpenQuantum, Schoeller1994MesoscopicQuantumTransport, Konig1995ResonantTunnelingCharging,Konig1995ResonantTunnelingCoulomb,Schoeller1994ResonantTunnelingCharge,Konig1996ResonantTunnelingUltrasmall}.

\subsection{General properties} 
\label{sec:background:basics} 
The system is assumed to be small, i.e., it is described by a finite Hamiltonian~\(H_\sys\) which we can solve exactly, e.g., via numerical techniques, providing eigenstates~\(\ket{i}_\sys\) and eigenenergies~\(\chi_i\).
The environment may be large/infinite, is described by the Hamiltonian~\(H_\env\), and is commonly assumed to be non-interacting such that it is exactly solvable.
It has eigenstates~\(\ket{u}_\env\) with eigenenergies~\(\xi_u\).
The combined unperturbed Hamiltonian describing the decoupled system and environment reads 
\begin{align}
\label{eq:H_0}
H_0 = \identity_\env \otimes H_\sys + H_\env\otimes \identity_\sys,
\end{align}
with eigenstates~\(\ket{u,i} = \ket{u}_\env\otimes \ket{i}_\sys\) and eigenenergies~\(\xi_u + \chi_i\).
Throughout this work, we omit system and environment subscripts at times when it improves readability (we use indices~\(i,j,f,g,n,m\) for system states whereas~\(u,v\) are reserved for environment states).
The environment is composed of multiple reservoirs~\(r\) that are individually at equilibrium (with corresponding temperatures~\(T_r\) and chemical potentials~\(\mu_r\)) but mutually out-of-equilibrium.
Each reservoir may be composed of fermions or bosons (or particles with exotic statistics).
They are coupled to the system by individual perturbations~\(V_r\), the sum of which makes up the total perturbing Hamiltonian 
\begin{align}
\label{eq:V}
V = \sum_r V_r.
\end{align}
The total Hamiltonian 
\(
H = H_0 + V
\)
which governs the physics of the full setup is the sum of the unperturbed part and the system--environment coupling.

\subsubsection{Unitary time evolution}

\begin{figure}
	\includegraphics[width=8.6cm]{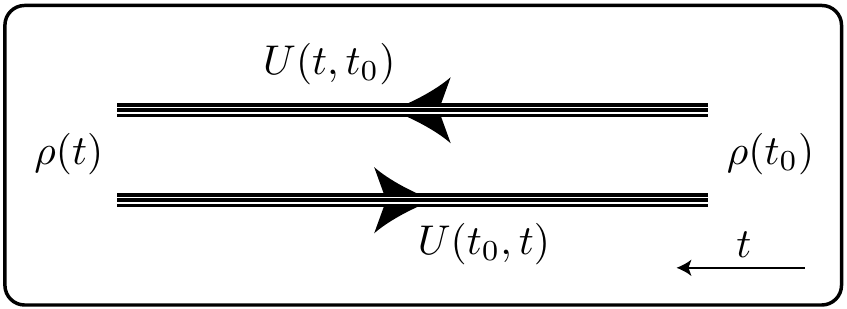}
	\caption{
		\label{Fig:Keldysh}
		The Keldysh contour; graphical representation of Eq.~\eqref{eq:urhou}.
		The density matrix~\(\rho(t_0)\) of the entire setup at time~\(t_0\) is evolved to a later (leftward) time~\(t\) by unitary time-evolution operators (triple line) for the coupled system--environment setup on each of the Keldysh branches.
		Note that, on the lower branch, the evolution is given by~\(U(t_0,t)=U^\dagger(t,t_0)\), corresponding to the different directions of propagation on the two branches.
		%
	}
\end{figure}

The state of the full setup is described by the density matrix~\(\rho\), which evolves according to the von Neumann equation~\cite{Neumann1955MathematicalFoundationsQuantum} 
\begin{align}
\label{eq:vonNeumann}
\pt{\rho}= -\frac{i}{\hbar}[H(t),\rho].
\end{align}
Here, we allow the Hamiltonian~\(H\) to be time dependent to account for the switch-on of the perturbation~\(V\).
Making use of the unitary time-evolution operator 
\begin{align}
\label{eq:unitary}
U(t,t_0) = \timeOrder \exp[\frac{-i}{\hbar}\int_{t_0}^{t} dt' H(t')],
\end{align}
with the time-ordering operator~\(\mathcal{T}\), the differential equation~\eqref{eq:vonNeumann} can be formally integrated, 
\begin{align}
\label{eq:urhou}
\rho(t) = U(t,t_0)\rho(t_0)U^\dagger(t,t_0).
\end{align}
With the density matrix composed of elements~\(\propto \ketbra{u,i}{v,j}\), the forward-evolution~\(U\) acts on the density matrix~\(\rho(t_0)\) from the left and propagates a state~\(\ket{u,i}\) from time~\(t_0\) to~\(t\); it is commonly referred to as the forward Keldysh branch of the time evolution.
Concurrently, the backward evolution~\(U^\dagger\) acts on~\(\rho(t_0)\) from the right and propagates a conjugate state~\(\bra{v,j}\) forward in time, commonly referred to as the backward Keldysh branch.
These two branches, shown pictorially in Fig.~\ref{Fig:Keldysh}, form the basis for our diagrammatic representation in Section~\ref{sec:formal}.
Combining the von Neumann equation~\eqref{eq:vonNeumann} with an initial condition~\(\rho(t_0)\) at a time~\(t_0\) fully specifies the density matrix of the system--environment setup at a later time~\(t\).

\subsubsection{Distant past}

We assume that in the distant past~\(t_0\to-\infty\), the system and environment were decoupled.
Such an initial condition implies that the density matrix 
\begin{align}
\label{eq:factorise}
\rho(t_0) =\rho_\env^0 \otimes \rho_\sys(t_0),
\end{align} 
at time~\(t_0\) can be decomposed into the product of the system's density matrix~\(\rho_\sys(t_0)\) and a locally-equilibrated distribution
of the environment 
\begin{align}
\label{eq:rho0env}
\rho_\env^0 \equiv \sum_{u}P_u^{\env}\ketbra{u}{u}_\env.
\end{align}
Here, the probability distribution~\(P_u^\env\) describes the thermal-equilibrium configuration of each reservoir in the environment.
The large size of the environment makes the direct calculation of the density matrix~\(\rho(t)\) from the initial condition~\eqref{eq:factorise} and the von Neumann equation~\eqref{eq:vonNeumann} intractable.
To tackle this problem, two steps are commonly applied: (i) expand the time evolution operator~\eqref{eq:unitary} as a perturbation series in~\(V\), and (ii) integrate out the environment dynamics to be left with only the system behaviour.
As we will see in Eqs.~\eqref{eq:recursiveU} and~\eqref{eq:freePropagator}, the expansion is incompatible with the~\(t_0\to-\infty\) limit for a time-independent perturbation~\(V\).
To circumvent this problem, we introduce the slow switch-on assumption 
\begin{align}
\label{eq:slow}
V\to \lim_{\eta\to0} e^{\eta t/\hbar}V,
\end{align}
where~\(\eta/\hbar\) is an infinitesimal switch-on rate.
The system and environment are in contact for an effective timescale~\(\hbar/\eta\to\infty\) before the present time~\(t= 0\), and thus if a steady-state exists it will have been reached.

\subsubsection{Expanding~\(U\)}

The expansion of the time evolution operator~\eqref{eq:unitary} can be written in the form 
\begin{align}
\label{eq:Uexpansion}
U=\sum_{\nu=0}^\infty U_\nu,
\end{align}
where the terms~\(U_\nu\) of order~\(V^\nu\) can be found using the recurrence relation 
\begin{align}
\label{eq:unitaryRecurrence}
U_\nu(t,t_0) = \frac{-i}{\hbar} \int_{t_0}^{t} 
\dd{t_1} U_0(t,t_1) V e^{\eta t_1 /\hbar}U_{\nu-1}(t_1,t_0),
\end{align}
with the first term~\(U_0\) specified by the free evolution of the unperturbed Hamiltonian~\(H_0\) in Eq.~\eqref{eq:H_0}, 
\begin{align}
U_0(t,t_0)=\exp[\frac{-i}{\hbar}(t-t_0)H_0].
\end{align}
We perform the integral in the recurrence relation~\eqref{eq:unitaryRecurrence}, keeping a finite switch-on rate~\(\hbar/\eta\) and assuming~\(t_0\to-\infty\) such that 
\begin{align}
\label{eq:timescales}
|t_0|\gg\hbar/\eta.
\end{align}
This relation is assumed to remain true when taking the second limit~\(\eta\to 0\) later on.
We work in the eigenbasis~\(\ket{u,i}\) of the unperturbed Hamiltonian~\(H_0\) and repeatedly insert identity operators to obtain 
\begin{align}
\label{eq:recursiveU}
\displaystyle U_\nu(t,\eta) \ket{u,i}=\!\frac {e^{\eta t /\hbar}}
{\chi_i\!+\!\xi_u\!-\!H_0\!+\!i\nu\eta}VU_{\nu-1}\ket{u,i},
\end{align}
where~\(U_\nu\) now depends on~\(t\) and~\(\eta\) (but not on~\(t_0\to-\infty\)) due to the slow switch-on.
There is no clear notion of intermediate times in Eq.~\eqref{eq:recursiveU}, however a specific operator ordering for the coupling events~\(V\) is inherited from the time-ordering in Eq.~\eqref{eq:unitaryRecurrence}.
Note that each time the recurrence~\eqref{eq:recursiveU} is applied, an additional factor of~\(\exp(\eta t/\hbar)\)  appears, which is the reason for the~\(\nu\)-fold enhancement of~\(\eta\)  in the denominator.
Furthermore, the rightmost operator in the recurrence is always an unperturbed time-evolution~\(U_0\).
The recurrence relation~\eqref{eq:recursiveU} is undefined at~\(\eta=0\) as the real part of the denominator of the free propagator 
\begin{align}
\label{eq:freePropagator}
\Pi_0^\pm(\omega,\nu \eta) = 
\frac{1}{\omega-H_0\pm i\nu \eta},
\end{align} 
vanishes for states~\(\ket{u,i}\) that fulfil~\(\chi_i+\xi_u=\omega\).
The resulting divergences motivate the introduction  of the slow switch-on~\eqref{eq:slow} with a finite~\(\eta>0\).
Making use of the perturbative expansion~\eqref{eq:Uexpansion}, we can now integrate out the environment.
The T-matrix, usually familiar from scattering theory~\cite{Bruus2004ManyBodyQuantumTheory}, provides a standard approach to achieve this goal:
It is commonly used to construct a rate equation for the {\it probabilities}~\(P_i\) of finding the system in a state~\(\ket{i}\), i.e., the {\it diagonal} elements of the density matrix.
Such a rate equation is also known as a Pauli master equation, and can be used to study {\it relaxation}, i.e., the time scales on which the system decoheres due to transitions between different states~\cite{Breuer2002TheoryOpenQuantum,Bruus2004ManyBodyQuantumTheory}.
In contrast, a full master equation can be used further to study {\it dephasing}, where the system state is preserved but the {\it off-diagonal} elements of the density matrix decay~\cite{Breuer2002TheoryOpenQuantum}.
Next, we sketch the derivation of the Pauli T-matrix rate equation as it will allow us to effectively highlight its pitfalls.
\subsection{Fermi's golden rule and the T-matrix}
\label{sec:background:TMat}
Let us consider the probability~\(P_{f|i}^\sys\) of finding the system in a state~\(\ket{f}\) at time~\(t\), given that in a far distant past the system was in an initial state~\(\ket{i}\).
The environment is in equilibrium at~\(t_0\) and its state at time~\(t\) is irrelevant.
To compute the probability~\(P_{f|i}^\sys\), we evolve the state~\(\ket{u,i}\) from~\(t_0\) to~\(t\), square its overlap with a final state~\(\ket{v,f}\), and sum over the environment states~\(u,v\), weighted by the distribution~\(P_u^\env\) of initial environment states.
This procedure propagates the system probabilities according to 
\begin{align}
\label{eq:Pif}
P_{f|i}^\sys = \sum_{u,v}P_u^\env|\bra{v,f}U(t,t_0)\ket{u,i}|^2,
\end{align}
and only depends on the original environment distribution, implying that we have integrated out the effect of the environment at all later times.
In order to obtain the T-matrix rate equation, we insert the expansion~\eqref{eq:recursiveU} into the expression~\eqref{eq:Pif} for the propagated system probability.
We differentiate the result with respect to time~\(t\) which we set to~\(t=0\) without loss of generality.
Taking the limit~\(\eta\to 0\), we obtain the transition rate 
\begin{align}
\label{eq:TMatrixRates}
\Gamma^{if}_{\rs T} = \sum_{u,v}\delta(\delta\chi_{if}
+\delta\xi_{uv})|\bra{v,f}T_M\ket{u,i}|^2P_u^\env,
\end{align}
where we have introduced~\(\delta\chi_{if}\equiv\chi_i-\chi_f\) and~\(\delta\xi_{uv}=\xi_u-\xi_v\) for compactness.
The delta function in Eq.~\eqref{eq:TMatrixRates} ensures energy conservation and arises from two occurrences of the leftmost denominator in Eq.~\eqref{eq:recursiveU} due to the two occurrences of~\(U\) in Eq.~\eqref{eq:Pif}.
Furthermore, we have introduced the T-matrix~\(T_M\) that is defined by the self-consistency relation 
\begin{align}
\label{eq:TMatrix}
T_M\ket{u,i} = V\ket{u,i}
+V\frac{1} {\chi_i\!+\!\xi_u\!-\!H_0\!+\!i0^+}T_M\ket{u,i},
\end{align}
with~\(0^+\) a positive infinitesimal.
Note that in the T-matrix rate equation the infinitesimals~\(0^+\) in each denominator are usually all taken equal, which is justified if~\(\Gamma_{\rs T}\) is convergent in the~\(\eta\to 0\) limit.
However, this convergence is not guaranteed and thus the prefactor~\(\nu\) of~\(\eta\) in Eq.~\eqref{eq:recursiveU} is a crucial ingredient that must be tracked in an exact method.
At lowest order~\(T_M\approx V\) and the rates~\eqref{eq:TMatrixRates} are identical to Fermi's golden gule.
At higher orders in~\(V\), the rates~\eqref{eq:TMatrixRates} are divergent in the limit~\(\eta \to 0\).
The rate matrix~\(\Gamma_{\rs T}\) and its elements~\(\Gamma_{\rs T}^{if}\) relate the change in the probability~\(P^\sys_f(0)\) (to find the system in state~\(\ket{f}\) at time~\(t=0\)) to the corresponding probabilities~\(P_i^\sys(-\infty)\) at time~\(t_0=-\infty\) 
\begin{align}
\label{eq:Trateeq}
\pt P_f^\sys(t=0) = \sum_{i}\Gamma^{if}_{\rs T} P_i^\sys(-\infty),
\end{align}
where~\(\Gamma^{ii}_{\rs T} = -\sum_f \Gamma^{if}_{\rs T}\) such that the total probability is conserved.
From here onwards we shall drop the "\(\sys\)" label on the system probabilities for brevity, while the "\(\env\)" label will remain to avoid ambiguity.
Equation~\eqref{eq:Trateeq} with its (time non-local) rates~\(\Gamma^{if}_{\rs T}\) does not generate a time-local rate equation, as pointed out in Ref.~\cite{Timm2008TunnelingMoleculesQuantum}.
The equivalence at lowest order between the T-matrix rates and Fermi's golden rule means that it is tempting to use the T-matrix rate equation~\eqref{eq:Trateeq}  as a higher-order generalisation of Fermi's golden rule~\cite{Bruus2004ManyBodyQuantumTheory}.
To do so, Eq.~\eqref{eq:Trateeq} is turned into an approximate time-local master equation
\begin{align} \label{eq:TrateeqApprox} 
\pt P_f(t) \approx \sum_{i}\Gamma^{if}_{\rs T} P_i(t), 
\end{align}
which at lowest non-vanishing order in~\(V\) is identical to Fermi's golden rule.
This path is fraught with difficulties though, even for the seemingly simple task of calculating the steady-state distribution of the system.
The standard procedure~\cite{Timm2008TunnelingMoleculesQuantum,Bruus2004ManyBodyQuantumTheory, Turek2002CotunnelingThermopowerSingle,Averin1994PeriodicConductanceOscillations,Koch2004ThermopowerSinglemoleculeDevices} to do this is to use~\eqref{eq:TrateeqApprox} and assume that the system has reached the steady state at~\(t=0\) such that 
\begin{align}
\label{eq:Tss}
\sum_{n}\Gamma^{nm}_{\rs T} P_n(0)= 0.
\end{align}
It is then claimed that solving for~\(P_n(0)\) in Eq.~\eqref{eq:Tss} gives us the steady-state system-distribution.
However, we have in fact solved for the distribution of system probabilities at~\(t_0=-\infty\) that leads to the steady-state at time~\(t=0\), cf. Eqs.~\eqref{eq:Trateeq} and~\eqref{eq:Tss} or see Ref.~\cite{Timm2008TunnelingMoleculesQuantum}.
Furthermore, in the limit~\(\eta \to 0\), every distribution at~\(t_0=-\infty\) leads to the steady-state at~\(t=0\), as an infinite amount of time has elapsed, and the constraint~\eqref{eq:Tss} becomes ill-defined.
This manifests as divergences in the T-matrix rates~\(\Gamma_{\rs T}\), which signal the breakdown in the approximation~\eqref{eq:TrateeqApprox}.
Consequently, physically motivated regularisation schemes for~\(\Gamma_{\rs T}\) have been developed~\cite{Averin1994PeriodicConductanceOscillations,Turek2002CotunnelingThermopowerSingle,Koch2004ThermopowerSinglemoleculeDevices} that remove the divergences, but produce results which differ from the exact expansion~\cite{Begemann2010InelasticCotunnelingQuantum}.
The {\TCL}, on the other hand, takes the step from Eq.~\eqref{eq:Trateeq} to~\eqref{eq:TrateeqApprox} in a rigorous manner, which is why we choose to use this approach in our present work.
\subsection{Superoperator formulation} \label{sec:background:super} 
Next, we introduce the mathematical tools, specifically the superoperators, which facilitate the calculations of the various master equations~\cite{Breuer2002TheoryOpenQuantum}.
We then briefly rederive the T-matrix in this more formal language and then provide a simplified derivation of the {\TCL} inspired by Refs.~\cite{Timm2011TimeconvolutionlessMasterEquation, Nestmann2019TimeconvolutionlessMasterEquation,Nestmann2020HowQuantumEvolution}.
Additionally, we discuss the relationships between the T-matrix, the {\TCL}, and the RT approaches.

\subsubsection{Setup evolution}

We start with the von Neumann equation~\eqref{eq:vonNeumann} in superoperator notation 
\begin{align}
\label{eq:vonNeumannL}
\pt\rho = -\frac{i}{\hbar}\liouvillian(t)\rho,
\end{align}
with the Liouvillian superoperator~\(\liouvillian\rho \equiv [H,\rho]\).
Every Hamiltonian (\(H_0,V\)) is associated with a Liouvillian (\(\liouvillian_0, \liouvillian_V\)) (we denote superoperators by caligraphic upper case letters~\footnote{The time-ordering operator~\(\mathcal{T}\) acts on both operators and superoperators and is thus also caligrahic.
}).
In analogy to~\eqref{eq:urhou}, we formally integrate~\eqref{eq:vonNeumannL} to obtain the time evolution 
\begin{align}
\label{eq:superU}
\rho(t) = \CU(t,t_0) \rho(t_0),
\end{align}
with the time evolution superoperator   
\begin{align}
\label{eq:superUorder}
\CU(t,t_0) = \timeOrder \exp[\frac{-i}{\hbar}
\int_{t_0}^{t} dt' \liouvillian(t')],
\end{align}
which can also be written in terms of the unitary evolution~\(\CU\rho = U\rho  U^\dagger\).
We expand the superoperator~\(\CU=\sum_{\alpha=0}^\infty \CU_\alpha\) in the Liouvillian~\(\liouvillian_V\) (equivalent to an expansion in~\(V\)) and perform the integrals, in the limit~\(t_0\to-\infty\), as for the unitary evolution, see Eqs.~\eqref{eq:unitary} and~\eqref{eq:unitaryRecurrence}--\eqref{eq:recursiveU}.
The resulting recurrence relation takes the form 
\begin{align}
\label{eq:superUexpansion}
\CU_\alpha&\ketbra{u,i}{v,j}= 
\\\nonumber&
\frac{e^{\eta t /\hbar}}
{\delta\chi_{ij}+\delta\xi_{uv}-\liouvillian_0+i\alpha\eta}
\liouvillian_V\CU_{\alpha-1} \ketbra{u,i}{v,j}.
\end{align}
which is similar to the one for the unitary time evolution~\eqref{eq:recursiveU}.
Each time the recurrence is applied an additional factor of~\(\exp(\eta t/\hbar)\) appears, the rightmost superoperator is an unperturbed evolution~\(\CU_0\), and the superoperator ordering is inherited from the time-ordering in Eq.~\eqref{eq:superUorder}.
By expanding the time evolution of the density matrix in the operator-~\eqref{eq:urhou} and superoperator-~\eqref{eq:superU} representations and comparing order by order, we obtain the identity 
\begin{align}
\label{eq:superUaUnumu}
\CU_\alpha\rho = \sum_{\mu+\nu=\alpha} 
U^\pdagger_\nu \,\rho \, U^\dagger_\mu, \qquad \mu,\nu \in [0,...,\alpha].
\end{align}
Henceforth, a sum over~\(\mu+\nu=\alpha\) implies that each of the indices~\(\nu,\mu\) runs from~\(0\) to~\(\alpha\) and in opposite direction for the partner.
Note that, the single term~\(\CU_\alpha\) in Eq.~\eqref{eq:superUexpansion} contains~\(\alpha\) Liouvillians~\(\liouvillian_V\), and hence~\(\alpha\) commutators with~\(V\).
When written explicitly, these commutators lead to~\(2^\alpha\) terms, significantly more than the~\(\alpha+1\) terms on the right hand side of Eq.~\eqref{eq:superUaUnumu}.
We thus conclude that Eq.~\eqref{eq:superUaUnumu} significantly reduces the complexity of computing series expansions of the evolution, and will use this feature in Sec.~\ref{sec:formal}.

\subsubsection{Projected space}

Having specified the evolution in the superoperator language, we proceed by integrating out the environment part of the evolution.
To this end, we define the projector~\(\projector\) through its action on a density matrix 
\begin{align}
\label{eq:superP}
\projector\rho(t) = \rho_\env^0\otimes\Tr_\env[\rho(t)],
\end{align}
where~\(\Tr_\env\) is the partial trace over the environment states~\(\ket{u}_\env\).
In effect, Eq.~\eqref{eq:superP} projects any density matrix to the space of valid initial conditions, such that 
\begin{align}
\projector\rho(t_0) = \rho(t_0).
\end{align}
The projector~\(\projector\) obeys the usual condition~\(\projector^2=\projector\), and it is straightforward to verify that~\([\liouvillian_0,\projector]=0\) as well as
\begin{align}
\label{eq:superTrace}
\Tr[\projector O]= \Tr\, O,
\end{align}
where~\(\Tr\) is the trace over all setup (system and environment) states~\(\ket{u,i}\) and~\(O\) is any setup operator.
We define the projected density matrix 
\begin{align}
\rhop(t) \equiv \projector\rho(t),
\end{align}
which carries only the system degrees-of-freedom but resides in the Hilbert space of the full setup, i.e., we can write
\begin{align}
\rhop(t) = \rho_\env^0\otimes\rho_\sys(t).
\end{align}
At this point, we can write down the projected time-evolution superoperator~\(\CUP = \projector\CU\projector\), which directly propagates the projected density matrix
\begin{align}
\label{eq:superUP}
\rhop(t) = \CUP(t,t_0) \rhop(t_0),
\end{align} 
from a decoupled initial time~\(t_0\) up to time~\(t\).
This projected evolution will be central to the derivation of master equations in the rest of this work.
%

%
To obtain the T-matrix master equation in the superoperator formalism, we differentiate~\eqref{eq:superUP} with respect to time~\(t\) using the identity 
\begin{align}
\label{eq:ptU}
\pt\,\CU(t,t_0) = -\frac{i}{\hbar}\left(\liouvillian_0
+e^{\eta t / \hbar}\liouvillian_{V}\right) \CU(t,t_0),
\end{align}
which follows from the definition of~\(\CU\) in Eq.~\eqref{eq:superUorder}.
We substitute the result into the projected evolution~\eqref{eq:superUP}, make use of the commutator~\([\liouvillian_0,\projector]=0\) and obtain the T-matrix master equation (describing both relaxation and dephasing) in the form 
\begin{align}
\label{eq:TMME}
\pt\rhop(t) = -\frac{i}{\hbar}\liouvillian_0\rho_p(t) + \CR(t,\eta)\rhop(-\infty),
\end{align}
see Fig.~\ref{Fig:masterEquations}.
Here, we have taken the limit~\(t_0\to-\infty\) and introduced the T-matrix generator in its superoperator form 
\begin{align}
\label{eq:superR}
\CR_\alpha(t,\eta)=- \frac{ie^{\eta t/\hbar}}{\hbar} 
\projector\liouvillian_V\CU_{\alpha-1}  \projector,
\end{align}
with~\(\CR=\sum_\alpha\CR_\alpha\).
Its explicit perturbation expansion can be obtained by inserting the expansion for the superoperator~\(\CU\), either from Eq.~\eqref{eq:superUexpansion}, or the one from Eq.~\eqref{eq:superUaUnumu} if the operator representation is preferable.
At higher orders this leads to divergences in~\(\eta\) as for the usual T-matrix formulation, see Section~\ref{sec:background:basics}.

\subsubsection{Pauli projected space}

The master equation~\eqref{eq:TMME} tracks both relaxation and dephasing, i.e., on- and off-diagonal elements of the system density matrix.
On the other hand, Pauli master equations describe only relaxation, and are thus sufficient to calculate occupation probabilities.
They require the introduction of the Pauli projector~\(\tilde{\projector}\), and Pauli projected space~\(\rhopt\), through
\begin{align}
\label{eq:superPauli}
\rhopt \equiv \tilde{\projector}\rho
= \rho_\env^0\otimes\mathrm{Diag}\{\Tr_\env[\rho(t)]\},
\end{align}
where Diag sets all non-diagonal elements of a matrix to zero.
The projector~\(\tilde{\projector}\) thus limits us to track only the system probabilities
\begin{align}
P_n(t) = \bra{n}\Tr_\env[\rho(t)]\ket{n},
\end{align}
which can in turn be used to reconstruct the Pauli projected density matrix
\begin{align}
\rhopt(t) = \rho^0_\env\otimes\sum_n P_n(t) \ketbra{n}.
\end{align}
Assuming that the initial system condition is completely dephased
\begin{align}
\rho_\sys(t_0) = \mathrm{Diag}\rho_\sys(t_0) 
\end{align}
we can make use of the Pauli projected evolution~\(\CU_{\tilde{\projector}} = \tilde{{\projector}} \CU \tilde{\projector} \) to construct Pauli master equations.

As an example, the Pauli T-matrix master equation takes on the form
\begin{align}
\label{eq:TMMEPauli}
\pt \rhopt(t) = \tilde{\CR}(t,\eta) \rhopt(-\infty),
\end{align} 
which is obtained by replacing each occurrence of the projector~\(\projector\) in Eq.~\eqref{eq:TMME} by a Pauli projector~\(\tilde{\projector}\), see Eq.~\eqref{eq:superPauli}.
Furthermore, we have used the property~\(\liouvillian_0\tilde{\projector}=0\), which can easily be verified, and introduced the Pauli T-matrix generator~\(\tilde{\CR}\), which is obtained by replacing~\(\projector \to \tilde{\projector}\) in Eq.~\eqref{eq:superR}.
This superoperator formulation of the T-matrix Pauli master equation is identical to the one in Section~\ref{sec:background:basics}, cf. Eqs.~\eqref{eq:Trateeq} and~\eqref{eq:TMMEPauli} with~\(t=0\).
%

%

%
\subsection{\Convolutionless approach} \label{sec:background:TCL}
The \convolutionless~\cite{Tokuyama1975StatisticalMechanicalApproachRandom, Tokuyama1976StatisticalMechanicalTheoryRandom, Breuer2002TheoryOpenQuantum,Timm2011TimeconvolutionlessMasterEquation, Nestmann2019TimeconvolutionlessMasterEquation, Gu2020DiagrammaticTimelocalMaster, Nestmann2020HowQuantumEvolution} master equation is of the form (see below for the derivation) 
\begin{align}
\label{eq:TCLME}
\pt{\rhop}(t) = -\frac{i}{\hbar}\liouvillian_0\rhop(t)
+\CS(t,\eta)\rhop(t),
\end{align}
with the superoperator~\(\CS(t,\eta)\) depending on~\(t\) and~\(\eta\) but not on~\(t_0\), that has been sent to the distant past,~\(t_0 \to-\infty\).
Its {\it time-local} property naturally sidesteps the issues that appear in the T-matrix approach.
As shown in Refs.~\cite{Timm2011TimeconvolutionlessMasterEquation, Nestmann2019TimeconvolutionlessMasterEquation}, each term~\(\CS_\alpha\) in the {\TCL} series expansion~\(\CS = \sum_{\alpha=1}^{\infty} \CS_\alpha\) is convergent for the setups we consider here and in the limit~\(\eta\to 0\).
The latter limit, in combination with~\(t_0\to-\infty\) implies that the system and environment have been in contact for an infinite amount of time, such that the asymptotic state has been reached.
Specifically, the projected density matrix may display an oscillating behaviour which is basis dependent.

The Pauli projected density matrix, however,  reaches a basis dependent but constant steady-state~\cite{Timm2011TimeconvolutionlessMasterEquation}.
As for the T-matrix scheme, the {\TCL} master equation has a Pauli equivalent 
\begin{align}
\label{eq:TCLMEPauli}
\pt{\rhopt}(t) = \tilde{\CS}(t,\eta)\rhopt(t),
\end{align}
which is of central importance to the present work.
Simultaneously, the Pauli {\TCL} superoperator is uniquely defined by its matrix elements~\(\tilde{\CS}^{if}\), which are the time-local transition rates between system states~\(\ket{i}\) and~\(\ket{f}\).
Written in terms of (diagonal) matrix elements, the Pauli {\TCL} master equation~\eqref{eq:TCLMEPauli} becomes
\begin{align}
\label{eq:TCLMEPauliME}
\pt P_f(t) = \sum_i \tilde{\CS}^{if}(t,\eta)P_i(t),
\end{align}
which we will use for practical calculations in sections~\ref{sec:quadratic} and~\ref{sec:currents}.
%
%
As the Pauli {\TCL}~\eqref{eq:TCLMEPauli} is time local, a steady-state constraint exists, and is obtained by setting the derivative in Eq.~\eqref{eq:TCLMEPauliME} to zero 
\begin{align}
\label{eq:TCLsteady}
\sum_{n}\tilde{\CS}^{nm}(0,\eta\to 0)P_n = 0,
\end{align}
where~\(P_n\) is the steady-state probability distribution of the system~\footnote{Strictly speaking and for~\(\eta>0\), Eq.~\eqref{eq:TCLsteady} gives rise to a quasi steady-state which is valid for a time window~\(\sim \hbar/\eta\) around~\(t=0\).
	We take the limit~\(\eta\to 0\) as a last step providing us with the exact steady-state.}  that also satisfies~\(\sum_n P_n=1\).
To avoid notational clutter, we denote the steady-state system probability by~\(P_n\), simply dropping the time-dependence.
By expanding Eq.~\eqref{eq:TCLsteady} and given an explicit representation for the perturbation series of~\(\tilde{\CS}\), we obtain a formally exact power series for~\(P_n\).
In the next subsection, inspired by Refs.~\cite{Timm2011TimeconvolutionlessMasterEquation, Nestmann2019TimeconvolutionlessMasterEquation, Nestmann2020HowQuantumEvolution}, we present a simplified derivation of the superoperator~\(\CS\).
\begin{figure}
	\includegraphics[width=8.6cm]{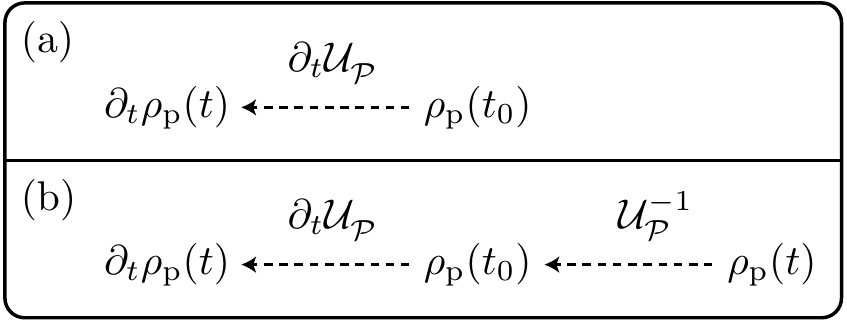}
	\caption{
		\label{Fig:masterEquations}
		Different master equations in a nutshell.
		(a) T-matrix master equation functionality. The full density matrix at time~\(t_0\), where~\(\rhop = \rho\), is propagated forward and differentiated with respect to time and then brought into the projected space using the time derivative of the projected evolution~\(\partial_t\CU_\projector\).
		(b) \Convolutionless master equation functionality. 
		The projected density matrix at time~\(t\) and the inverse projected evolution~\(\CUP^{-1}\) are used to construct the projected density matrix at time~\(t_0\), where~\(\rhop = \rho\).
		The result is then fed through the same procedure as in the T-matrix master equation.
	}
\end{figure}
\subsubsection{Derivation of~\(\CS\)}    
In order to find an expression for the {\TCL} superoperator~\(\CS\), we invert the projected time evolution~\eqref{eq:superUP} and insert it back into the superoperator formulation of the T-matrix~\eqref{eq:TMME} approach; the result 
\begin{align}
\label{eq:superRUP}
\pt \rhop(t) = -\frac{i}{\hbar}\liouvillian_0 \rhop(t) + \CR\CUP^{-1} \rhop(t),
\end{align}
then provides the identification~\cite{Nestmann2020HowQuantumEvolution}
\begin{align}
\label{eq:S=RUP}
\CS= \CR \CUP^{-1},
\end{align}
see Fig.~\ref{Fig:masterEquations}.
Starting directly from Eq.~\eqref{eq:S=RUP} provides a significant technical advantage in deriving an expression for the generator~\(\CS\) as compared to previous derivations~\cite{Timm2011TimeconvolutionlessMasterEquation,Nestmann2019TimeconvolutionlessMasterEquation}.
As pointed out in Ref.~\cite{Buzek1998ReconstructionLiouvillianSuperoperators}, the operator~\(\CUP\) is analytic in time and finite dimensional as it evolves only the projected space density matrix
.
It is thus invertible at all but isolated points in time, see Ref~\cite{FergusonThesis}.
This mathematical property guarantees the existence of the {\TCL} for any non-zero finite value of~\(\eta\) or~\(t_0\).
In the limit~\(|t_0|\gg \hbar/\eta \to \infty\), the system and environment have been in contact for an infinite time and the steady-state has been reached.
The {\TCL} we use in this work is therefore limited to steady-state calculations and cannot describe dynamical properties.

To obtain a series expansion of the inverse~\(\CUP^{-1}\), it is convenient to introduce the time-local propagator 
\begin{align}
\label{eq:superG}
\CG = \CUP \CU_{0}^{-1} - \mathcal{I},
\end{align} 
where~\(\mathcal{I}\) is the identity superoperator, and its Pauli equivalent~\(\tilde{\CG}\) obtained by replacing all occurrences of~\(\projector\)  by~\(\tilde{\projector}\).
The time-local propagator~\(\CG\) encodes the non-trivial evolution in the projected space and its series expansion 
\(\CG=\sum_{\alpha=1}^{\infty}\CG_\alpha\), is defined by
\begin{align}
\label{eq:superGa}
\CG_\alpha = \projector\CU{}_\alpha\CU_0^{-1}\projector.
\end{align} 
The terms~\(\CG_\alpha\) are divergent in the limit~\(\eta\to 0\) and must therefore be computed as a Laurent series in~\(\eta\), i.e., a power series that includes terms of negative powers of~\(\eta\).

In the projected space, we can recast the projected time evolution~\eqref{eq:superUP} in terms of the time-local propagator
\begin{align}
\CU_\projector=(\mathcal{I}+\CG)\,\CU_0,
\end{align}
Note that this representation of~\(\CUP\) is only appropriate when acting on projected density matrices as we have omitted the projector~\(\projector\) in the unperturbed evolution.
It is then straightforward to write down the inverse of~\(\CUP\) as a power series  
\begin{align}
\label{eq:superUPExpansion}
\CU_\projector^{-1}
= \CU_0^{-1}
(\mathcal{I}+\CG)^{-1}=\CU_0^{-1}(\mathcal{I}-\CG+\CG^2-...).
\end{align}
We insert Eq.~\eqref{eq:superUPExpansion} and the T-matrix generator~\eqref{eq:superR} into Eq.~\eqref{eq:superRUP} and compare the result, order by order, with the {\TCL} master equation~\eqref{eq:TCLME}.
As a result, we obtain the recurrence relation 
\begin{align}
\label{eq:superS}
\CS_\alpha= -\frac{i e^{\eta t /\hbar}}{\hbar}
\projector\liouvillian_V\CU_{\alpha-1} \CU^{-1}_0 \projector 
- \sum_{\beta=1}^{\alpha-1}\CS_{\alpha-\beta}\CG_\beta,
\end{align}
for the expansion of the {\TCL} generator.
Both the T-matrix and {\TCL} generators,~\(\CR\) and~\(\CS\), conserve probability (but not necessarily positivity) at each order, i.e.,
\begin{align}
\label{eq:conservedAbstract}
\Tr[\CS_\alpha \rhop] = 0,
\end{align}
which can be seen by first taking the trace over Eqs.~\eqref{eq:TCLME} or~\eqref{eq:TMME} (or their Pauli equivalents) and then performing an expansion on both sides.
The fact that probability must be conserved~\eqref{eq:conservedAbstract} at each order, puts constraints on each term~\(\CS_\alpha\), which we will use to reduce the number of diagrams we compute later on in section~\ref{sec:quadratic}.
From equations~\eqref{eq:superS} and~\eqref{eq:superR}, we immediately infer that at lowest non-vanishing order the Pauli (\(\tilde{\projector}\CU_0 = \tilde{\projector}\)) {\TCL} and T-matrix superoperators coincide~\cite{Timm2011TimeconvolutionlessMasterEquation, Nestmann2019TimeconvolutionlessMasterEquation}.
Hence, at the lowest order the Pauli {\TCL} is identical to Fermi's golden rule.
At higher orders, the T-matrix includes divergent terms, which in the {\TCL} are correctly subtracted by the recursion formula 
\begin{align}
\label{eq:superRecursive}
\CA_\alpha^{\reg} 
= \CA_\alpha\CU_0^{-1}-\sum_{\beta=1}^{\alpha-1}
\CA_{\alpha-\beta}^{\reg}\CG_\beta,
\end{align} 
that generates a time-local version~\(\CA^\reg(t) = \sum_\alpha\CA_\alpha^\reg\) from the expansion~\(\CA_\alpha\) of a propagating projected-space superoperator~\(\CA(t,t_0)\).
The recursion~\eqref{eq:superRecursive} appears in~\eqref{eq:superS} and was derived in Ref.~\cite{Nestmann2019TimeconvolutionlessMasterEquation}.
In the specific case of the T-matrix and {\TCL} master equations, it can be thought of as a \textit{regularisation} procedure, which subtracts \textit{reducible} contributions, and guarantees that each term~\(\CS_\alpha\) in the expansion of the {\TCL} generator is convergent in the limit~\(\eta\to 0\)~\cite{Timm2011TimeconvolutionlessMasterEquation, Nestmann2019TimeconvolutionlessMasterEquation}.
We stress that the regularisation~\eqref{eq:superRecursive} leads to exact results for the power series of steady-state system properties that are different from those obtained by the more common, physically motivated, regularisation scheme~\cite{,Averin1994PeriodicConductanceOscillations,Turek2002CotunnelingThermopowerSingle,Koch2004ThermopowerSinglemoleculeDevices,Timm2011TimeconvolutionlessMasterEquation,Timm2008TunnelingMoleculesQuantum,Koller2010DensityoperatorApproachesTransport}.
\subsubsection{Relation to real-time method}  
\label{sec:background:RT} 
The real-time~\cite{Schoeller1994MesoscopicQuantumTransport,Schoeller1994ResonantTunnelingCharge,Konig1995ResonantTunnelingCharging,Konig1995ResonantTunnelingCharging,Konig1996ResonantTunnelingUltrasmall}, Bloch-Redfield~\cite{Bloch1957GeneralizedTheoryRelaxation, Wangsness1953DynamicalTheoryNuclear,Redfield1965TheoryRelaxationProcesses} or Nakajima-Zwanzig~\cite{Nakajima1958QuantumTheoryTransport, Zwanzig1960EnsembleMethodTheory} master equations are equivalent, formally exact, and time-non-local, formulations of the dynamics of the projected-space density matrix.
They encode the evolution with
\begin{align}
\label{eq:NZME}
\pt{\rhop}(t) =-\frac{i}{\hbar}\liouvillian_0\rhop(t)
+\int_{t_0}^{t}\kernel(t,t_1)\rhop(t_1)\dd{t_1},
\end{align}
and are embodied by the kernel~\(\kernel\).
This approach is conceptually more complex than the T-matrix or TCL master equations as it involves an integral over all previous times; it is often combined with physical assumptions to generate a Lindblad master equation~\cite{Breuer2002TheoryOpenQuantum} or other more complex often non-Markovian master equations~\cite{Breuer2016ColloquiumNonMarkovianDynamics}.
Furthermore, a set of powerful tools, including resummation schemes~\cite{Konig1995ResonantTunnelingCharging,Konig1995ResonantTunnelingCoulomb,Konig1996ResonantTunnelingUltrasmall,Konig1996ZeroBiasAnomaliesBosonAssisted,Konig1997CotunnelingResonanceSingleElectron} and renormalisation group methods~\cite{Schoeller2000RealTimeRenormalizationGroup,Schoeller2009PerturbativeNonequilibriumRenormalization,Pletyukhov2012NonequilibriumKondoModel,Becker2012NonequilibriumQuantumDynamics,Saptsov2012FermionicSuperoperatorsZerotemperature,Reimer2019DensityoperatorEvolutionComplete} have been developed for these master equations.
As for the T-matrix and {\TCL} approaches, Eq.~\eqref{eq:NZME} has a Pauli equivalent 
\begin{align}
\label{eq:NZMEPauli}
\pt{\rhopt}(t) =\int_{t_0}^{t}\tilde{\kernel}(t,t_1)\rhopt(t_1)\dd{t_1},
\end{align} 
where~\(\tilde{\kernel}\) is the Pauli Kernel.
The derivation of either the full~\(\kernel\) or Pauli~\(\tilde{\kernel}\) kernels, can be found in a breadth of pedagogical texts on the topic, see for example Refs.~\cite{Breuer2002TheoryOpenQuantum, Koller2010DensityoperatorApproachesTransport, FergusonThesis}.
The RT method is commonly used to compute the asymptotic-state of the projected-space density matrix.
Here, we focus on the Pauli case where the probabilities reach a steady-state and refer the interested reader to Ref.~\cite{FergusonThesis} for a treatment of the off-diagonal terms.
We combine Eq.~\eqref{eq:NZMEPauli} with a steady-state assumption 
\begin{align}
\label{eq:trueSteady}
\rhopt(t) = \rhoptb, \quad \forall \quad t_0\leq t\leq0,
\end{align} 
and move the limit~\(t_0\to -\infty\) of the integral to the far distant past.
Making use of the steady-state assumption~\eqref{eq:trueSteady} (also known as the Markov approximation), we add the qualifier steady-state to the RT method (SRT), as we already did for the {TCL}.
We substitute Eq.~\eqref{eq:trueSteady} into the (Pauli) real-time master equation~\eqref{eq:NZME} and obtain the condition 
\begin{align}
\label{eq:NZsteady}
\tilde{\CZ} \rhoptb
\equiv
\left[
\int_{-\infty}^{0}\tilde{\kernel}(0,t_1)\dd{t_1}\right]{\rhoptb} = 0,
\end{align}
which when solved for, order-by-order in~\(V\), leads to the same steady-state as the {\TCL}  condition~\eqref{eq:TCLsteady}.
In Eq.~\eqref{eq:NZsteady}, we introduce the Pauli SRT generator~\(\tilde{\CZ}\), which is obtained by performing the integral over time~\cite{Schoeller1994MesoscopicQuantumTransport}.
The series expansion~\(\tilde{\CZ} = \sum_\alpha\tilde{\CZ}_\alpha\) of the Pauli SRT generator 
\begin{align}
\label{eq:superZ}
\tilde{\CZ}_\alpha = -\frac{i}{\hbar}\tilde{\projector} \liouvillian_V
\tilde{{\CU}}_{\alpha-1}\CU_0^{-1}\tilde{\projector},
\end{align} 
is closely linked to the expansions of the T-matrix and {\TCL} generators, see Refs.~\cite{Breuer2002TheoryOpenQuantum,  Schoeller1994MesoscopicQuantumTransport, Koller2010DensityoperatorApproachesTransport,FergusonThesis} for a derivation.
Here,~\(\tilde{{\CU}}_\alpha\) is the Pauli version (\(u=v\) and~\(i=j\)) of the expansion of the full evolution minus \textit{secular} contributions (\(\mathcal{I} \to \mathcal{I}-\tilde{\projector}\)), such that 
\begin{align}
\label{eq:superUbarExpansion}
\tilde{{\CU}}_\alpha&
= \frac{\mathcal{I}-\tilde{\projector}}
{-\liouvillian_0+i\alpha\eta}
\liouvillian_V\tilde{{\CU}}_{\alpha-1},
\end{align}
with the initial condition~\(\tilde{\CU}_0 = \CU_0\), cf.~Eq.~\eqref{eq:superUexpansion} with~\(t=0\).
The Pauli SRT generator~\(\tilde{\CZ}\) is thus composed of an unperturbed backward propagation~\(\CU_0^{-1}\) and a full forward propagation minus the \textit{secular} contributions.
The same way the \textit{regularising} recurrence~\eqref{eq:superRecursive} guarantees that~\(\CS\) is finite at each order, the subtraction of \textit{secular} contributions (through~\(\mathcal{I} \to \mathcal{I}-\tilde{\projector}\)) guarantees that~\(\tilde{\CZ}\) is finite at each order.
The full SRT generator~\(\CZ\) is not as easy to obtain as in the {\TCL} case due to the oscillating off-diagonal elements, see for example Ref.~\cite{FergusonThesis}.
Both the SRT and {\TCL} methods lead to identical power series for steady-state observables, through related but different generators.
This difference between the {\TCL} and SRT methods occurs because the {\TCL} generator~\(\CS\) carries a remnant of the dynamics within it.
The SRT master equation, instead, includes an explicit steady-state (or Markov) assumption~\eqref{eq:trueSteady} and therefore cannot be straightforwardly extended to an exact time-local master equation.
We choose to work with the {\TCL} over the SRT for three main reasons: (i) its conceptually simpler form, (ii) its direct formulation in terms of~\(\CUP\), which we will use to considerably simplify the computation of rates, and (iii) its potential to be extended to the TCL, a time-local master-equation that can describe the dynamics after a quench.
Next, we proceed with the derivation of a diagrammatic expansion of the {\TCL} generator~\(\CS\).
\section{Diagrammatic expansion}
\label{sec:formal}
Computing the rates in the superoperator~\(\CS\) of the {\TCL} master equation~\eqref{eq:TCLME} requires evaluating the expressions~\(\projector\liouvillian_V\CU\CU_0^{-1}\projector\) and~\(\CG\) order by order in~\(V\), see Eq.~\eqref{eq:superS}.
The derivation in~\ref{sec:background:TCL} does not rely on the limits~\(t_0 \to -\infty\) or~\(\eta\to 0\), and Eq.~\eqref{eq:superS} is valid for any initial time as long as~\(t_0\) is adapted correspondingly in~\(\CG\) and~\(\CU\).
We will now show that, in the~\(t_0\to-\infty\) limit, the combination~\(\projector\liouvillian_V\CU\CU_0^{-1}\projector\) can be recast in terms of~\(\CG\) and thus finding the latter is sufficient for computing~\(\CS\).
We then construct a novel diagrammatic approach to compute~\(\CG\), inspired by T-matrix calculations~\cite{Koch2004ThermopowerSinglemoleculeDevices,Koch2006TheoryFranckCondonBlockade,Averin1994PeriodicConductanceOscillations,Bruus2004ManyBodyQuantumTheory,Turek2002CotunnelingThermopowerSingle} and the simplifications in Ref.~\cite{Koller2010DensityoperatorApproachesTransport}.
Specifically, we will perform the expansion of~\(\CG\) using operators, as it leads to a significant reduction in the number of terms compared with the superoperator formulation, see Appendix~\ref{app:timeorder}.
As~\(\CG\) contains terms that diverge in~\(1/\eta\) and higher powers thereof, it is necessary to perform the computations for finite~\(\eta\) and then express the result as a Laurent series (including negative powers of~\(\eta\)).
Our diagrammatic representation of~\(\CG\) can then be used to construct a diagrammatic expansion for~\(\CS\).
The recurrence~\eqref{eq:superRecursive} then guarantees that negative powers of~\(\eta\) are subtracted from the expressions for~\(\CS\).
%

%

%
\subsection{\(t_0\to -\infty\) limit}
Our first goal is to simplify the expression~\eqref{eq:superS} for~\(\CS_\alpha\), and more specifically the term~\(\projector\liouvillian_V\CU_{\alpha-1}\CU_0^{-1}\projector\), by
recasting Eq.~\eqref{eq:superS} in terms of~\(\CG_\alpha\) and~\(\liouvillian_0\) only.
To do so, we compare two alternative representations of the derivative~\(\partial_t \CU\) of the full time evolution.
To obtain the first expression, we expand both sides of the identity~\eqref{eq:ptU} order by order in~\(V\) to arrive at the equality 
\begin{align}
\label{eq:superUdot1}
i\hbar\pt{\CU}_\alpha = \liouvillian_0\CU_\alpha 
+ e^{\eta t/\hbar}\liouvillian_V\CU_{\alpha-1}.
\end{align}
The second form, valid only in the~\( t_0\to -\infty\) limit, is found by differentiating~\eqref{eq:superUexpansion} with respect to time and expanding to obtain 
\begin{align}
\label{eq:superUdot2}
i\hbar \, \pt{\CU}_\alpha
=\CU_\alpha (\liouvillian_0 +i \alpha \eta ).
\end{align}
Combining the two expressions~\eqref{eq:superUdot1} and~\eqref{eq:superUdot2}, we find the relation 
\begin{align}
\label{eq:LVU1}
e^{\eta t/\hbar}\liouvillian_V \CU_{\alpha-1} 
= i\alpha\eta\,\CU_\alpha +[\CU_\alpha,\liouvillian_0],
\end{align}
which can easily be verified using~\eqref{eq:superUexpansion}.
Substitution into Eq.~\eqref{eq:superS} provides us with the result 
\begin{align}
\label{eq:simpleSuperS}
\CS_\alpha=
\frac{\alpha\eta}{\hbar}\CG_\alpha-\frac{i}{\hbar}
[\CG_\alpha,\liouvillian_0] -
\sum_{\beta=1}^{\alpha-1}\CS_{\alpha-\beta}\CG_\beta,
\end{align}
that only involves the expansion of~\(\CG\) and and the Liouvillian~\(\liouvillian_0\).
The recursion in~\eqref{eq:simpleSuperS} is of the form~\eqref{eq:superRecursive} that provides us with a \textit{regularised} expression and guarantees that each order~\(\CS_\alpha\) is convergent in the~\(\eta\to 0\) limit.
The first two terms in Eq.~\eqref{eq:simpleSuperS} are generated by the time derivative; the first one 
\begin{align}
\label{eq:superGprime}
\CGH_\alpha\equiv\frac{\alpha \eta}{\hbar}\CG_\alpha,
\end{align}
arises from the derivative of the explicit time dependence of the perturbation~\(V\propto\exp(\eta t/\hbar)\).
The second term 
\begin{align}
\label{eq:superGdot}
\CGL_\alpha \equiv -\frac{i}{\hbar}[\CG_\alpha,\liouvillian_0],
\end{align}
is similar to the von Neumann equation~\eqref{eq:vonNeumann} but formulated in the superoperator space; it can be thought of as the evolution of the propagator~\(\CG\) under the influence of the unperturbed evolution~\(\liouvillian_0\).
The above derivation tells us that the expansion~\eqref{eq:superGa} of the generator~\(\CG\) is the key to computing the {\TCL} generator order by order.
We use the Hamiltonian representation~\eqref{eq:superUaUnumu} of~\(\CU\) to reduce the number of terms in our computation by defining
\begin{align}
\label{eq:superUasuperUnumu}
\CU_\alpha = \sum_{\nu+\mu=\alpha}\CU_{\nu\mu},
\end{align}
where~\(\CU_{\nu\mu}\rho = U_\nu^\pdagger \rho U^\dagger_\mu\) is the contribution to the time evolution of~\(\rho\) with~\(\nu\) (\(\mu\)) occurrences of the perturbation on the upper (lower) Keldysh branches.
We apply this same procedure to the expansion~\eqref{eq:superGa} of the generator~\(\CG\), such that 
\begin{align}
\label{eq:superGnumu}
\CG_\alpha=\sum_{\mu+\nu=\alpha} \CG_{\nu \mu}, 
\end{align}
where~\(\CG_{\nu\mu} = \projector \CU^{\phantom 1}_{\nu\mu}\CU_{0}^{-1}\projector\) and~\(\nu+\mu \geq 1\).
Similarly, the terms~\(\CA_\alpha\) in the expansion of any superoperator~\(\CA\) generated from~\(\CG\), can be decomposed into terms~\(\CA_{\nu \mu}\).
This decomposition will reduce the number of terms that we have to compute, from~\(2^\alpha\) to~\((\alpha+1)\), in complete analogy to the diagrammatic grouping in Ref.~\cite{Koller2010DensityoperatorApproachesTransport}, see also Appendix~\ref{app:timeorder}.
Next, we construct the terms~\(\CS_{\nu \mu}\) by generalising the regularising recursion~\eqref{eq:simpleSuperS} to account for the two Keldysh branches 
\begin{align}
\label{eq:superSnumu}
\CS_{\nu \mu} = 
\CGH_{\nu \mu} +
\CGL_{\nu \mu} -\!\!\!\!
\sum_{\substack{{\nu'+\nu''=\nu}\\{\mu'+\mu''=\mu}}}\!\!\!\!\CS_{\nu'\mu'}\CG_{\nu''\mu''}.
\end{align}
Here, each occurrence of the subscript~\(\alpha\) in the definitions~\eqref{eq:superGprime} and~\eqref{eq:superGdot} of~\(\CGH\) and~\(\CGL\) is replaced by~\(\alpha\to\nu\mu\).
Simultaneously, the prefactor~\(\alpha\) in Eq.~\eqref{eq:superGprime} is replaced by~\(\alpha \to \nu+\mu\).
\subsection{Diagrammatic rules}
We develop our diagrammatic technique for the elements 
\begin{align}
\label{eq:superMatrixElements}
\CS^{ijfg} = \Tr_\env \{\bra{f}[\,\CS
\,(\,\rho_\env^0\otimes\ketbra{i}{j}\,)\,]\ket{g}\},
\end{align}
of the generator~\(\CS\).
While these are in fact elements of a four dimensional tensor, we will refer to them as matrix elements to keep the nomenclature consistent between the full and Pauli generators.
They are indexed by four system indices~\(i,j,f,g\), and are the rates that can then be used to compute the steady-state.
Furthermore, they can be used to reconstruct the superoperator through the identity 
\begin{align}
\label{eq:reconstructSuperOp}
\CS\left( \rho_\env^0\otimes\ketbra{i}{j}\right)
=\sum_{fg}\CS^{ijfg}
\left(\rho_\env^0\otimes\ketbra{f}{g}\right).
\end{align}
The matrix elements for other projected-space superoperators, such as~\(\CG\),~\(\CR\), and~\(\liouvillian_0\), are defined in the same way~\eqref{eq:superMatrixElements} as those for~\(\CS\).
Simultaneously, these matrix elements can be used to reconstruct their corresponding superoperators with Eq.~\eqref{eq:reconstructSuperOp}.
%
\begin{figure}
	\includegraphics[width=8.6cm]{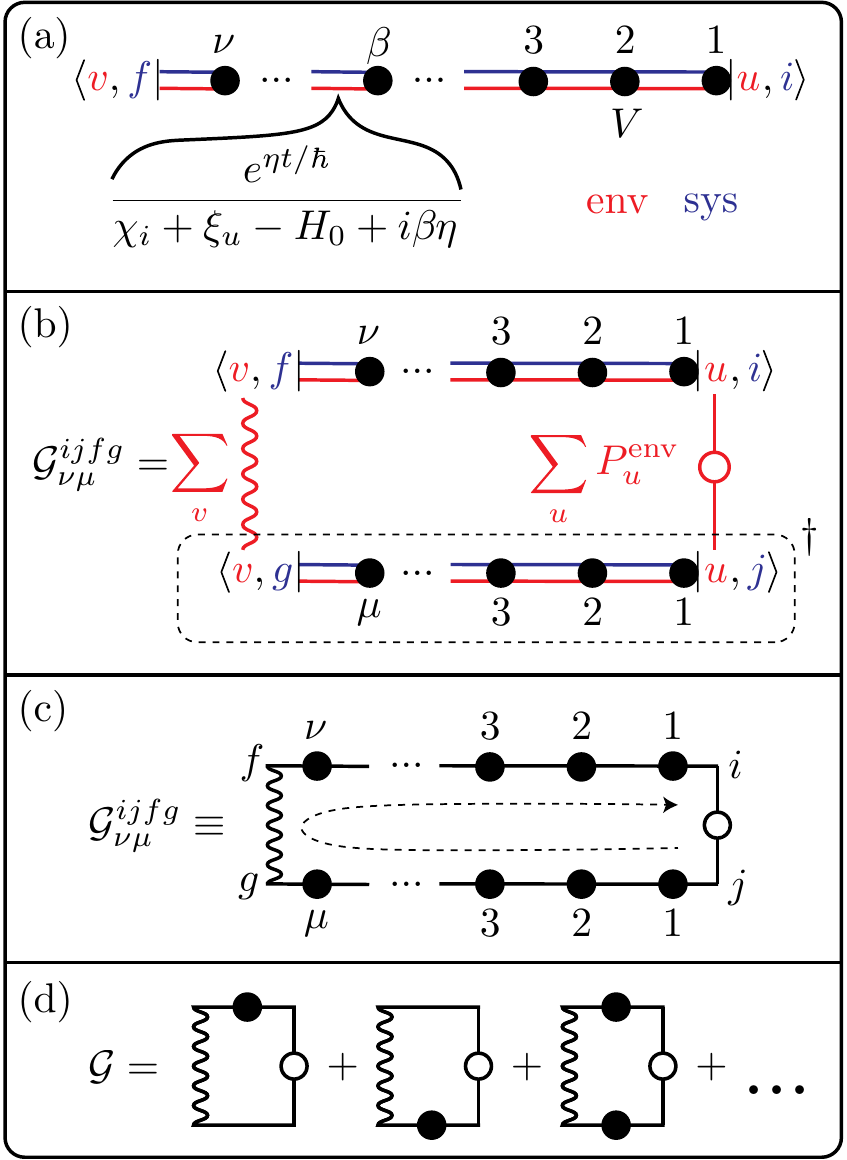}
	\caption{
		\label{Fig:genericDiagram}
		Diagrammatic rules for the propagator~\(\CG\) of arbitrary open setups.
		(a) Diagrammatic representation of the perturbative evolution of a state~\(\ket{u,i}\).
		The system (red) and environment (blue) evolve freely and independently (lines) between applications of the system--environment coupling~\(V\) (black dots).
		(b) Diagram for the matrix elements~\(\CG_{\nu\mu}^{ijfg}\), see Eq.~\eqref{eq:diagramaticG}, which includes the initial environment density matrix (open red circle) and trace over final environment states (wiggly red line).
		The total operator on the lower Keldysh branch (dashed box) appears as a Hermitian conjugate, see Eq.~\eqref{eq:diagramaticG}.
		The states on the upper (lower) branch scatter a total of~\(\nu\) (\(\mu\)) times.
		(c) Simplified form of the diagram (b) for the matrix
		elements~\(\CG_{\nu\mu}^{ijfg}\), where we have omitted the
		double lines and references to the environment for clarity.
		The dashed arrow indicates the operator ordering which appears in Eq.~\eqref{eq:diagramaticG}, once the Hermitian conjugation has been performed.
		(d) Series expansion for~\(\CG\) in the diagrammatic representation, see Eqs.~\eqref{eq:superGa} and~\eqref{eq:superGnumu}. 
		Note that the lowest order terms contain one occurrence of the coupling~\(V\) (black dot).
	}
\end{figure}
The generator~\(\CS\) is constructed from the propagator~\(\CG\), see Eq.~\eqref{eq:simpleSuperS}, and we have to consider the expansion of the latter first.
We use the decomposition~\eqref{eq:superUasuperUnumu} of~\(\CU_\alpha\) in terms of unitary evolutions~\eqref{eq:recursiveU} in the definition~\eqref{eq:superGnumu} of the expansion of~\(\CG\) and insert the result into Eq.~\eqref{eq:superMatrixElements} to obtain 
\begin{align}
\label{eq:diagramaticG}
\CG^{ijfg}_{ \nu \mu} &= \Tr_\env
\left(
\bra{g}  U^\pdagger_\mu U_0^\dagger\ket{j}^\dagger
\bra{f} U_\nu^\pdagger U_0^\dagger \ket{i}
\rho_\env^0
\right).
\end{align}
Here, we immediately notice the strength of the operator formulation, where~\(\CG_\alpha\) contains~\(\alpha+1\) terms, each with~\(\alpha\) occurrences of~\(V\), see Eqs.~\eqref{eq:superGnumu} and~\eqref{eq:diagramaticG}. 
On the other hand, the superoperator formulation~\eqref{eq:superUexpansion} contains a single term with~\(\alpha\) occurrences of the Liouvillian~\(\liouvillian_V\), which leads to~\(2^\alpha\) terms once the commutators in the Liouvillians are written explicitly, see also Appendix~\ref{app:timeorder}.
\begin{figure}
	\includegraphics[width=8.6cm]{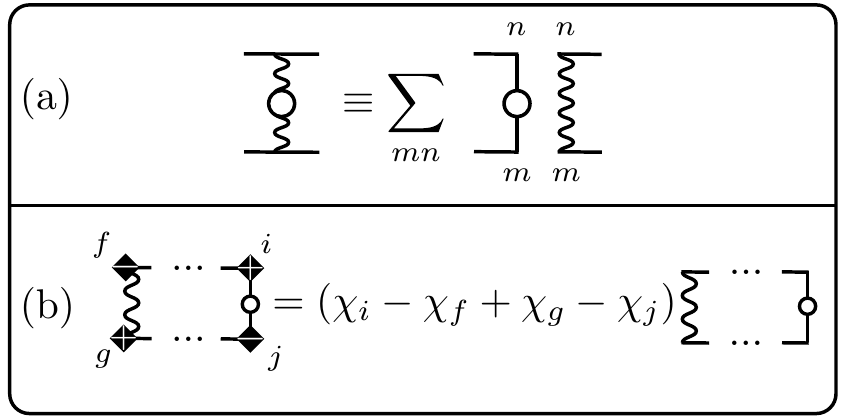}
	\caption{
		\label{Fig:compositionRule}
		Common diagrammatic operations.
		(a) Internal wiggly lines (or cut) appearing in the multiplication of two projected-space operators, see Eq.~\eqref{eq:composition}.
		The cut signals that there are actually two distinct diagrams which must be multiplied.
		Alternatively, it is possible to formulate a set of rules for them and retain a single diagram, see main text.
		(b)~The commutator of~\(\CG\) with~\(\liouvillian_0\), see Eq.~\eqref{eq:commutatorMatrixElements}, is represented by the diagram for~\(\CG\) with diamonds at each of the corners, marking the four system energies~\(\chi\).
		The~\(\pm\) signs mark positive/negative contributions of the energies~\(\chi\) and are dropped henceforth.
	}
\end{figure}
The operators~\(G_{\nu}=U_{\nu}^\pdagger U_0^{\dagger}\) in Eq.~\eqref{eq:diagramaticG}, propagate a hybrid system--environment state~\(\ket{u,i}\) backwards in time with an unperturbed evolution~\(U_0^\dagger\), before propagating it forward in time with~\(\nu\) system--environment scattering events.
We use the recurrence relation~\eqref{eq:recursiveU} for~\(U_\nu\) to define the diagrammatic rules (in energy space) for the matrix elements~\(\bra{v,f}G_\nu\ket{u,i}\), see Fig.~\ref{Fig:genericDiagram}(a): 
\begin{enumerate}
	\item Act with a perturbation~\(V\), which mixes the system and environment.
	\item Evolve forward freely using the decoupled system--environment Green's function~\eqref{eq:freePropagator}.
	%
	\item Repeat the steps 1 and 2 a total of~\(\nu\) times.
\end{enumerate}
The operator~\(G_\nu^\pdagger\) (\(G_\mu^\dagger\)) appears on the upper (lower) Keldysh branches, see Eq.~\eqref{eq:diagramaticG} and Figs.~\ref{Fig:Keldysh} and~\ref{Fig:genericDiagram}(b).
The initial environment state~\(\ket{u}\) is the same on each of the Keldysh branches and is summed over, weighted by the probability~\(P_u^\env\) (encoded in~\(\rho^0_\env\)) of finding the unperturbed environment in the state~\(\ket{u}\), see Fig.~\ref{Fig:genericDiagram}(b) and Eq.~\eqref{eq:diagramaticG}.
As the environment is traced out in~\eqref{eq:diagramaticG}, the final environment state~\(\ket{v}\) is the same on the upper and lower Keldysh branches as well, see Fig.~\ref{Fig:genericDiagram}(b).
We define a simpler diagrammatic form for~\(\CG_{\nu\mu}^{ijfg}\) in Fig.~\ref{Fig:genericDiagram}(c), where we merge the system and environment lines for simplicity.
Note that the decomposition~\eqref{eq:superUaUnumu}, means that scattering events on each Keldysh branch have an operator order inherited from the temporal order, however, there is no inherited ordering relation between the upper and lower branches.
\begin{figure}
	\includegraphics[width=8.6cm]{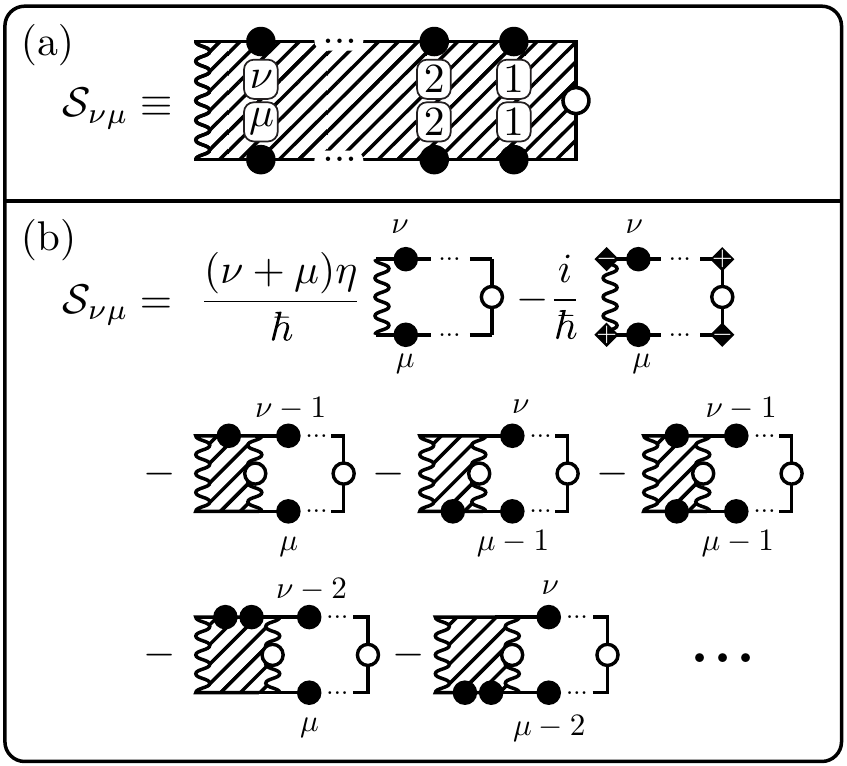}
	\caption{
		\label{Fig:genericSmoreGeneral}
		Diagramatic representation of the {\TCL} generator~\(\CS\), see Eqs.~\eqref{eq:S=RUP},~\eqref{eq:superS} and~\eqref{eq:simpleSuperS}.
		(a) Definition of the diagram for~\(\CS_{\nu\mu}\) exhibiting~\(\nu\) (\(\mu\)) system-bath scattering events~\(V\) on the upper (lower) branch of the Keldysh contour.
		Compared to~\(\CG_{\nu\mu}\), see Fig.~\ref{Fig:genericDiagram}, the hatched area indicates that the recursion~\eqref{eq:superSnumu} and its associated cuts, see Fig.~\ref{Fig:compositionRule}, have not been applied. 
		Furthermore, the hatched area has dimension [time\(^{-1}\)] as~\(\CS\) is a rate whereas~\(\CG\) is a propagator.
		(b) Rules of the recursion~\eqref{eq:superSnumu} for~\(\CS_{\nu\mu}\), where the first two terms refer to~\(\CGH_{\nu\mu}\) and~\(\CGL_{\nu\mu}\) respectively.
		The remaining lines encode the recursive regularisation~\eqref{eq:superSnumu}.
		The hatched parts again denote lower-order versions of~\(\CS_{\nu'\mu'}\), which themselves must be evaluated using the recursion, leading to diagrams with an increasing number of cuts.
	}
\end{figure}
To construct a diagrammatic expansion of the {\TCL} generator~\(\CS\), see Eq.~\eqref{eq:superSnumu}, we require two further elements.
The first is the composition~\(\CA\CB\) of two projected-space superoperators~\(\CA\) and~\(\CB\), which in matrix element form is 
\begin{align}
\label{eq:composition}
\left(\CA \CB\right)^{ijfg} = \sum_{nm}\CA^{nmfg}\CB^{ijnm}.
\end{align}
Diagrammatically, we replace this product by a \textit{cut}, see Fig.~\ref{Fig:compositionRule}(a), which can be understood as reset
of the free propagators~\(\Pi_0^\pm\) from Eq.~\eqref{eq:freePropagator}. 
Namely, the counter~\(\nu\) for the prefactor of~\(\eta\) in the propagators~\(\Pi_0^\pm(\omega,\nu\eta)\) returns to one, and the energy~\(\omega\) becomes that of the system state at the cut~\(\chi_n\) or~\(\chi_m\) plus the environment energy~\(\xi_u\) of the unperturbed environment state~\(\ket{u}\) (which are summed over, weighted by the unperturbed distribution~\(P_u^\env\)).
The second new element involves the commutator of a projected-space operator, e.g.,~\(\CG\) with the unperturbed Liouvillian~\(\liouvillian_0\).
In the projected space, the matrix elements of the unperturbed Liouvillian are 
\begin{align}
\label{eq:liouvilianME}
(\projector\liouvillian_0)^{ijfg} = (\chi_i-\chi_j)\delta_{if}\delta_{jg},
\end{align}
which follows from the definitions of~\(\liouvillian_0\) in~\eqref{eq:vonNeumannL}, of the projector~\(\projector\) in~\eqref{eq:superP}, and of the matrix elements~\eqref{eq:superMatrixElements}.
To obtain the matrix elements of the superoperator commutator~\([\CG,\liouvillian_0]\) from Eq.~\eqref{eq:superGdot}, we combine Eqs.~\eqref{eq:liouvilianME} and~\eqref{eq:composition}, leading to 
\begin{align}
\label{eq:commutatorMatrixElements}
[\CG,\liouvillian_0]^{ijfg} = (\chi_i-\chi_f+\chi_g-\chi_j)\CG^{ijfg},
\end{align}
see Fig.~\ref{Fig:compositionRule}(b) for a simple diagrammatic representation.
The diagrams for~\(\CS_{\nu \mu}\) are composed of three parts, see Eq.~\eqref{eq:simpleSuperS} and Fig.~\ref{Fig:genericSmoreGeneral}.
The first two are~\(\CGH_{\nu\mu}\) and~\(\CGL_{\nu\mu}\), which are obtained by multiplying~\(\CG_{\nu\mu}\) by~\((\nu+\mu)\eta\) and~\(-i(\delta\chi_{ij}-\delta\chi_{fg})/\hbar\) respectively.
The last part is the recursion scheme, which for convenience, we rearrange according to the depth of the recursion, see also Ref.~\cite{Gu2020DiagrammaticTimelocalMaster}.
\begin{figure*}
	\includegraphics[width=17.8cm]{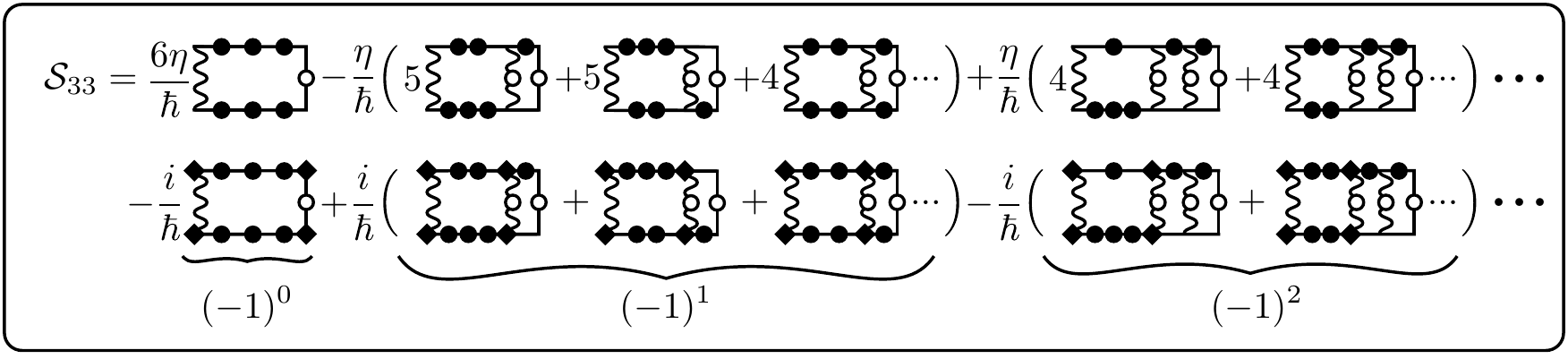}
	\caption{
		\label{Fig:genericSDiagram}
		Diagrammatic construction of the {\TCL} generator, exemplified on~\(\CS_{33}\).
		By introducing every possible set of cuts in the diagrams, we implement the recursion~\eqref{eq:superSnumu}.
		At first the number of diagrams in~\(\CS_{33}\) appears large, however we note that all contributions that contain a cut are obtained from products of lower order diagrams, such that a single sixth order object (\(\CG_{33}\)) must be computed.
		The prefactor is negative (positive) for diagrams with an odd (even) number of iterations of the recursion, i.e., the number of introduced cuts.
	} 
\end{figure*}
We thus arrive at rules to generate~\(\CS_{\nu\mu}\) directly from~\(\CG_{\nu\mu}\): 
\begin{enumerate}
	\item Create every distinct composite diagram by introducing cuts into the diagram.
	\item The prefactor of the subdiagrams is multiplied by~\(1\) (\(-1\)) for an even (odd) number of cuts.
	\item Transform the leftmost term~\(\CG\) in each diagram into the sum of its time differentiated versions~\(\CGL\) and~\(\CGH\).
\end{enumerate}
Combining these three rules and the rules for~\(\CG\) in Fig.~\ref{Fig:genericDiagram}, we generate all terms which contribute to~\(\CS_{\nu\mu}\).
We exemplify this scheme with a representation of~\(\CS_{33}\) in Fig.~\ref{Fig:genericSDiagram}.
Its generalisation provides us with the diagrammatic expansion of the full {\TCL} generator~\(\CS\).
\subsection{Pauli {\TCL}}
\begin{figure}
	\includegraphics{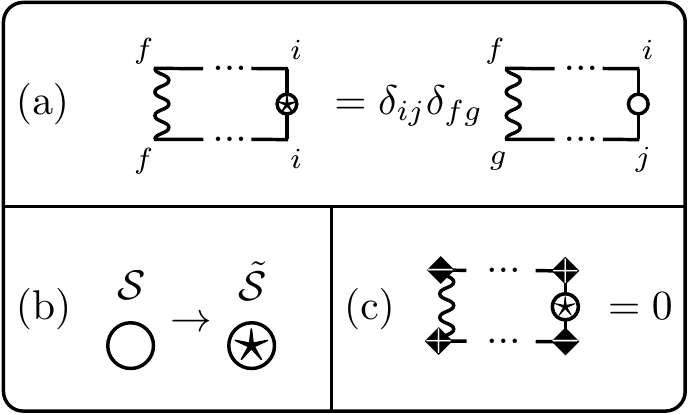}
	\caption{
		\label{Fig:PauliRules}
		Diagrammatic rules for the Pauli steady-state master equation.
		(a) For the Pauli version~\(\tilde{\CG}\) of the superoperator~\(\CG\) the initial and final system states must be the same on both Keldysh branches, see Eq.~\eqref{eq:matrixElementConstraint}.
		We introduce this property diagrammatically with a symbol~\(\star\), which is only shown in the open circle but is understood to also act at the wiggly line.
		(b) The Pauli {\TCL} generator~\(\tilde{\CS}\) diagrams are obtained by adding a star to each and every open circle in the diagrams obtained from Fig.~\ref{Fig:genericSDiagram}.
		(c) The commutator of the Pauli propagator~\(\tilde{\CG}\) with the unperturbed Liouvillian~\(\liouvillian_0\) vanishes, see Eqs.~\eqref{eq:simpleSuperS} versus~\eqref{eq:simpleSuperSPauli}.
	}
\end{figure}

The Pauli {\TCL} master equation~\eqref{eq:TCLMEPauli} does not track the off-diagonal elements of the system density matrix in~\(\rhop\).
Its generator~\(\tilde{\CS}\) is obtained by replacing all occurrences of the projector~\(\projector \to \tilde{\projector}\) in the {\TCL} generator Eq.~\eqref{eq:superSnumu}, leading to 
\begin{align}
\label{eq:simpleSuperSPauli}
\tilde{\CS}_\alpha = \frac{\alpha\eta}{\hbar}\tilde{\CG}_\alpha 
-\sum_{\beta=1}^{\alpha-1}\tilde{\CS}_{\alpha-\beta}\tilde{\CG}_\beta,
\end{align}
where we have used that~\(\tilde{\projector}\liouvillian_0=0\).
As for the full {\TCL} generator, we decompose the Pauli generator~\(\tilde{\CS}\) according to the number of perturbations~\(\nu,\mu\) on each of the Keldysh branches 
\begin{align}
\label{eq:PauliSuperSnumu}
\tilde{\CS}_{\nu \mu} = 
(\nu+\mu)\frac{\eta}{\hbar}\tilde{\CG}_{\nu \mu} + \!\!\!\!
\sum_{\substack{{\nu'+\nu''=\nu}\\{\mu'+\mu''=\mu}}} \!\!\!\!
\tilde{\CS}_{\nu'\mu'}\tilde{\CG}_{\nu''\mu''}.
\end{align}
The Pauli {\TCL} generator~\eqref{eq:PauliSuperSnumu} should be contrasted with the Pauli T-matrix generator 
\begin{align}
\label{eq:PauliSuperRnumu}
\tilde{\CR}_{\nu\mu} = (\nu+\mu)\frac{\eta}{\hbar}\tilde{\CG}_{\nu\mu},
\end{align}
which is not convergent in the~\(\eta\to 0\) limit.
Note that the matrix elements~\(\tilde{\CG}_{\nu\mu}^{if}\) of the Pauli superoperator~\(\tilde{\CG}\) are indexed by only two system indices~\(i\) and~\(f\), i.e., they are obtained by enforcing 
\begin{align}
\label{eq:matrixElementConstraint}
{{\tilde{\CG}}}^{if}_{ \nu \mu} =
\delta_{ij}\delta_{fg}{{\CG}}^{ijfg}_{ \nu \mu},
\end{align}
see Eqs.~\eqref{eq:superPauli} and~\eqref{eq:superGa}, as well as Fig.~\ref{Fig:PauliRules}(a).
In contrast, the matrix elements~\(\tilde{\CS}^{if}\) of the Pauli {\TCL} generator contain multiple projectors and are obtained by combining the Pauli propagators~\(\tilde{\CG}\) according to Eq.~\eqref{eq:PauliSuperSnumu} and cannot be immediately obtained from~\(\CS^{ijfg}\).
In the diagrammatic representation, we differentiate between~\(\CG\) and its Pauli counterpart~\(\tilde{\CG}\) by a single star~\(\star\) in the starting environment distribution.
On the other hand,~\(\CS\) and~\(\tilde{\CS}\) additionally differ by a star at every cut, see Fig.~\ref{Fig:PauliRules}(b).
The additional constraint~\eqref{eq:matrixElementConstraint} immediately implies that the~\(\CGL\) contributions vanish for the Pauli master equation, see Fig.~\ref{Fig:PauliRules}(c) and Eqs.~\eqref{eq:commutatorMatrixElements} and~\eqref{eq:matrixElementConstraint}.
There are several properties of the Pauli {\TCL} generator that can be used to reduce the number of diagrams that need to be computed.
First, note that matrix elements related by an inversion of the number of scatterings on the two Keldysh branches are complex conjugates of each other~\(\tilde{\CS}_{\nu\mu}^{if}=\tilde{\CS}_{\mu\nu}^{if*}\).
Next, we show how the conservation of probability can be used to avoid computing  the diagrams with a vanishing index~\(\nu=0\) or~\(\mu = 0\).
Just as for the full master equation, the conservation of probability~\eqref{eq:conservedAbstract} implies that 
\begin{align}
\label{eq:conservedAlpha}
\tilde{\CS}^{ii}_\alpha = -\sum_{f\neq i} \tilde{\CS}^{if}_\alpha.
\end{align}
Here,~\(\tilde{\CS}^{ii}\) has the usual interpretation as the inverse lifetime of the system state~\(\ket{i}\).
We use Eqs.~\eqref{eq:diagramaticG} and~\eqref{eq:PauliSuperSnumu} to conclude that~\(\tilde{\CS}^{if}_{\alpha 0}\propto\delta_{if}\) (and similarly for~\(\tilde{\CS}^{if}_{0\alpha}\)), and use this conclusion to split~\eqref{eq:conservedAlpha}, such that
\begin{align}
\label{eq:conserved}
\tilde{\CS}_{0\alpha}^{ii}+\tilde{\CS}_{\alpha 0}^{ii} 
= -\sum_{f}\sum_{\nu=1}^{\alpha-1} \tilde{\CS}_{\nu\mu}^{if},\quad\mu 
= \alpha-\nu.
\end{align}
Note that here, unlike in~\eqref{eq:conservedAlpha}, the sum over the system states~\(f\) is arbitrary and does not exclude diagonal elements.
Hence, the sum of diagrams with a zero index follows from the diagrams without vanishing indices, and at lowest order we obtain the simplification~\(\tilde{\CS}_{01}+\tilde{\CS}_{10}=0\).
The rates of the form~\(\tilde{\CS}_{\alpha0}^{ii}\) or~\(\tilde{\CS}_{0\alpha}^{ii}\) are purely diagonal and, as we will see in section~\ref{sec:currents}, they are not associated with a physical process, such that we denote them as \textit{probability conserving} rates.
\subsection{Discussion}
\begin{figure}
	\includegraphics[width=8.6cm]{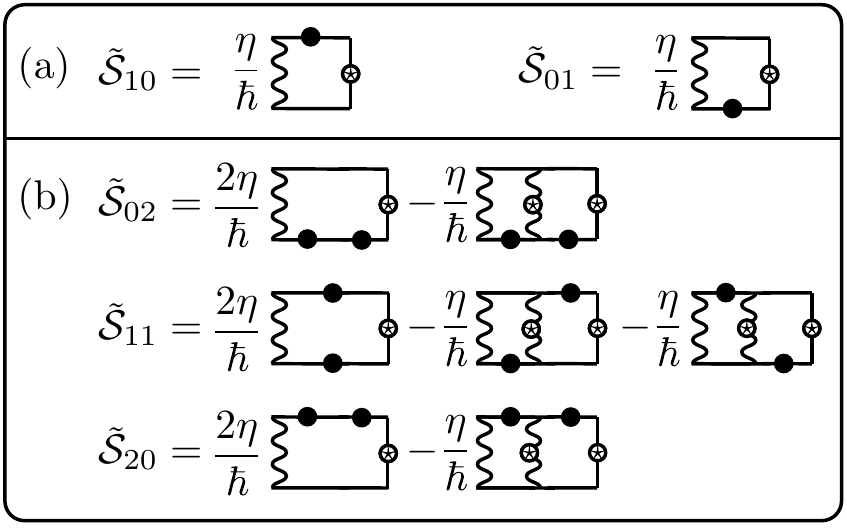}
	\caption{
		\label{Fig:genericSexamples}
		Diagrammatic construction of the Pauli {\TCL} generator.
		(a) The first-order Pauli diagrams (\(\CS_{00}\) does not exist), which due to conservation of probability~\eqref{eq:conserved} vanish when summed.
		(b) Second-order Pauli diagrams, which contain products of lower order terms.
		%
	}
\end{figure}

We have developed a diagrammatic method for the matrix elements of the {\TCL} generator~\(\CS\) and its Pauli counterpart~\(\tilde{\CS}\), which extends upon results from Refs.~\cite{Timm2011TimeconvolutionlessMasterEquation, Nestmann2019TimeconvolutionlessMasterEquation, Gu2020DiagrammaticTimelocalMaster, Schoeller1994MesoscopicQuantumTransport,
	Koller2010DensityoperatorApproachesTransport,
	Schoeller1994ResonantTunnelingCharge,Konig1995ResonantTunnelingCharging,Konig1995ResonantTunnelingCoulomb,Konig1996ResonantTunnelingUltrasmall,Konig1999QuantumFluctuationsSingleElectron}.
Within our scheme, there is no inherited time-order between the two Keldysh branches, i.e., the vertices on each branch are ordered but are free to move left or right with respect to the other branch.
This should be contrasted with the more common approach, where scattering events on different Keldysh branches obey a specific (left-right) ordering with respect to each other~\cite{Schoeller1994MesoscopicQuantumTransport, Schoeller1994ResonantTunnelingCharge}.
This simplification drastically reduces the number of diagrams for~\(\CG_\alpha\) to be evaluated, from~\(2^\alpha\) to~\(\alpha+1\).
These results for~\(\CG_\alpha\) can then be used in combination with products of lower order diagrams to generate the expansion~\(\CS_\alpha\) of the {\TCL}.
We achieved this simplification by using Eq.~\eqref{eq:superUaUnumu}, which recasts the evolution~\(\CU\) of the full density matrix in terms of pairs of the unitary evolution~\(U\), see Appendix~\ref{app:timeorder}.
We thus develop a generalised {\TCL} equivalent of the diagrammatic simplification that was presented in Ref.~\cite{Koller2010DensityoperatorApproachesTransport} at fourth-order for the RT method.
To illustrate the method, we draw the first- and second-order Pauli diagrams in Fig.~\ref{Fig:genericSexamples}.
In the next section, we apply this formalism to setups with quadratic environments, where Wick's theorem allows for further simplifications, and compute explicit rates up to fourth-order for the Pauli {\TCL}.
\section{Quadratic environments} \label{sec:quadratic}
In the following, we focus on setups with quadratic environments.
We restrict ourselves to the case of system--environment couplings that involve a single environment particle, though our diagrams straightforwardly generalise to multi-particle couplings (such as in the Kondo model~\cite{Bruus2004ManyBodyQuantumTheory}).
We use the Pauli {\TCL} as it is sufficient to calculate the steady-state distribution in- and the current through- the system.
The strength of the (Pauli) {\TCL} and our diagrammatic approach is demonstrated at fourth-order in~\(V\), with a natural and correct regularisation and only five diagrams that need to be computed.
The fourth-order rates~\(\tilde{\CS}_4^{if}\) are convergent in the limit~\(\eta\to 0\), as opposed to the T-matrix rates~\(\tilde{\CR}_4^{if}\) which diverge, see, e.g., Eq.~\eqref{eq:S22a} below.
The rates~\(\tilde{\CS}_2^{if}\) and~\(\tilde{\CS}_4^{if}\) provide the exact results for the first two terms~\(P_n^{(0)}\) and~\(P_n^{(2) }\) in the expansion of the steady-state system probabilities~\(P_n\).
We validate our analysis through a comparison with the exact solution of the non-interacting resonant level.
\subsection{Setup}
\label{sec:quadratic:setup}
Our setup consists of an arbitrary system, a quadratic environment, and a system--environment coupling that involves only a single environment operator at each vertex.
This last assumption does not affect the diagrams from Sec.~\ref{sec:formal}. It does, however, influence the specifics of how Wick's theorem is applied, as seen later in this subsection~\footnote{ In the case of multi-particle scattering the coupling~\(V\) must be split into two (or more) consecutive single-particle scattering events that happen immediately one after another.
This procedure, known as point splitting, guarantees that the correct sign will be maintained in the case of fermionic environments~\cite{vonDelft1998BosonizationBeginnersRefermionization}.}.
While we work with a fermionic example, our approach is generic.
To keep the notation concise, we group all indices of the environment into the index~\(\kappa = (\lambda_\kappa,k_\kappa)\).
Here, the indices of all discrete degrees of freedom, such as particle/hole, reservoir~\(r\) and spin~\(\sigma\) are encoded in the discrete multi-index~\(\lambda=\pm( r, \sigma, \dots)\), while~\(k\) indexes momentum.
The sign of the index~\(\lambda\) (\(\sign \kappa \equiv \sign \lambda_\kappa\)) indicates whether we are considering particle creation (\(+\)) or annihilation (\(-\)) as viewed from the reservoir.
The environment Hamiltonian reads 
\begin{align}
\label{eq:quadratic:env}
H_\env=\sum_{\kappa>0} \epsilon_\kappa c_{\kappa}c_{-\kappa},
\end{align}
where~\(c_\kappa\) (\(c_{-\kappa}\)) creates (annihilates) a particle with quantum numbers~\(\kappa\) and energy~\(\epsilon_\kappa\).
Note that the creation and annihilation operators obey~\(c_{\kappa}^{\dagger} = c_{-\kappa}^\pdagger\), and we define~\(\epsilon_{-\kappa} = -\epsilon_\kappa\).
A many-body environment eigenstate~\(\ket{u}_\env\) is constructed by repeatedly applying creation operators~\(\kappa>0\) (in proper order) on the vacuum 
\begin{align}
\label{eq:manyBodyEnvironment}
\ket{u}_\env = \prod_{\kappa\in u} c_{\kappa} \ket{0}_\env,
\end{align}
where~\(\ket{0}_\env\) is the environment vacuum, and~\(u\) is a set of positive indices~\(\kappa\).
In a non-interacting environment, the single-particle energies are additive, with the eigenenergy of the many-body state~\(\ket{u}\) given by 
\begin{align}
\label{eq:additive}
\xi_u = \sum_{\kappa\in u} \epsilon_\kappa.
\end{align}
In contrast to the environment, we keep the system general, i.e., its Hamiltonian is 
\begin{align}
\label{eq:quadratic:sys}
H_\sys=\sum_{n=1}^{N_\sys}\chi_n\ketbra{n}_\sys,
\end{align}
where~\(N_\sys\) is the number of eigenstates in the system.
Furthermore, for simplicity, we assume that the microscopic coupling between the system and environment involves a single environment particle
, allowing us to write 
\begin{align}
\label{eq:quadratic:V}
V = \sum_{nm,\kappa}V_{nm\kappa} c_\kappa\otimes\ketbra{n}{m}_\sys,
\end{align}
where the amplitudes~\(V_{nm\kappa}^{*}=V_{mn-\kappa}^{\phantom *}\) of the perturbation guarantee Hermiticity.
Last, we assume that the amplitudes~\(V_{nm\kappa}\) do not depend on the continuous (momentum) part of the subscript~\(\kappa\).
We can thus interchange~\(V_{nm\kappa}\) and~\(V_{nm\lambda}\) whenever it is convenient.
\begin{figure}
	\includegraphics[width=8.6cm]{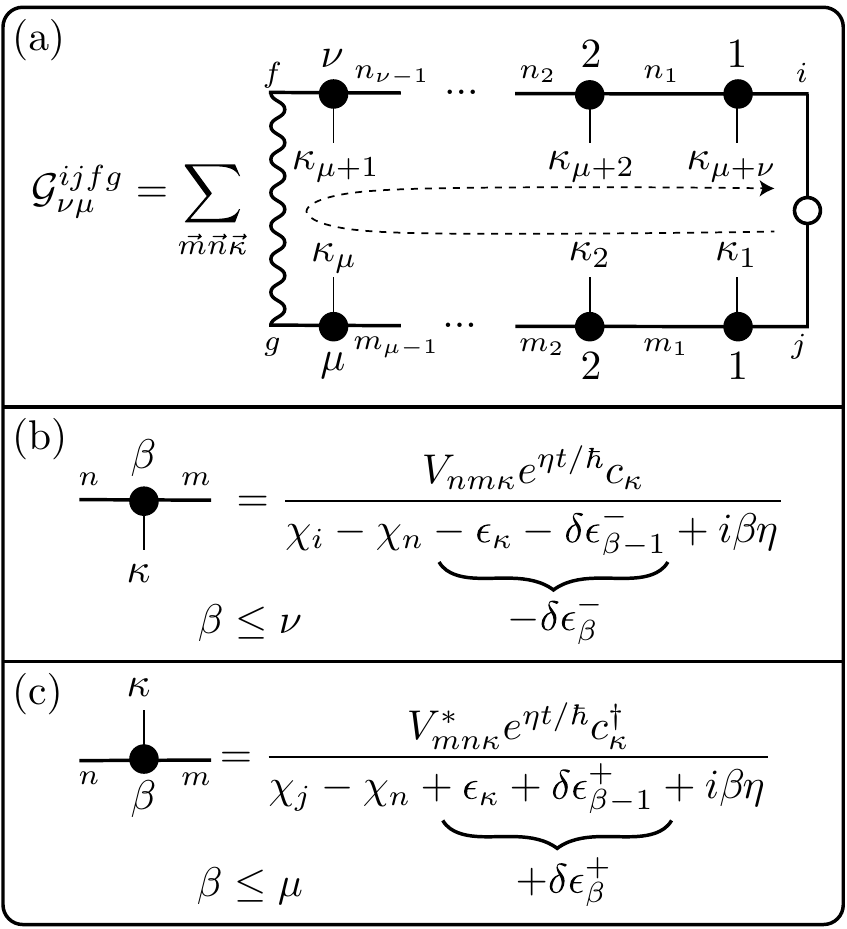}
	\caption{
		\label{Fig:specificDiagram}
		Diagrammatic rules applied to the specific system--environment setup described in~\eqref{eq:quadratic:env}--\eqref{eq:quadratic:V}.
		(a) The diagrams remain structurally the same as those in Fig.~\ref{Fig:genericDiagram} and we have explicitly included the system states~\(i,j,f,g,\vec{n},\vec{m}\) for clarity.
		For each application of the perturbation~\(V\), a (anti-) particle with quantum numbers~\(\kappa\) is added to the environment.
		The environment indices~\(\kappa_1 \dots \kappa_{\mu+\nu}\) are chosen to run clockwise (dashed arrow) through the diagram.
		(b) A factor in the perturbative series on the upper branch, that adds a particle with quantum numbers~\(\kappa\) to the environment, while changing the system state from~\(\ket{m} \to \ket{n}\).
		The operator~\( c_\kappa\) acts on the environment, whereas all other factors are numbers.
		The energy~\(\delta\epsilon_{\beta}^{-}\) tracks the total number of particles (on the upper branch) that have been added to the environment.
		(c) Similar to (b) but for the lower branch of the Keldysh contour.
		We use the fact that the perturbation is Hermitian~\(\sum_{\kappa} V_{nm\kappa}c_{\kappa} = \sum_{\kappa} V^*_{mn\kappa}c^\dagger_{\kappa} \).
		Note, that the Hermitian conjugation on the lower branch (see Fig.~\ref{Fig:genericDiagram}) both flips the order of the environment operators and conjugates them individually 
		\(c_\kappa^\dagger\to c_\kappa\).
		As a result, the environment correlator assumes the form shown in Eq.~\eqref{eq:envCorrelator}.
	}
\end{figure}
We insert the explicit representation of the setup~\eqref{eq:quadratic:env}--\eqref{eq:quadratic:V} into the expansion~\eqref{eq:diagramaticG} for the propagator~\(\CG\), see Fig.~\ref{Fig:specificDiagram}.
Each system--environment scattering event~\(V\) now changes the system state and describes the emission or absorption of an environment particle, i.e., adding or removing a particle in the environment, respectively.
As usual~\cite{Schoeller1994MesoscopicQuantumTransport,Konig1999QuantumFluctuationsSingleElectron, Koller2010DensityoperatorApproachesTransport}, we embody this emission (absorption) process by a thin line attached to each vertex~\(V\) in the diagrams~\ref{Fig:specificDiagram}(a).
The additive property~\eqref{eq:additive} of the single-particle environment energies~\(\epsilon_\kappa\) leads to a simplifications in the matrix elements~\(\CG^{ijfg}_{\nu \mu}\), cf.
Figs.
~\ref{Fig:specificDiagram}(b)--(c) and~\ref{Fig:genericDiagram}(b).
The denominator of the recurrence relation~\eqref{eq:recursiveU} for the time evolution, which appears in~\eqref{eq:diagramaticG}, now changes by one single-particle environment energy~\(\epsilon_\kappa\) after each application of the perturbation~\(V\), see Fig.~\ref{Fig:specificDiagram}(b).
We thus use~\(\epsilon_\kappa\) to update the energy denominator due to the environment~\(\delta\epsilon_\beta^{\pm}\), which tracks the total number of particles added to the environment after~\(\beta\) applications of the perturbation~\(V\) on the upper (\(-\)) or lower (\(+\)) Keldysh branches.
We use a clockwise labelling of the environment indices, see Fig.~\ref{Fig:specificDiagram}(a), and use~\(V\) (\(V^\dagger = V\)) on the upper (lower) Keldysh branch, cf.
Figs.~\ref{Fig:specificDiagram}(b) and~(c).
These last steps give rise to an environment operator expectation value 
\begin{align}
\label{eq:envCorrelator}
\!\!\expval{c_{\kappa_1}c_{\kappa_2}\dots c_{\kappa_{\nu+\mu}}} \!
\equiv\! \sum_{u}P_u^\env\bra{u}c_{\kappa_1}c_{\kappa_2}\dots c_{\kappa_{\nu+\mu}}\ket{u},
\end{align}
of~\(\nu+\mu\) creation or annihilation operators~\(c_\kappa\), where~\(\sum_uP_u^\env\) corresponds to the open circle in our diagrammatic formulation.
\subsubsection{Wick's Theorem} \label{sec:quadratic:setup:Wick} 
\begin{figure}
	\includegraphics[width=8.6cm]{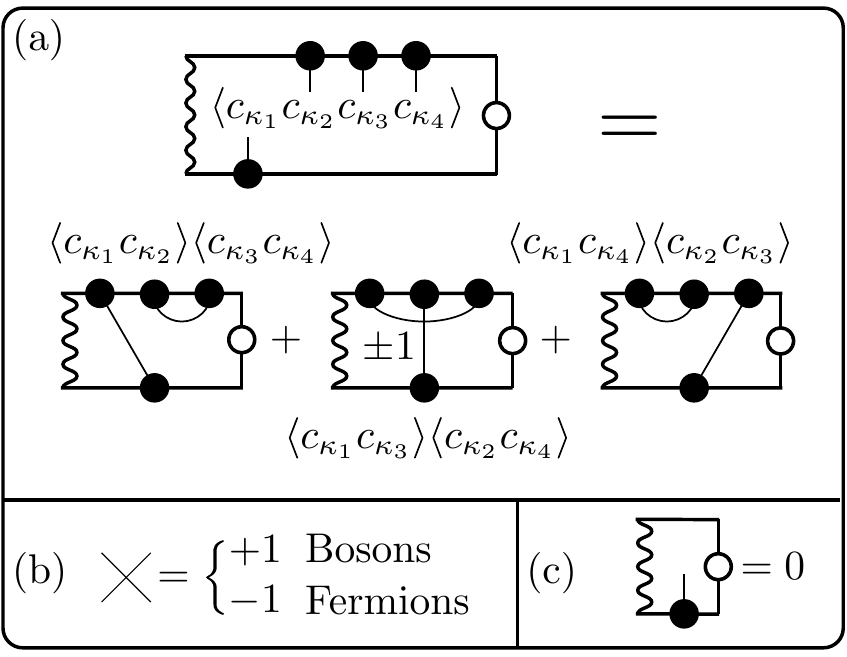}
	\caption{
		\label{Fig:WickTheorem}
		Wick's theorem applied to the diagrams of Fig.~\ref{Fig:specificDiagram} for~\(\CG_{31}\).
		(a) Expectation value of a four-point environment correlator.
		The environment operators~\(c_\kappa\) are arranged clockwise starting in the bottom right due to the hermitian conjugation of the lower branch, see Figs.~\ref{Fig:genericDiagram} and~\ref{Fig:specificDiagram}.
		Wick's theorem~\eqref{eq:WickTheorem} implies that we can decompose the diagram by summing over every possible set of contractions (connected thin lines).
		(b) A number~\(p\) of crossed fermionic environment lines produces a~\((-1)^p\) sign.
		(c) Odd numbers of environment legs result in a vanishing diagram because the unperturbed environment density matrix is diagonal, see Eq.~\eqref{eq:rho0env}.
	}
\end{figure}

The evaluation of locally equilibrated environment correlators to describe the steady-state, is a great strength of the {\TCL} methodology. 
It is a consequence of the combination of the back and forth propagation in Eq.~\eqref{eq:S=RUP}.
We can therefore apply Wick's theorem~\cite{Wick1950EvaluationCollisionMatrix} when evaluating Eq.~\eqref{eq:envCorrelator}, which is thus recursively reduced by
\begin{align}
\label{eq:WickTheorem}
\expval{c_{\kappa_1}c_{\kappa_2}...c_{\kappa_\alpha}}  
= \sum_{p = 2}^\alpha (\pm1)^p\expval{c_{\kappa_1}c_{\kappa_p}}
\big\langle{\prod_{l\neq 1,p} c_{\kappa_l}} \big\rangle.
\end{align}
The sign in Eq.~\eqref{eq:WickTheorem} indicates fermionic (\(-\)) or bosonic (\(+\)) environment modes~\footnote{In mixed systems the sign is negative if, and only if,~\(c_{\kappa_1}\) is fermionic and the number of fermionic operators between~\(c_{\kappa_1}\) and~\(c_{\kappa_p}\) is odd.
}.
Note that the normal-order contribution appearing in Wick's theorem vanishes because the environment is locally-equilibrated in the distant past~\cite{Wick1950EvaluationCollisionMatrix,Evans1996WickTheoremFinite}.
We thus reduce the~\(\alpha\)-point correlator~\eqref{eq:envCorrelator} into a sum over products of two point correlators 
\begin{align}
\label{eq:fermiDirac}
\expval{c_{\kappa}c_{\kappa'}} 
= \delta_{-\kappa, \kappa'} n_{\rs A}(\epsilon- \mu_\lambda, T_\lambda),
\end{align}
where~\(n_{\rs A}\) is the Fermi-Dirac (A\(=\)F) or Bose-Einstein (A\(=\)B) distribution
, with~\(\mu_\lambda\) the chemical potential (\(0\) for massless bosons) and~\(T_\lambda\) the temperature  of the discrete degree of freedom~\(\lambda\),~\(\delta\) is the Kronecker delta, and we again use the fact that the environment is locally equilibrated at~\(t=-\infty\).
Furthermore, in the far distant past, the environment is in an incoherent superposition~\(P_u^\env\) of states~\(\ket{u}_\env\) which each have well defined particle numbers, see Eq.~\eqref{eq:manyBodyEnvironment}.
There is thus no coherent superposition of states with different numbers of particles, and expectation values~\(\expval{c_\kappa}=0\) vanish (as well as other odd correlators).
Within a diagrammatic formulation, Eqs.~\eqref{eq:WickTheorem} and~\eqref{eq:fermiDirac} are implemented by: 
\begin{enumerate}
	\item Create all possible sets of pairs of scattering events and connect the vertices within each pair by an environment line.
	For a connected pair of scattered environment particles replace~\(c_{\kappa} \dots c_{\kappa'} \to \dots \expval{c_{\kappa}c_{\kappa'}} \), i.e., the appropriate equilibrium distribution function as given by Eq.~\eqref{eq:fermiDirac}.
	\item Crossing fermionic lines produce additional minus signs.
\end{enumerate}
In Fig.~\ref{Fig:WickTheorem}, we provide a specific example for the calculation of~\(\CG_{31}\).
Pairs that reside on the same Keldysh branch do not influence the environment state, when considered together.
On the other hand, a pair connecting the upper and lower branch is a physical process where the system emits a particle (or anti-particle) into the environment, i.e., these diagrams will contribute to the current flow in the steady-state.
Diagrammatically, the (thin) Wick contraction lines act in the same way as the (weighted) trace line with an open circle that appears on the right of every diagram, see also Fig.~\ref{Fig:genericDiagram}.
Using the diagrammatic rules for the Wick decomposition, we build each superoperator~\(\tilde{\CG}_{\nu\mu} \) (or~\(\CG_{\nu\mu}\)) as the sum over its Wick contributions
\begin{align}
\label{eq:WickDecomposition}
\tilde{\CG}_{\nu\mu} = \sum_{w}\tilde{\CG}_{\nu\mu w}, 
\end{align} 
where we have introduced the Wick index~\(w\), see below for an example.
The Wick index~\(w\) shows up in the same way in the calculation of other superoperators such as~\(\tilde{\CR}_{\nu\mu}\)~\eqref{eq:PauliSuperRnumu} or~\(\tilde{\CS}_{\nu\mu}\)~\eqref{eq:PauliSuperSnumu}.
At second order, there is only one contraction~\(\expval{c_{\kappa_1}c_{\kappa_2}}\) and the Wick index can be dropped.
At fourth (second non-vanishing) order the four-point correlator decomposes as 
\begin{align}
\label{eq:WickExample}
\expval{c_{\kappa_1}c_{\kappa_2}c_{\kappa_3}c_{\kappa_4}}
=&\expval{c_{\kappa_1}c_{\kappa_4}}\expval{c_{\kappa_2}c_{\kappa_3}}
\\ \nonumber
&
\pm 
\expval{c_{\kappa_1}c_{\kappa_3}}\expval{c_{\kappa_2}c_{\kappa_4}}
+
\expval{c_{\kappa_1}c_{\kappa_2}}\expval{c_{\kappa_3}c_{\kappa_4}},
\end{align}
where the~\(\pm\) sign accounts for the bosonic (\(+\)) or fermionic (\(-\)) nature of the particles, see Fig.~\ref{Fig:WickTheorem}(b).
We denote the three contractions in Eq.~\eqref{eq:WickExample} by the Wick indices~\(w=a,b,c\), respectively.
Note that, the linear additivity of energies~\eqref{eq:additive}  coupled with Wick's theorem implies 
\begin{align}
\left( 
\epsilon_{\kappa_1}
+
\epsilon_{\kappa_2}
+
...
+
\epsilon_{\kappa_\alpha}
\right)
\expval{c_{\kappa_1}c_{\kappa_2}... c_{\kappa_\alpha}}
 = 0,
\end{align}
which we will use to simplify expressions for~\(\CG_{\nu\mu}\) later.
At this point it is useful to recall the hierarchy 
\begin{align}
\label{eq:superGConstruction}
\CG 
= \!\sum_\alpha \CG_{\alpha} 
= \!\sum_{\alpha}\sum_{\nu+\mu=\alpha} \CG_{\nu\mu} 
= \!\sum_{\alpha}\sum_{\nu+\mu=\alpha}\sum_w \CG_{\nu\mu w},
\end{align}
of the indices~\(\alpha,\nu,\mu\), and~\(w\) as we will use them extensively for explicit calculations of the rates.
Furthemore, we note that in the Pauli {\TCL} rates, complex conjugation is equivalent to flipping the upper and lower Keldysh branches.
This means that~\(\tilde{\CG}_{\nu\mu}^{if}=\tilde{\CG}_{\mu\nu}^{if*}\), and that for each contraction~\(w\) there exists a contraction~\(w'\) such that~\(\tilde{\CG}_{\nu\mu w}^{if}=\tilde{\CG}_{\mu\nu w'}^{if*}\).
In some cases we have~\(w=w'\), e.g., at second order where there is a single contraction.
\subsection{Explicit rates}
We focus on the Pauli {\TCL} as it is sufficient for the purpose of calculating steady-state occupations and currents.
First, we show how to recover Fermi's golden rule as encoded in the leading (or second) order in~\(V\) Pauli {\TCL} master equation with its associated rates~\(\tilde{\CS}^{if}_2\), see Eq.~\eqref{eq:FGR} and Fig.~\ref{Fig:mesoDiagramsSeq}.
These rates involve the exchange of a single particle between the system and an environment reservoir, and are commonly dubbed \textit{sequential tunnelling rates} in mesoscopic research~\cite{Ihn2009SemiconductorNanostructuresQuantum,Bruus2004ManyBodyQuantumTheory}, see Figs.~\ref{Fig:mesoDiagramsSeq}(b) and~(c).
We proceed with the fourth-order in~\(V\) Pauli {\TCL}, where the rates~\(\tilde{\CS}_4^{if}\) are associated with the exchange of up to two particles (potentially from different reservoirs) between the system and the environment.
These rates include co- and pair-tunnelling~\cite{Bruus2004ManyBodyQuantumTheory}, see Figs.~\ref{Fig:mesoDiagrams22}(b) and~(c), as well as virtually-assisted sequential- tunnelling which are responsible for level broadening and renormalisation~\cite{Schoeller1994ResonantTunnelingCharge,Koller2010DensityoperatorApproachesTransport}, see Fig.~\ref{Fig:mesoDiagrams31}(b).
The strength of the STCL already manifests at this order: while such fourth-order rates diverge in the T-matrix formulation~\cite{Turek2002CotunnelingThermopowerSingle,Timm2008TunnelingMoleculesQuantum}, the STCL, by construction, systematically removes these divergences~\cite{Timm2011TimeconvolutionlessMasterEquation}.
For illustration, we will demonstrate a perfect agreement of the STCL rates with exact results for a single-particle level, see Fig.~\ref{Fig:nonInteracting}.
\subsubsection{Fermi's golden rule} \label{sec:rates:FGR} 
\begin{figure}
	\includegraphics[width=8.6cm]{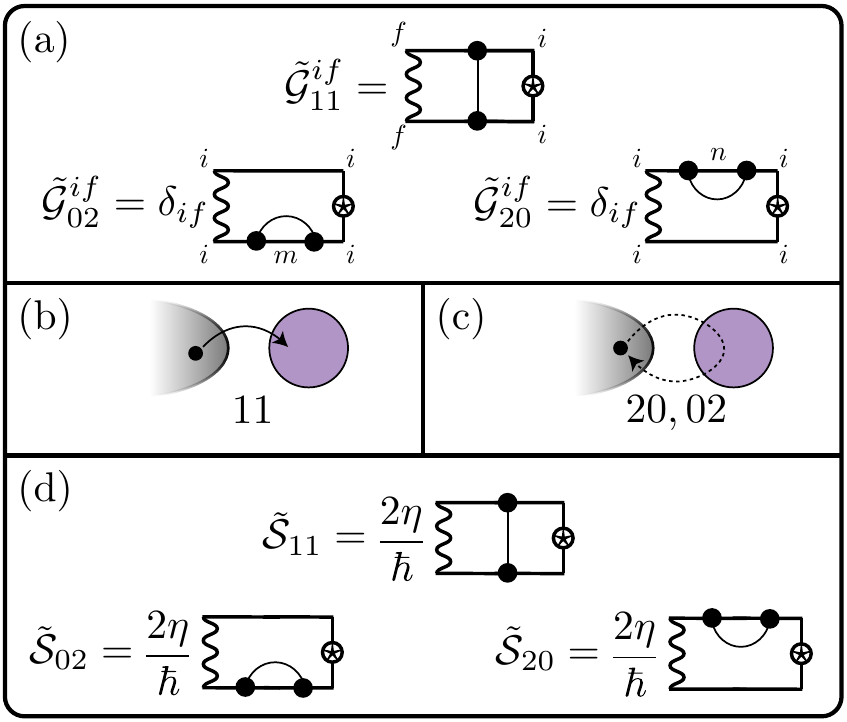}
	\caption{
		\label{Fig:mesoDiagramsSeq}
		Second-order diagrams.
		(a) The second order~\(\tilde{\CG}_2\) diagrams, where we have dropped the sums over system (\(n,m\)) and environment (\(\kappa\)) variables for clarity.
		(b)--(c) Illustration of the various processes occurring in the diagrams of (a).
		The solid line in (b) indicates a sequential tunnelling process, where the initial and final states of the system differ by one (anti-) particle that has tunnelled out of the environment.
		The dotted line in (c) indicates a probability conserving process, where a (anti-) particle tunnels back and forth between the environment and system, leaving the setup unchanged.
		(d) Second-order rates~\(\tilde{\CS}_2\), where we have dropped all sums and indices for clarity.
		Unlike the more general case in Fig.~\ref{Fig:genericSexamples}, due to Wick's theorem [see Fig.~\ref{Fig:WickTheorem}(c)], all odd-order diagrams (\(\tilde{\CG}_{01}\) and~\(\tilde{\CG}_{10}\)) vanish and there are therefore no subtractions of products of lower orders.
	}
\end{figure}

Fermi's golden rule~\cite{Sakurai1985ModernQuantumMechanics} successfully describes transitions in open quantum systems with weak couplings to large environments; it commonly includes solely the diagonal of the system's density matrix (Pauli) but can be extended to describe the system's coherences as well.
We consider the Pauli {\TCL}~\eqref{eq:TCLMEPauli} at lowest non-vanishing (second) order in the perturbation~\(V\) given in Eq.~\eqref{eq:quadratic:V}.
The first order term has an odd number of environment operators~\(c_\kappa\) in the correlator~\eqref{eq:envCorrelator} and thus vanishes, see Fig.~\ref{Fig:WickTheorem}(c).
At second order, we have three diagrams  for~\(\tilde{\CG}\).
We apply the rules from Figs.~\ref{Fig:genericDiagram},~\ref{Fig:compositionRule},~\ref{Fig:specificDiagram}, and~\ref{Fig:WickTheorem} to find 
\begin{align}
\label{eq:PauliG2}
\tilde{\CG}^{if}_{11}&=
\sum_{\kappa}
\frac{|V_{if\kappa}|^2 \expval{c_\kappa c_{-\kappa}} }
{(\epsilon_\kappa+\delta\chi_{if})^2+\eta^2},
\\ \nonumber
\tilde{\CG}^{if}_{20}&=
\delta_{if}\frac{1}{2i\eta}
\sum_{m\kappa}
\frac{|V_{im\kappa}|^2 \expval{c_\kappa c_{-\kappa}}}
{\epsilon_\kappa+\delta\chi_{im}+i\eta},
\qquad
\tilde{\CG}^{if}_{02} = (\tilde{\CG}^{if}_{20})^*,
\end{align}
that we display in Fig.~\ref{Fig:mesoDiagramsSeq}(a) (recall~\(\delta\chi_{nm} \equiv \chi_n-\chi_m\)).
The~\(\tilde{\CG}_{11}\) diagrams are associated with a sequential tunnelling event, see Fig.~\ref{Fig:mesoDiagramsSeq}(b), where a particle tunnels in or out of the system.
The~\(\tilde{\CG}_{20}\) and~\(\tilde{\CG}_{02}\) diagrams on the other hand are associated with probability conserving events, see Eq.~\eqref{eq:conserved}, where an excitation tunnels back and forth between the system and the environment along one of the Keldysh branches, with the setup finally returning to the initial state.
Note that in our discussion, we will alternate between the superoperators~\(\tilde{\CS}\),~\(\tilde{\CG}\) and their matrix elements~\(\tilde{\CS}^{if}\),~\(\tilde{\CG}^{if}\), according to convenience.
We substitute Eq.~\eqref{eq:PauliG2} into Eqs.~\eqref{eq:PauliSuperSnumu} and~\eqref{eq:matrixElementConstraint}  to compute explicit values for the second-order Pauli {\TCL} (equivalently, Fermi's golden rule) rates.
We recall that~\(\kappa=(\lambda_\kappa,k_\kappa)\) to convert the sum over momentum~\(k_\kappa\) in Eq.~\eqref{eq:PauliG2} into an integral over energy 
\begin{align}
\label{eq:sumToInt}
\sum_\kappa  
\to \sum_\lambda \int \dd{\epsilon} \DOS_{\lambda}\Delta_\lambda(\epsilon)%
,
\end{align}
while maintaining the sum over discrete degrees of freedom~\(\lambda\).
The density of states~\(\DOS_{\lambda}\Delta_\lambda(\epsilon)\)  associated with~\(\lambda\) is composed of two parts.
The first~\(\DOS_\lambda\) is energy independent and is a typical scale of the density of states, while the second~\(\Delta_\lambda(\epsilon)\) is a dimensionless function that encodes the dependence on energy.
Furthermore, we evaluate the expectation value of the (fermionic) environment operators using~\eqref{eq:fermiDirac} and introduce the real and dimensionless spectral function
\begin{align}
\label{eq:spectral}
\spectral_\lambda(\epsilon) 
= \Delta_\lambda(\epsilon) n_{\rs F}(\epsilon- \mu_\lambda, T_\lambda).
\end{align}
We can then write the second-order Pauli {\TCL} (Fermi's golden rule) rates in the form 
\begin{align}
\label{eq:FGR}
\tilde{\CS}^{if}_{2}&=\frac{2\pi}{\hbar} \sum_{\lambda}
|V_{if\lambda}|^2\DOS_\lambda C_{\lambda}(\delta\chi_{fi})
\\ \nonumber
&\qquad-\frac{2\pi}{\hbar}\delta_{if} \sum_{m\lambda}
|V_{im\lambda}|^2 \DOS_\lambda C_{\lambda}(\delta\chi_{mi}),
\end{align}
which is the sum of the three diagrams shown in Fig.~\ref{Fig:mesoDiagramsSeq}(d).
The first term (\(\tilde{\CS}_{11}\)) describes sequential tunnelling and is identical to the rates obtained from Fermi's golden rule.
The second term 
\begin{align}
\tilde{\CS}^{if}_{20}+\tilde{\CS}^{if}_{02} = -\delta_{if} \sum_m\tilde{\CS}_{11}^{im},
\end{align} 
is a manifestation of conservation of probability, see Eq.~\eqref{eq:conserved}.
The explicit rates for the lowest-order full (as opposed to Pauli) {\TCL} for quadratic environments is not required for our discussion but can be found in Ref.~\cite{FergusonThesis}.
\subsubsection{Co- and pair-tunnelling} \label{sec:rates:cot} 
Fourth-order processes are those that arise from matrix elements with~\(\nu+\mu =\alpha= 4\).
In contrast to the T-matrix rates~\(\tilde{\CR}_{4}\sim1/\eta\) associated with the same physical processes, the STCL rates~\(\tilde{\CS}_{4}\propto\eta^0\) are convergent in the limit~\(\eta\to 0\).
We start by computing Pauli {\TCL} co- and pair-tunnelling rates~\(\tilde{\CS}_{22}\), before moving to the rates~\(\tilde{\CS}_{31}\) and~\(\tilde{\CS}_{13}\) which renormalise the lower-order sequential-tunneling rate~\(\tilde{\CS}_{11}\) from Fermi's golden rule.
We avoid computing the rates~\(\tilde{\CS}_{40}\) and~\(\tilde{\CS}_{04}\), as they are merely a manifestation of conservation of probability~\eqref{eq:conserved}.
\begin{figure}
	\includegraphics[width=8.6cm]{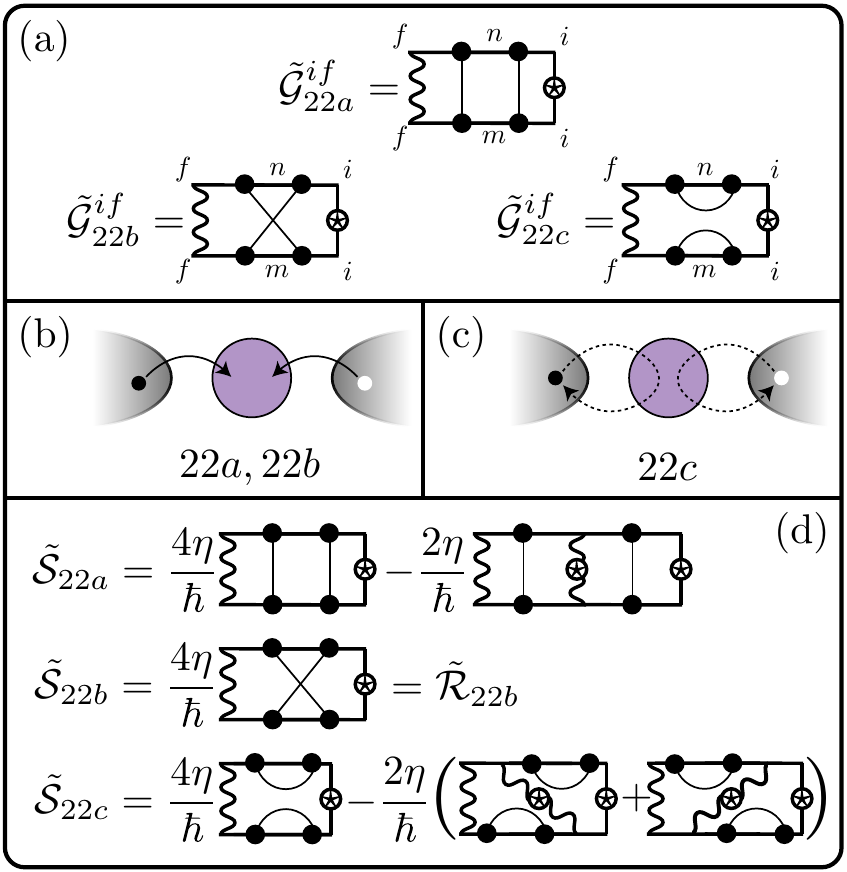}
	\caption{
		\label{Fig:mesoDiagrams22}
		Co- and pair-tunnelling.
		(a) Fourth-order Pauli diagrams for~\(\tilde{\CG}_{22}\), see Eq.~\eqref{eq:PauliG22},  where we have dropped the sums over system (\(n,m\)) variables for clarity.
		(b) Illustration of the co- and pair-tunnelling processes associated with the diagrams~\(\tilde{\CG}_{22a}\) and~\(\tilde{\CG}_{22b}\).
		A particle or antiparticle (black dot) tunnels into the system from the environment simultaneously with a second particle or anti-particle (white dot) from the same or a different reservoir.
		If both particles tunnelling to the system are electrons (or holes) this process is known as pair tunnelling.
		On the other hand, if one particle is a hole and the other is an electron, the process is known as a cotunnelling event.
		An elastic cotunnelling event is one which leaves the system state unchanged while in an inelastic process the initial and final system states differ.
		(c) Illustration of processes associated with the diagram~\(\tilde{\CG}_{22c}\).
		A particle or anti-particle tunnels back and forth between the system and the environment along each of the Keldysh branches.
		This process leaves the environment unchanged, while it can modify the state of the system.
		(d) Diagrammatic representation of the three contractions for~\(\tilde{\CS}_{22}\) with all indices dropped.
	}
\end{figure}
The matrix elements 
\begin{align}
\label{eq:PauliG22}
&\tilde{\CG}^{if}_{22}=
\sum_{nm \vec{\kappa}} 
\frac{V_{in\kappa_1}V_{nf\kappa_2}V_{fm\kappa_3}V_{mi\kappa_4}}
{(\delta\chi_{if}+\epsilon_{\kappa_1}+\epsilon_{\kappa_2})^2+4\eta^2}
\times
\\ \nonumber
&\qquad\qquad\qquad
\frac{ \expval{c_{\kappa_1}c_{\kappa_2}c_{\kappa_3}c_{\kappa_4}}}
{(\delta\chi_{in}+\epsilon_{\kappa_1}-i\eta)(\delta\chi_{im}-\epsilon_{\kappa_4}+i\eta)},
\end{align}
required to compute~\(\tilde{\CS}_{22}\), contains two occurrences of the perturbation on each of the Keldysh branches.
We combine Eq.~\eqref{eq:PauliG22} with the contractions~\eqref{eq:WickExample} to obtain the matrix elements of the superoperators~\(\tilde{\CG}_{22a}\),~\(\tilde{\CG}_{22b}\), and~\(\tilde{\CG}_{22c}\), see Fig.~\ref{Fig:mesoDiagrams22}(a).
These matrix elements correspond to new processes beyond sequential tunnelling, where the initial and final state differ by an even (possibly 0) number of particles.
Co- and pair-tunnelling arise from the diagrams~\(\tilde{\CG}_{22a}\) and~\(\tilde{\CG}_{22b}\), see Fig.~\ref{Fig:mesoDiagrams22}(b).
In a cotunnelling process, an electron or hole tunnels onto the system while at the same time a second particle of the same kind tunnels out of the system.
These two particles may originate from the same or different reservoirs, and a cotunnelling process may change the system state, but not its charge.
If the initial and final system states are identical such a process is termed \textit{elastic} cotunnelling.
A pair-tunnelling process on the other hand, occurs when two electrons (or holes) from the same or different reservoirs simultaneously tunnel into the system, thus changing the state and charge of the system.
The~\(\tilde{\CG}_{22c}\) diagrams gives rise to processes where two particles (electrons or holes), one on each Keldysh branch, tunnel in and out of the system, leaving the environment unchanged, see Fig.~\ref{Fig:mesoDiagrams22}(c).
The rates~\(\tilde{\CS}_{22c}\) associated with these processes vanish, as we show in Eq.~\eqref{eq:S22c}.
\textit{Non-crossing co- and pair-tunnelling.}
We first consider the type~\(a\) Wick contraction from Eq.~\eqref{eq:WickExample}.
We note that the associated T-matrix co- and pair-tunnelling rates, evaluated by substituting Eq.~\eqref{eq:PauliSuperRnumu} into Eq.~\eqref{eq:PauliG22} for the~\(a\) contraction, scale as 
\begin{align}
\label{eq:R22a}
\tilde{\CR}^{if}_{22a} = \frac{4\eta}{\hbar}\tilde{\CG}^{if}_{22a} 
\propto \frac{1}{\eta} +\bigO(\eta^0),
\end{align}
and thus diverge in the limit~\(\eta \to 0\)~\cite{Timm2011TimeconvolutionlessMasterEquation}.
On the other hand, combining the recurrence relation~\eqref{eq:PauliSuperSnumu} for the {\TCL} with the same contraction provides us with the rates 
\begin{align}
\label{eq:S22ashort}
\tilde{\CS}_{22a}^{if}=\frac{4\eta}{\hbar}\tilde{\CG}_{22a}^{if}
&-
\frac{2\eta}{\hbar}\sum_m 
\tilde{\CG}_{11}^{im}
\tilde{\CG}_{11}^{mf}
\propto \eta^0,
\end{align}
which contains corrections from two consecutive sequential-tunnelling events~\(\tilde{\CG}_{11}^{im} \tilde{\CG}_{11}^{mf}\), see Fig.~\ref{Fig:mesoDiagrams22}(d).
These corrections cancel the diverging part of~\eqref{eq:R22a} and leave the Pauli {\TCL} rates~\(\tilde{\CS}_{22a}^{if}\) convergent in the limit~\(\eta\to 0\).
Diagrammatically, corrections due to products of lower order occur whenever a diagram for~\(\tilde{\CG}\) can be cut vertically without slicing a contraction line.
We substitute the matrix elements~\(\tilde{\CG}^{if}_{22a}\) [combining Eqs.~\eqref{eq:PauliG22} and~\eqref{eq:WickExample}] and~\(\tilde{\CG}^{if}_{11}\) from~\eqref{eq:PauliG2} into Eq.~\eqref{eq:S22ashort}, transform the sums over momenta to integrals over energy and assume constant density of states~\eqref{eq:sumToInt}. We then evaluate the expectation values~\eqref{eq:fermiDirac} for environment operators and take the~\(\eta\to 0\) limit to obtain the expression 
\begin{widetext}
	\begin{subequations}
		\label{eq:S22a}
		\begin{align}
		\label{eq:S22anneqm}
		\tilde{\CS}_{22a}^{if}
		=&\frac{2\pi}{\hbar}\sum_{\substack{n\neq m\\\lambda_1\lambda_2}}
		\frac{V_{in\lambda_1}V_{nf\lambda_2}V_{fm-\lambda_2}V_{mi-\lambda_1}}{\delta\chi_{nm}}\DOS_{\lambda_1}\DOS_{\lambda_2}
		\left[
		I_{\lambda_1 \lambda_2}^-(\delta\chi_{in},\delta\chi_{fn})
		-
		I^+_{\lambda_1 \lambda_2}(\delta\chi_{im},\delta\chi_{fm})
		\right]
		\\
		\label{eq:S22ameqnMat}
		&+
		\frac{2\pi}{\hbar}\sum_{\substack{m\\\lambda_1\lambda_2}} |V_{im\lambda_1}|^2|V_{mf\lambda_2}|^2 \DOS_{\lambda_1}\DOS_{\lambda_2}  \partial_{\chi_m} \Re I^+_{\lambda_1 \lambda_2}(\delta\chi_{im},\delta\chi_{fm}) \\ \label{eq:S22ameqnTCLCor} 
		&+\frac{2\pi}{\hbar}\sum_{\substack{m\\\lambda_1\lambda_2}}
		|V_{im\lambda_1}|^2|V_{mf\lambda_2}|^2 
		\DOS_{\lambda_1}\DOS_{\lambda_2}
		\Re
		\big[
		C_{\lambda_2}(\delta\chi_{fm}) 
		\partial_{\chi_i}
		{J_{\lambda_1}^+}(\delta\chi_{im})
		+C_{\lambda_1}(\delta\chi_{mi})
		\partial_{\chi_f}
		{J_{\lambda_2}^+}(\delta\chi_{mf})
		\big],
		\end{align}	
	\end{subequations}
\end{widetext}
with dimensionless integrals~\(I\) and~\(J\) defined in Eqs.~\eqref{eq:IntegralIpm} and~\eqref{eq:IntegralJpm}.
To arrive at this formula, several lengthy but straightforward steps are required.
These are outlined in detail in Appendix~\ref{app:rates}.
Eq.~\eqref{eq:S22ashort} is a typical example of the subtraction of reducible contributions~\eqref{eq:superRecursive}, where two cancelling diverging terms in~\(1/\eta\) appear (one from~\(\tilde{\CG}_{22a}\) and one from the product~\(\tilde{\CG}_{11}\tilde{\CG}_{11}\)).
In the course of this calculation, we have expanded the expressions in Eq.~\eqref{eq:S22a} as Laurent series in~\(\eta\) before taking the~\(\eta\to 0\) limit; in doing so, we have replaced the derivatives with respect to~\(\eta\) arising from this expansion by derivatives with respect to~\(\chi\).
The first line~\eqref{eq:S22anneqm} in the cotunnelling rate~\(\tilde{\CS}_{22a}^{if}\) is a convergent contribution which is attributed to tunnelling through two different system states~\(n\neq m\) on each of the Keldysh branches, see Fig.~\ref{Fig:mesoDiagrams22}(a).
The same contribution appears in the T-matrix rate~\(\tilde{\CR}_{22a}^{if}\).
The second and third lines~\eqref{eq:S22ameqnMat} correspond to a co- or pair-tunelling process where the  intermediate states on both Keldysh branches are the same, i.e.,~\(n=m\).
The second line coincides with the contribution obtained in the T-matrix approach using the phenomenological regularisation scheme developed in Ref.~\cite{Turek2002CotunnelingThermopowerSingle}, see Ref.~\cite{FergusonThesis}.
The third line~\eqref{eq:S22ameqnTCLCor} contains additional corrections which are of the same order as~\eqref{eq:S22ameqnMat} but are missing in the regularisation of Ref.~\cite{Turek2002CotunnelingThermopowerSingle}, see also Refs.~\cite{Averin1994PeriodicConductanceOscillations,Koch2006TheoryFranckCondonBlockade,Timm2008TunnelingMoleculesQuantum,Koller2010DensityoperatorApproachesTransport}.
For a detailed discussion comparing Eqs.~\eqref{eq:S22a} to the corresponding result using the SRT approach, we refer the reader to Ref.~\cite{FergusonThesis}.
In Eqs.~\eqref{eq:S22a}, we introduced two dimensionless integrals~\(I\) and~\(J\).
To evaluate them, we assume constant densities of states~\(\Delta_\lambda(\epsilon)=1\) for each environment degree of freedom and further assume a constant temperature~\(T\) across the entire environment in the distant past.
Under these assumptions, the first integral~\(I\) is (see Appendix~\ref{app:integrals}) 
\begin{widetext}
	\begin{align}
	\label{eq:IntegralIpm}
	I^\pm_{\lambda_1 \lambda_2}(\delta_1,\delta_2) &= \lim_{\eta\to 0}
	\int
	\frac{C_{\lambda_1}(\epsilon-\delta_1)C_{\lambda_2}(\delta_2-\epsilon)}
	{\epsilon\pm i\eta} \dd{\epsilon}
	=n_{\rs B}(\delta_2-\delta_1
	-\mu_{\lambda_1}\!-\mu_{\lambda_2})
	\left[
	n_\psi\left(\mp\delta_2\pm\mu_{\lambda_2}\right)
	\!-\!
	n_\psi\left(\mp\delta_1\mp\mu_{\lambda_1}\right)
	\right],
	\end{align}
\end{widetext}
	where~\(n_{\rs B}\) is the Bose-Einstein distribution at temperature~\(T\) and~\(n_\psi(\delta) = \psi(1/2+i\delta/2\pi  T)\) is a generalised distribution in terms of the digamma function~\(\psi\).
	Note that, the latter is related to both the Fermi-Dirac and Bose-Einstein distributions, see Appendix~\ref{app:integrals}.
	Furthermore, we have dropped the explicit dependence on  temperature~\(T\) in all distribution functions~\(n_{\rs A}\), as we assume all temperatures in the setup to be equal.
	The second integral~\(J\) is (see Appendix~\ref{app:integrals}) 
	\begin{align}
	\label{eq:IntegralJpm}
	&J^\pm_\lambda(\delta) 
	=\lim_{\eta\to 0}\int\frac{n_{\rs F}(\epsilon-\delta-\mu_\lambda)}
	{\epsilon\pm i\eta} \dd{\epsilon}
	\\\nonumber
	&=\mp i\pi n_{\rs F}(-\delta-\mu_\lambda)
	\!+\Re n_\psi\left(\delta + \mu_\lambda\right)
	\!+\ln(\Lambda_{\lambda})+\bigO(\Lambda_\lambda^{-1}),
	\end{align}
where~\(\Lambda_\lambda\) is an ultraviolet cutoff for the continuum variable~\(k\) in the environment reservoir~\(\lambda\).
Typically, in electronic systems, this is the bandwidth of the Fermi reservoirs.
Due to the derivatives with respect to~\(\chi\) acting on the integrals~\(J\) in Eq.~\eqref{eq:S22ameqnTCLCor}, the ultraviolet cutoffs do not affect the rates~\(\tilde{\CS}_{22a}\) in the wide band limit~\(\Lambda \to \infty\), see Appendix~\ref{app:integrals}.
\textit{Crossing co- and pair-tunnelling.} Next, we consider the~\(b\) contraction, see Eq.~\eqref{eq:WickExample}, which we combine with Eqs.~\eqref{eq:PauliSuperSnumu} and~\eqref{eq:PauliSuperRnumu} to obtain 
\begin{align}
\label{eq:S22bshort}
\tilde{\CS}_{22b}^{if}
=\tilde{\CR}_{22b}^{if}
=\frac{4\eta}{\hbar}\tilde{\CG}_{22b}^{if}.
\end{align}
This process includes crossing contraction lines, see Fig.~\ref{Fig:mesoDiagrams22}, and thus depends on the particle statistics (in the present case fermions).
For the~\(b\) contraction it is not possible to cut the~\(\tilde{\CG}_{22b}\) diagram vertically without cutting a contraction line, see Figs.~\ref{Fig:mesoDiagrams22}(a) and (d).
Thus, there are no corrections from second-order products in Eq.~\eqref{eq:S22bshort} and the associated T-matrix, {\TCL} and SRT rates are identical and convergent as~\(\eta\to 0\).
Combining Eqs.~\eqref{eq:PauliG22}, and~\eqref{eq:S22bshort}, the evaluation of this term yields 
\begin{widetext}
	\begin{align}
	\label{eq:S22b}
	\tilde{\CS}_{22b}^{if} = \frac{4\eta}{\hbar}\tilde{\CG}_{22b}^{if}
	=\frac{2\pi}{\hbar}\sum_{mn\lambda_1\lambda_2}
	\frac{V_{in\lambda_1}V_{nf\lambda_2}V_{fm-\lambda_1}V_{mi-\lambda_2}}
	{\delta\chi_{mf}+\delta\chi_{ni}}
	\left[
	I_{\lambda_1\lambda_2}^-(\delta\chi_{in},\delta\chi_{fn})
	-
	I_{\lambda_1\lambda_2}^-(\delta\chi_{mf},\delta\chi_{mi})
	\right].
	\end{align}
\end{widetext}
\textit{No-tunnelling.} We now turn to the third contraction~\(c\) for the rates associated with~\(\tilde{\CG}_{22}\).
These are associated with processes that do not change the state of the environment.
Substituting the contraction~\(c\) into the expression for the matrix elements of~\(\tilde{\CG}_{22}\), see Eq.~\eqref{eq:PauliG22}, we find that the associated T-matrix rates~\eqref{eq:PauliSuperRnumu} scale as 
\begin{align}
\label{eq:R22c}
\tilde{\CR}^{if}_{22c} = \frac{4\eta}{\hbar}\tilde{\CG}^{if}_{22c} 
\propto \frac{\delta_{if}}{\eta} + \bigO(\eta).
\end{align}
This expression diverges in the~\(\eta\to 0\) limit for equal initial and final states and vanishes otherwise.
It therefore does not contribute to the rate equation (or current rates see Section~\ref{sec:currents}) and was therefore discarded in the phenomenological regularisation scheme of Ref.~\cite{Turek2002CotunnelingThermopowerSingle}.
The corresponding {\TCL} rates, see Fig.~\ref{Fig:mesoDiagrams22}(d), are properly regularised, 
\begin{align}
\label{eq:S22c}
\tilde{\CS}_{22c}^{if} = 4\eta\tilde{\CG}_{22c}^{if}
- 4\eta\delta_{if}\tilde{\CG}^{ii}_{20}\tilde{\CG}^{ii}_{02} \propto \eta,
\end{align}
and vanish in the~\(\eta\to 0\) limit, further justifying the omission of~\(\tilde{\CR}^{if}_{22c}\) in Ref.~\cite{Turek2002CotunnelingThermopowerSingle}.
The two contributions~\(\tilde{\CS}_{22a}\) and~\(\tilde{\CS}_{22b}\) to the {\TCL} rates are associated with co- and pair-tunnelling, which are distinct from the sequential-tunnelling process at second order.
In the following, we study further contributions at fourth-order, that renormalise the lower-order rates and are interpreted as virtually-assisted sequential tunnelling.
These contributions arise from the~\(\tilde{\CG}_{31}\) and~\(\tilde{\CG}_{13}\) superoperators~\eqref{eq:superGnumu} and are as important as the cotunnelling rates~\eqref{eq:S22a} and~\eqref{eq:S22b} in an exact expansion.
\subsubsection{Virtually-assisted sequential tunnelling} 
\begin{figure}
	\includegraphics[width=8.6cm]{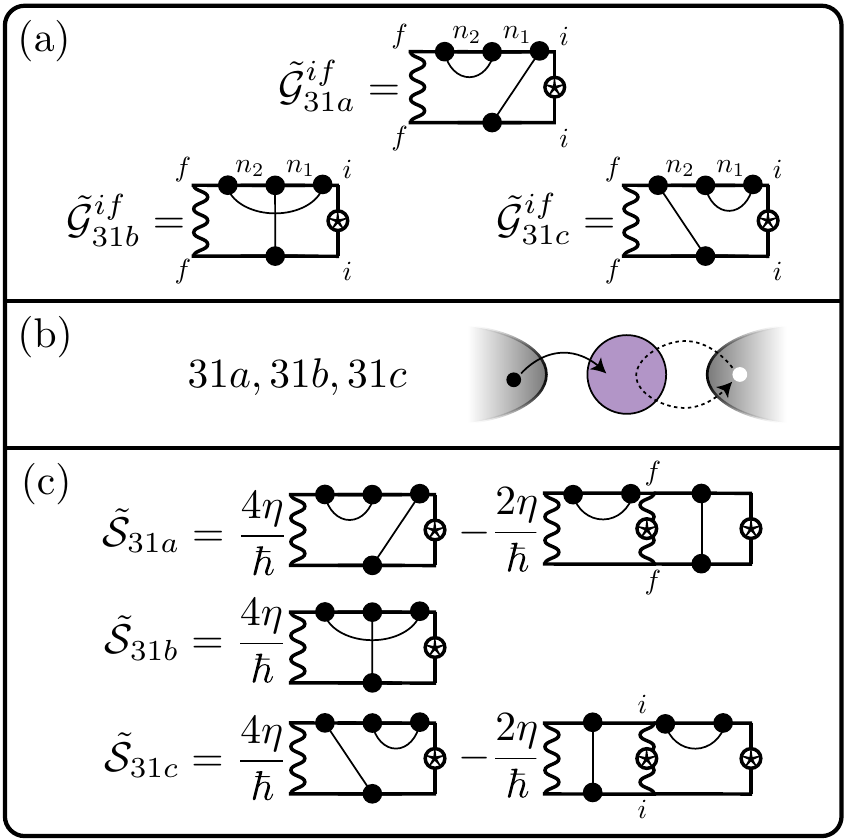}
	\caption{
		\label{Fig:mesoDiagrams31}
		Virtually-assisted sequential-tunnelling processes.
		(a) Fourth-order diagrams for~\(\tilde{\CG}_{31}\).
		(b) Illustration of the physical processes associated with the diagrams in~(a): A particle or anti-particle (black dot) tunnels into (full arrow) the system on each of the Keldysh branches.
		Simultaneously, a second (anti-) particle (white dot) enters (dotted arrow) the system before returning to the environment (along the upper Keldysh branch).
		These processes renormalise the sequential tunnelling~\eqref{eq:FGR} and are thus often ignored~\cite{Koch2004ThermopowerSinglemoleculeDevices}.
		(c) Diagrams for~\(\tilde{\CS}_{31}\).
	}
\end{figure}
We now compute the virtually-assisted sequential-tunnelling rates~\(\tilde{\CS}_{31}\) and~\(\tilde{\CS}_{13}\) in the Pauli {\TCL} master equation, see Eqs.~\eqref{eq:PauliSuperSnumu} and~\eqref{eq:matrixElementConstraint}.
To this end, we use our diagrammatic rules for~\(\tilde{\CG}\) from Sec.~\ref{sec:quadratic:setup} to write 
\begin{align}
\label{eq:PauliG31}
&\tilde{\CG}^{if}_{31}=
\sum_{nm\vec{\kappa}} 
\frac{V_{if\kappa_1}V_{fn\kappa_2}V_{nm\kappa_3}V_{mi\kappa_4}}
{(\delta\chi_{if}+\epsilon_{\kappa_1}-i\eta)
	(\delta\chi_{if}+\epsilon_{\kappa_1}+3i\eta)}
\\\nonumber
&
\qquad\qquad
\times\frac{ \expval{c_{\kappa_1}c_{\kappa_2}c_{\kappa_3}c_{\kappa_4}}}
{(\delta\chi_{in}-\epsilon_{\kappa_3}-\epsilon_{\kappa_4}+2i\eta)
	(\delta\chi_{im}-\epsilon_{\kappa_4}+i\eta)},
\end{align}
with~\(\tilde{\CG}^{if}_{13}=\tilde{\CG}^{if*}_{31}\).
We combine~\eqref{eq:PauliG31} with the contractions~\eqref{eq:WickExample} to obtain the three propagators~\(\tilde{\CG}_{31a}\),~\(\tilde{\CG}_{31b}\), and~\(\tilde{\CG}_{31c}\), see Figs.~\ref{Fig:mesoDiagrams31}(a) and (b), which can be understood as sequential-tunnelling processes where one of the Keldysh branches takes an indirect path to the final state.
We use the recurrence relation~\eqref{eq:PauliSuperSnumu} and the contractions~\eqref{eq:WickExample} to obtain 
\begin{subequations}
	\label{eq:S31short}
	\begin{align}
	\label{eq:S31ashort}
	\tilde{\CS}_{31a}^{if}&=
	\frac{4\eta}{\hbar}\tilde{\CG}_{31a}^{if}
	-\frac{2\eta}{\hbar}\tilde{\CG}_{11}^{if}
	\tilde{\CG}_{20}^{ii},
	\\
	\label{eq:S31bshort}
	\tilde{\CS}_{31b}^{if}&=
	\frac{4\eta}{\hbar}\tilde{\CG}_{31b}^{if},
	\\
	\label{eq:S31cshort}
	\tilde{\CS}_{31c}^{if}
	&=\frac{4\eta}{\hbar}\tilde{\CG}_{31c}^{if}
	-\frac{2\eta}{\hbar}\tilde{\CG}_{11}^{if}
	\tilde{\CG}_{20}^{ff},
	\end{align}
\end{subequations}
see Fig.~\ref{Fig:mesoDiagrams31}(c).
For each contraction~\(w\) there is a~\(w'\) such that~\(\tilde{\CS}_{13w}^{if}=\tilde{\CS}_{31w'}^{if*}\).
For the~\(31\leftrightarrow13\) diagrams the pairs~\((w,w')\) are~\((a,c)\),~\((b,b)\) and~\((c,a)\).
As for~\(\tilde{\CS}_{22b}\)~\eqref{eq:S22bshort}, the~\(b\) contraction~\eqref{eq:S31bshort} does not involve products of the lower-order terms as this diagram cannot be split.
The explicit calculations of the rates~\eqref{eq:S31short} involve similar methods as the calculation of the cotunnelling rates~\(\tilde{\CS}_{22}\) and can be found in Appendix~\ref{app:rates}.
We have now computed all fourth-order rates that are associated to physical processes.
The remaining fourth-order contributions~\(\tilde{\CS}_{40}\) and~\(\tilde{\CS}_{04}\) ensure conservation of probability.
Their sum can be obtained from the superoperators~\(\tilde{\CS}_{22}\),~\(\tilde{\CS}_{31}\), and~\(\tilde{\CS}_{13}\) or calculated explicitly as for the other fourth-order rates.
We take the former route though, for completeness, we now briefly discuss the general structure of~\(\tilde{\CS}_{40}\) and~\(\tilde{\CS}_{04}\).
\subsubsection{Probability conservation at fourth-order} 
\begin{figure}
	\includegraphics[width=8.6cm]{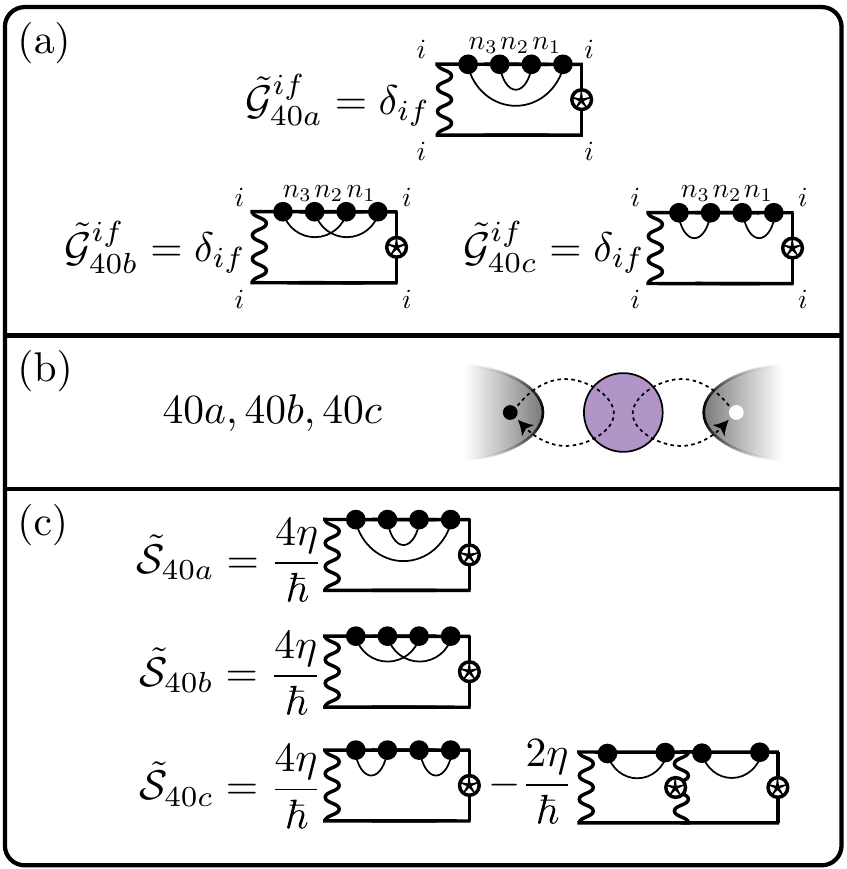}
	\caption{
		\label{Fig:mesoDiagrams40}
		Probability conserving diagrams~\(\CS_{\alpha0}\) at fourth-order~\(\alpha = 4\).
		(a) Fourth-order diagrams for~\(\tilde{\CG}_{40}\).
		These terms each contain a Kronecker delta, which enforces identical initial (\(i\)) and final (\(f\)) system states.
		(b) Illustration of processes described by the diagrams in (a) resembling those for~\(\tilde{\CG}_{22c}\) in Fig.~\ref{Fig:mesoDiagrams22}(c), except for the fact that now the entire setup, rather than just the environment, initial and final states must be the same.
		The different diagrams in (a) correspond to different time orderings of the events in (b).
		(c) Diagrams for~\(\tilde{\CS}_{40}\).
	}
\end{figure}
The probability conserving processes at fourth-order involve the propagator 
\begin{align}
\label{eq:PauliG40}
\tilde{\CG}^{if}_{40}= 
\delta_{if}&\sum_{nmo\vec{\kappa}} 
\frac{V_{io\kappa_1}V_{on\kappa_2}V_{nm\kappa_3}V_{mi\kappa_4}}
{(4i\eta)(\delta\chi_{io}+\epsilon_{\kappa_1}+3i\eta)}
\\
\nonumber
\times&\frac{ \expval{c_{\kappa_1}c_{\kappa_2}c_{\kappa_3}c_{\kappa_4}}}
{(\delta\chi_{in}-\epsilon_{\kappa_3}-\epsilon_{\kappa_4}+2i\eta)
	(\delta\chi_{im}-\epsilon_{\kappa_4}+i\eta)},
\end{align}
with~\(\tilde{\CG}_{04}^{if}=\tilde{\CG}_{40}^{if*}\), see Fig.~\ref{Fig:mesoDiagrams40}(a).
Combined with the contractions~\eqref{eq:WickExample}, the matrix elements~\eqref{eq:PauliG40} correspond to physical processes where the system goes through three intermediate states on one of the branches of the Keldysh contour, before returning to the initial state.
This can be thought of as two particles, that tunnel in and out of the environment only to return to the original state, see Fig.~\ref{Fig:mesoDiagrams40}(b).
We combine these fourth-order terms with the pairs of second-order terms according to the recurrence rule~\eqref{eq:superSnumu}, see Fig.~\ref{Fig:mesoDiagrams40}, to obtain 
\begin{subequations}
	\label{eq:S40short}
	\begin{align}
	\label{eq:S40ashort}
	\tilde{\CS}_{40a}^{ii}&=\frac{4\eta}{\hbar}\tilde{\CG}_{40a}^{ii},
	\\
	\label{eq:S40bshort}
	\tilde{\CS}_{40b}^{ii}&=\frac{4\eta}{\hbar}\tilde{\CG}_{40b}^{ii},
	\\
	\label{eq:S40cshort}
	\tilde{\CS}_{40c}^{ii}&=\frac{4\eta}{\hbar}\tilde{\CG}_{40c}^{ii}-
	\frac{2\eta}{\hbar}\tilde{\CG}_{20}^{ii}\tilde{\CG}_{20}^{ii},
	\end{align}
\end{subequations}
where we note that~\(\tilde{\CS}_{40w}^{if} = \delta_{if} \tilde{\CS}_{40w}^{ii}\) for each contraction~\(w\).
Instead of computing the terms in~\eqref{eq:S40short}, we make use of the conservation of probability~\eqref{eq:conserved} to compute the sum 
\begin{align}
\label{eq:S4004Probability}
\tilde{\CS}_{40}^{if}+\tilde{\CS}_{04}^{if} = -\delta_{if} \sum_m
\left(\tilde{\CS}_{22}^{im}
+\tilde{\CS}_{31}^{im}
+\tilde{\CS}_{13}^{im}\right).
\end{align}
\subsubsection{Steady-state system probability distribution} 
We now have all expressions needed for the computation of the Pauli {\TCL} rates up to fourth-order.
The latter then give us access to the steady-state probability distribution
\begin{align}
\label{eq:probabilityExpansion}
P_n=\sum_{\alpha = 0}^\infty P_n^{(2\alpha)},
\end{align}
of finding the system in state~\(\ket{n}\).
Here,~\(P_n^{(2\alpha)}\) denote the contributions of order~\(2\alpha\) to the probabilities~\(P_n\) (with odd terms vanishing, see Fig.~\ref{Fig:WickTheorem}).
This is the first observable quantity that we can compute and showcases the strength of the {\TCL}.
We insert the formal expansions
for~\(P_n\) and~\(\tilde{\CS}^{nm}\) into the steady-state condition~\eqref{eq:TCLMEPauli} and solve it order by order (all odd orders vanish) to obtain the conditions 
\begin{subequations}
	\label{eq:probabilityExpansion02}
	\begin{align}
	\label{eq:probabilityExpansion0}
	&0 = \sum_n \tilde{\CS}_{2}^{nm}P_n^{(0)} , 
	\\
	\label{eq:probabilityExpansion2}
	&0 =\sum_n \left(
	\tilde{\CS}_{4}^{nm}P_n^{(0)}+\tilde{\CS}_{2}^{nm}P_n^{(2)} 
	\right).
	\end{align}
\end{subequations}
In order to test our scheme and for illustration, we calculate the {\TCL} rates up to fourth-order for a non-interacting model and compare the results of Eq.~\eqref{eq:probabilityExpansion02} with the expansion of the exact solution.
\subsection{Non-interacting system}
\label{sec:quadratic:nonInteracting}
\begin{figure}
	\includegraphics[width=8.6cm]{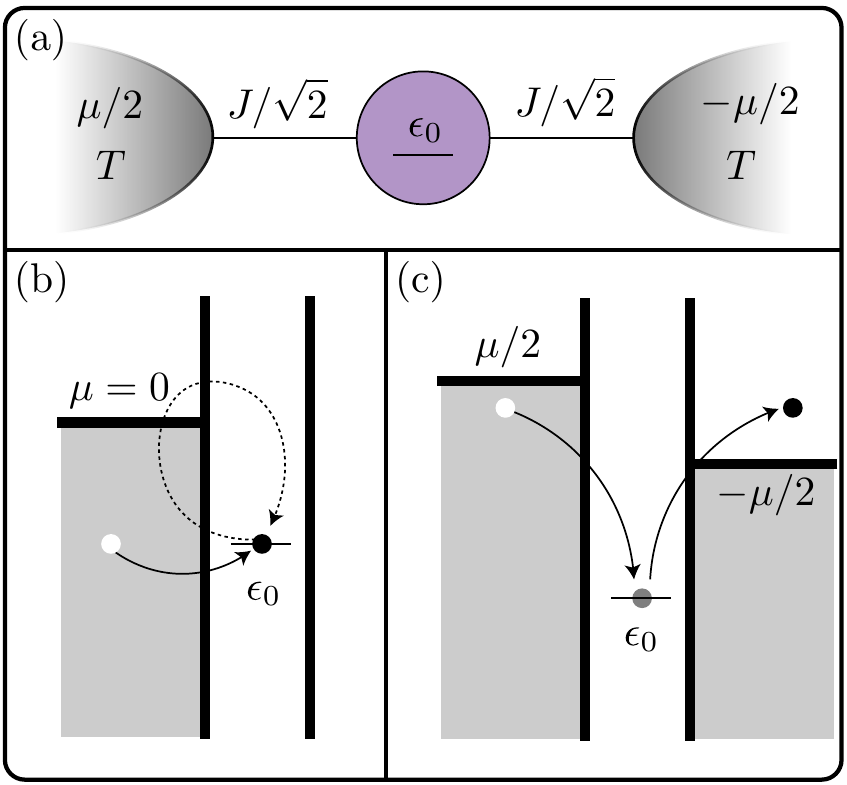}
	\caption{ 
		\label{Fig:nonInteractingModel}
		Non-interacting lead-dot-lead setup and processes.
		(a) The setup involves two reservoirs/leads at temperature~\(T\) and chemical potentials~\(\pm\mu/2\), that are connected by a hopping amplitude~\(\hop/\sqrt{2}\) to a single non-interacting fermionic mode (the dot-level) with energy~\(\epsilon_0\), see the discussion around Eqs.~\eqref{eq:NINEsys}--\eqref{eq:NINEV}.
		(b) Addition spectrum of relevant processes at equilibrium, combining two reservoirs into a single one (grey shaded area).
		In a sequential tunnelling process, a particle (or hole) enters the system (thin full arrow) and leaves a hole (particle) behind.
		In the probability conserving trivial processes (\( \tilde{\CS}_{20} \) and~\( \tilde{\CS}_{02} \)), a particle (or hole) visits the environment (dotted arrow) and returns back.
		A fourth-order, virtually-assisted sequential tunnelling combines the two preceding processes coherently.
		(c) Cotunnelling process in an out-of-equilibrium setup.
		A particle or hole exits the system on one side, while another enters on the other (or same) side.
		This process leaves the system unchanged and its contribution to probabilities is compensated by a probability conserving rate~\eqref{eq:S4004Probability}.
	}
\end{figure}

In non-interacting setups the steady-state probability distribution of the system states and current between the reservoirs through the system can be computed exactly, e.g., using equations of motion for Green's functions or scattering matrices~\cite{Landauer1957SpatialVariationCurrents, Buttiker1985GeneralizedManychannelConductance, Meir1992LandauerFormulaCurrent, Bruus2004ManyBodyQuantumTheory}, see also Appendix~\ref{app:exact}.
The non-interacting setup, see Fig.~\ref{Fig:nonInteractingModel}(a), is thus an ideal platform to test the {\TCL}.

\subsubsection{Setup}  
The non-interacting resonant level, i.e., a single fermionic level (or quantum dot) coupled to Fermi leads, see Fig.~\ref{Fig:nonInteractingModel}, can be realised in gate-defined electronic devices, see Fig.~\ref{Fig:openSystem}(a).
Its Hamiltonian is composed of the same three parts as our generic model in Sec.~\ref{sec:background:basics} with a quadratic environment as described in Sec.~\ref{sec:quadratic}, see Fig.~\ref{Fig:nonInteractingModel}(a) for a sketch.
The system involves a single fermionic mode (with energy~\(\epsilon_0\)) and is described by the Hamiltonian 
\begin{align}
\label{eq:NINEsys}
H_\sys^{\rs NI } &= \epsilon_0 \, d_0^\dagger d_0^\pdagger,
\end{align}
where~\(d_0^\dagger\) (\( d_0^\pdagger\)) creates (annihilates) an electron in the system.
The environment is composed of left (\(l = \mathrm{L}\)) and right (\(l = \mathrm{R}\)) leads and is described by 
\begin{align} \label{eq:NINEenv}
H_\env^{\rs NI }=\sum_{l,  k} \epsilon_k \, c_{l k}^\dagger
c_{lk}^\pdagger,
\end{align}
where~\(c_{lk}^\dagger\) (\(c_{lk}^\pdagger\)) creates (annihilates) a particle with momentum~\(k\) and energy~\(\epsilon_k\) in lead~\(l\).
The left and right environments are assumed to be at the same temperature~\(T\) and have chemical potentials~\(\mu/2\) and~\(-\mu/2\), respectively, with the total bias  given by~\(\mu\).
The system--environment coupling is given by the tunnelling Hamiltonian 
\begin{align}
\label{eq:NINEV}
V^{\rs NI } &=  \frac{\hop}{\sqrt{2}}\sum_{l,k} 
(d_0^\dagger c_{lk}^\pdagger + c_{lk}^\dagger d_0^\pdagger),
\end{align}
that couples the system to each reservoir in the environment with hopping amplitude~\(\hop/\sqrt{2}\).
We choose this tunnelling amplitude such that in the case~\(\mu=0\) the setup is equivalent to a single mode coupled to a single reservoir with amplitude~\(\hop\).

\subsubsection{Diagrams}

\begin{figure}
	\includegraphics[width=8.6cm]{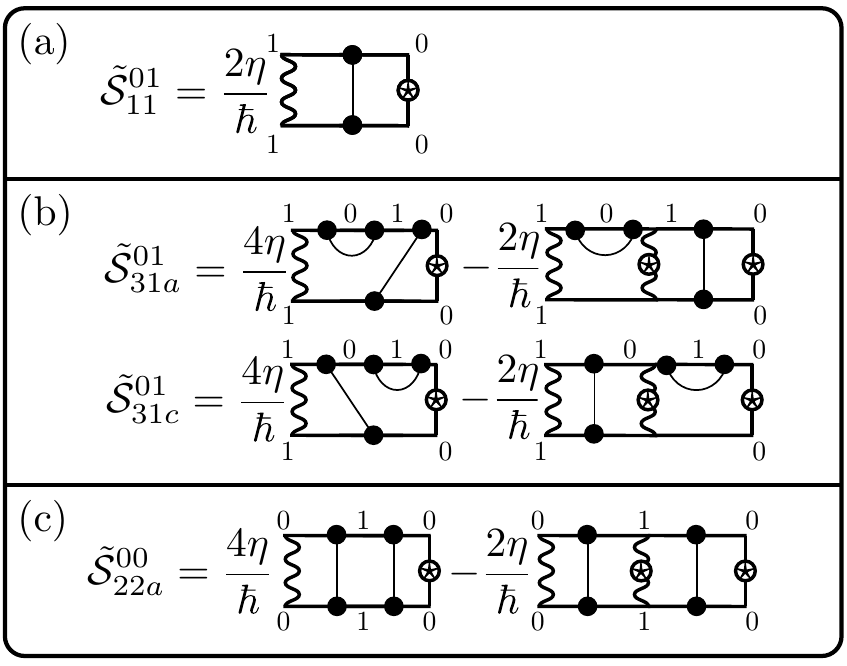}
	\caption{ 
		\label{Fig:nonInteractingDiagrams}
		Relevant diagrams for the non-interacting setup.
		(a) {\TCL} diagram for a sequential tunnelling event, where an electron from the environment enters the system, changing the latter from empty~\(\ket{0}\) to full~\(\ket{1}\).
		(b) {\TCL} diagram for virtually-assisted sequential tunnelling events where an electron from the environment enters the system.
		Simultaneously this electron leaves and returns to the environment (potentially in a different location).
		The~\(b\) contraction does not contribute for the non-interacting setup.
		(c) Diagram for an elastic cotunnelling rate, where an electron enters the system before leaving again.
		These rates do not contribute to the rate equation as they leave the system invariant (elastic process) and are compensated by a probability conserving rate~\eqref{eq:S4004Probability}.
		They do however contribute to the current as we will see in Section~\ref{sec:currents}.
		As for the virtually-assisted processes, the~\(b\) contraction does not contribute to the non-interacting setup rates, while the~\(\tilde{\CS}_{22c}\) contributions always vanish.
	}
\end{figure}
We consider the {\TCL} rates up to fourth-order in the perturbation, including the expressions~\eqref{eq:FGR} for Fermi's golden rule, co- and pair-tunnelling~\eqref{eq:S22a},~\eqref{eq:S22b},~\eqref{eq:S22c}, virtually assisted sequential tunnelling~\eqref{eq:S31a}--\eqref{eq:S31c}, and the fourth-order probability conserving rates~\eqref{eq:S4004Probability}, see Fig.~\ref{Fig:nonInteractingModel}.
As there are only two system states for the non-interacting level, empty~\(\ket{0}\) or full~\(\ket{1}\), there are only a small number of diagrams  contributing to the {\TCL} rates, see Fig.~\ref{Fig:nonInteractingDiagrams}.
The sequential tunnelling rates~\(\tilde{\CS}_{11}\), familiar from Fermi's golden rules change the occupation of the dot from full (\(\ket{1}\)) to empty (\(\ket{0}\)) or vice-versa, see Figs.~\ref{Fig:nonInteractingModel}(b) and~\ref{Fig:nonInteractingDiagrams}(a).
During a virtually-assisted sequential-tunnelling process, an (anti) particle is removed from the environment and changes the system state on both Keldysh branches, while another (anti) particle is removed from the environment before returning on one of the two Keldysh branches, see Figs.~\ref{Fig:nonInteractingModel}(b) and~\ref{Fig:nonInteractingDiagrams}(b).
There are only elastic cotunnelling rates in the non-interacting setup, i.e., rates that leave the system unchanged and therefore do not contribute to the rate equation~\eqref{eq:TCLMEPauli}, see Figs.~\ref{Fig:nonInteractingModel}(c) and~\ref{Fig:nonInteractingDiagrams}(c).
These elastic processes do, however, contribute to the steady-state current through the system as we will see in Sec.~\ref{sec:currents:nonInteracting}.
\subsubsection{Equilibrium} 
\begin{figure}
	\includegraphics[width=8.6cm]{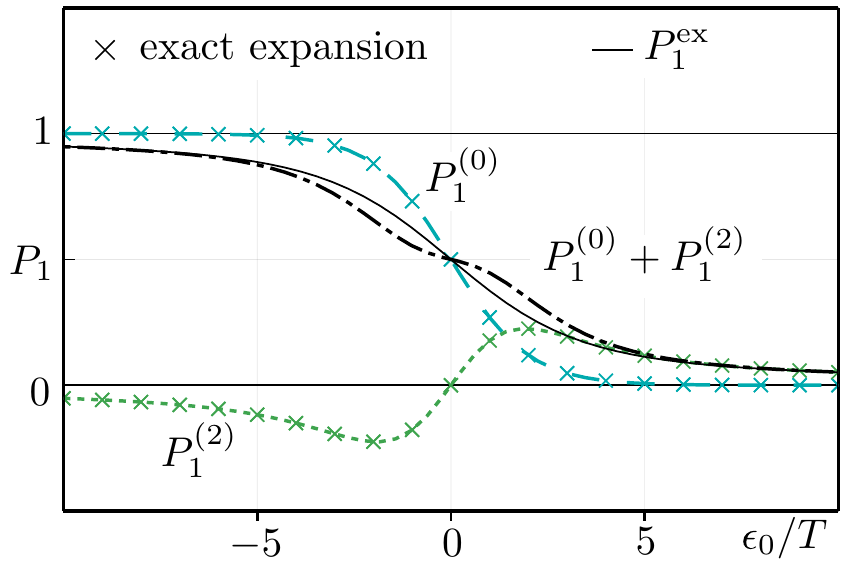}
	\caption{
		\label{Fig:nonInteracting} Probability~\(P_1\) of finding the non-interacting single level occupied, see Eqs.~\eqref{eq:probabilityExact},~\eqref{eq:nonInteractingPopulation0}, and~\eqref{eq:nonInteractingPopulation2}.
		The lowest order (blue dashed), and first correction (green dotted), for~\(\Gamma_0/T = \pi\) were calculated with the {\TCL}.
		The corresponding (blue and green) crosses are obtained from an expansion of the exact result and must coincide with any formally-exact method.
		For completeness we show the sum of the first two terms (black dotted-dashed) and the exact result (full black).
	}
\end{figure}

We start with the equilibrium situation~\(\mu=0\).
In the steady-state, the exact probability~\(P_1\) of the level being occupied is~\cite{Ryndyk2015TheoryQuantumTransport} (see also Appendix~\ref{app:exact})
\begin{align}
\label{eq:probabilityExact}
P_1 &= \frac{1}{2}+\frac{1}{\pi}\Im \psi\left(\frac{1}{2}
+\frac{\Gamma_0-2i\epsilon_0}{4\pi T}\right),
\end{align}
where we have introduced the width~\(\Gamma_0=2\pi|\hop|^2\DOS\).
The exact probability~\eqref{eq:probabilityExact} can be expanded in powers of~\(V \propto J\), or equivalently~\(\Gamma_0\propto|J|^2\) as odd terms vanish, and thus serves as a benchmark for the {\TCL} results, which we now compute.
We insert the model~\eqref{eq:NINEsys}--\eqref{eq:NINEV} into our expressions for the sequential rates~\eqref{eq:FGR} and obtain 
\begin{align}
\label{eq:nonInteractingS2}
\tilde{\CS}^{01}_{2} = \Gamma_0 
n_{\rs F} (\epsilon_{0})/\hbar 
, \qquad
\tilde{\CS}^{10}_{2} = \Gamma_0
n_{\rs F} (-\epsilon_{0})/\hbar,
\end{align}
where the temperature in the Fermi-Dirac distribution~\(n_{\rs F}\) is implicit.
These rates are familiar from Fermi's golden rule and correspond to an electron entering (leaving) the system on both the upper and lower Keldysh branches, see Fig.~\ref{Fig:nonInteractingDiagrams}(a).
The second-order diagonal elements of the rate matrix~\(\tilde{\CS}^{00}_{2}=-\tilde{\CS}^{01}_{2}\) and~\(\tilde{\CS}^{11}_{2}=-\tilde{\CS}^{10}_{2}\) are obtained directly from the conservation of probability~\eqref{eq:conservedAlpha}.
We solve the lowest-order steady-state constraint~\eqref{eq:probabilityExpansion0} and obtain
\begin{align}
\label{eq:nonInteractingPopulation0}
P_1^{(0)}=n_{\rs F}(\epsilon_0),\qquad P_0^{(0)}=n_{\rs F}(-\epsilon_0),
\end{align} 
where~\(P_1^{(0)}\) (\(P_0^{(0)}\)) is the zeroth-order probability of finding the system in the occupied (empty) state~\(\ket{1}\) (\(\ket{0}\)).
At this order, the results of the {\TCL} and T-matrix approaches coincide and both give rise to the same result as the one obtained from Fermi's golden rule~\eqref{eq:nonInteractingPopulation0}, see Fig.~\ref{Fig:nonInteracting}(a).
Furthermore, the result coincides with the expansion of the exact result~\eqref{eq:probabilityExact}, see Appendix~\ref{app:exact}.
For an infinitely sharp non-interacting level, the result~\eqref{eq:nonInteractingPopulation0} is exact, however, the coupling to the environment broadens the levels~\cite{Bruus2004ManyBodyQuantumTheory}, which manifests at fourth-order in the rates.
At fourth-order there are three types of rates,
see Figs.~\ref{Fig:mesoDiagrams22},~\ref{Fig:mesoDiagrams31}, and~\ref{Fig:mesoDiagrams40}.
The virtually-assisted sequential tunnelling rates~\(\tilde{\CS}_{31}\) and~\(\tilde{\CS}_{13}\), change the configuration of the system, see Fig.~\ref{Fig:nonInteractingDiagrams}(b).
We use the formulas~\eqref{eq:S31a}--\eqref{eq:S31c} in Appendix~\ref{app:rates} to compute the virtually-assisted sequential-tunnelling diagrams, see Fig.~\ref{Fig:nonInteractingDiagrams}(b), for the non-interacting setup.
We obtain the fourth-order correction 
\begin{align}
\label{eq:nonInteractingS4}
\tilde{\CS}_{4}^{01}
= - \frac{\Gamma_0^2}{4\pi^2\hbar T} 
\Im \psi'\left(\frac{1}{2}+i\frac{\epsilon_0}{2\pi  T}\right),
\end{align}
to the rate for changing the system state from empty to occupied, where~\(\psi'\) is the trigamma function.
The rate~\(\tilde{\CS}_{4}^{10}\) for the reverse process can be found by replacing~\(\epsilon_0 \to -\epsilon_0\) in Eq.~\eqref{eq:nonInteractingS4}.
The probability conserving (\(\tilde{\CS}_{40}+\tilde{\CS}_{04}\)) rates at fourth-order are again found by enforcing Eq.~\eqref{eq:conserved}.
We insert~\eqref{eq:nonInteractingS4} and~\eqref{eq:nonInteractingS2} into the constraint~\eqref{eq:probabilityExpansion02}  for the next-to-leading order steady-state probability distribution correction~\(P^{(2)}\) and obtain 
\begin{align}
\label{eq:nonInteractingPopulation2}
P_1^{(2)} = 
\frac{\Gamma_0}{4\pi^2  T} 
\Im\psi'\left(\frac{1}{2}-i\frac{\epsilon_0}{2\pi T}\right),
\end{align}
%
as well as~\(P_0^{(2)}=-P_1^{(2)}\) as required by conservation of probability.
The expression~\eqref{eq:nonInteractingPopulation2}, is identical to the one obtained from an expansion of the exact result~\eqref{eq:probabilityExact} and is plotted in Fig.~\ref{Fig:nonInteracting} for the specific case~\(\Gamma_0/T=\pi\).
The result~\eqref{eq:nonInteractingPopulation2} decays as~\(\propto\Gamma_0 /\epsilon_0\) for large onsite energies (such that~\(\epsilon_0\exp(-\epsilon_0/T)\ll\Gamma_0\)) and thus embodies the broadening of the level due to its coupling to the environment, see for example Refs.~\cite{Bruus2004ManyBodyQuantumTheory,Ryndyk2015TheoryQuantumTransport} for a detailed overview.
We conclude that the {\TCL} approach provides a useful tool to perturbatively compute the steady-state distribution of open systems.
It defines a time-local master equation with proper rates~\(\tilde{\CS}^{if}\) that are convergent in the~\(\eta\to 0\) limit.
In a next step, we take the setup out of equilibrium~\(\mu\neq 0\) and investigate the current flowing through the system.
%

%
\section{Currents}
\label{sec:currents}
%

%
%
%

%
The most common transport observable in mesoscopic research is the electrical current\cite{Ihn2009SemiconductorNanostructuresQuantum, Rossler2015TransportSpectroscopySpincoherent,Bischoff2015MeasurementBackActionStacked,Nicoli2018CavityMediatedCoherentCoupling,Ferguson2020QuantumMeasurementInduces}, where electric charge is flowing in or out of leads attached to the system.
Other types of currents can be defined that are associated with the environment reservoirs, e.g., in the case of a spin-full electronic system, each lead~\(l\) is further split into two reservoirs~\(r=l,\sigma\) by the spin degree of freedom~\(\sigma\).
It is then possible to consider both electrical and spin currents.
Here, we extend the diagrammatic formulation of the  {\TCL} rates~\(\CS\) from sections~\ref{sec:formal} and~\ref{sec:quadratic} to include current flow out-of-equilibrium.
\subsection{Definition and Derivation}
In our discussion, we consider currents to (from) the reservoirs that contain a mean number of particles~\(\Tr (\rho N_\lambda)\), where 
\begin{align}
N_\lambda = \sum_{k} c_{\lambda k}c_{-\lambda k},
\end{align}
denotes the reservoir number operator and we use~\(\lambda =+r \), and~\( -\lambda = -r\) (with the~\(+\) describing particles) instead of just~\(r\) to keep the notation compatible with previous sections.
The particle current~\(\pt\Tr (\rho N_\lambda)\) can always be thought of as an anti-particle current flowing in the other direction~\(\pt\Tr (\rho N_{-\lambda})= -\pt\Tr (\rho N_\lambda)\).
Furthermore, the particle currents~\(\pt\Tr (\rho N_\lambda)\) in and out of the different reservoirs, multiplied by the charge~\(q\) carried by each particle,  give rise to physical and measurable currents~\(I_\lambda\).
For electrical currents, the charge of each particle (electron) is~\(-e\), with the elementary charge~\(e\), while for spin currents the `charge' of each particle is~\(\pm \hbar/2\), depending on the reservoir, i.e., the charge~\(q_r\) may depend on the reservoir (or even the momentum~\(k\) in the case of a heat current).
This leads us to define the charge operator 
\begin{align}
\label{eq:charges}
\Qr =  q_{\lambda} N_\lambda,
\end{align}
which is the charge operator associated with the reservoir~\(\lambda\) and the single particle charge~\(q_\lambda\).
Note that the number (and charge) operators commute with the unperturbed Hamiltonian~\([H_0,\Nr] = 0\).
In fact, our considerations apply for any environment operator~\(Q_\env\) that satisfies~\([Q_\env,H_\env]=0\).
This criterion, allows us to unambiguously promote the operator~\(\Nr\) to a superoperator when desired, see Appendix~\ref{app:mixing}.
Furthermore, the vanishing commutator~\([H_0,\Nr]=0\) implies that no currents flow when the system and environment are decoupled, i.e., when the Hamiltonian~\(V\rightarrow 0\) vanishes.
The current~\(\Ir(t)\) is defined as the time derivative of the  expectation value of the charge  
\begin{align}
\label{eq:currentExp}
\Ir(t) = \Tr [\Qr \partial_t \rho(t)],
\end{align}
where the density matrix~\(\rho\) describes the coupled system--environment setup.
Working in the Schr\"odinger picture, the time derivative acts solely on~\(\rho(t)\).
An alternative path, particularly common in the RT approach, uses the Heisenberg equation of motion, and thus~\([H,\Qr]\) to encode the time dependence with a static~\(\rho\).
In the limit~\(\eta\to 0\) and~\(t_0\to -\infty\) the system is in the steady-state at~\(t\approx 0\) (\(\pt \rhop = 0\)), though out-of-equilibrium with a finite current flow across the system (\(\pt \rho \neq 0\)).
The large environment inhibits a direct solution of Eq.~\eqref{eq:currentExp}, cf. the similar challenge in solving the von Neumann equation~\eqref{eq:vonNeumann}.
Ostensibly, we would like to replace~\(\rho\) by the projected density matrix~\(\rhop\) in Eq.~\eqref{eq:currentExp}.
However, the latter does not include the evolution of the environment as it has been projected out, i.e.,~\(\Tr[Q_\lambda\rhop]= Q_\lambda^0\) is constant for all~\(\rhop\).
Furthermore, the projected space has reached the steady-state~\(\rhopb\) for~\(t\approx 0\) and therefore does not encode any dynamics.
Either of the latter two arguments is sufficient to show that a simple substitution~\(\rho\to\rhop\) in Eq.~\eqref{eq:currentExp} leads to vanishing currents and contains no information
\begin{align}
\pt\Tr [\Qr \rhop(t)]  =  0 = \pt\Tr[\Qr \rhopb].
\end{align}
The determination of currents thus starts from the full density matrix~\(\rho\) which includes the non-trivial evolution of the environment.
Even though the projected space has reached a steady-state, the full setup has not.
Common treatments of currents (as in the T-matrix approach) rely on transition rates, in the form of a rate matrix, acting on steady-state populations~\cite{Bruus2004ManyBodyQuantumTheory}.
These transition rates change the environment charge and produce finite currents into and out of the reservoirs.
In the following, we obtain a similar description for the {\TCL}.
Specifically, we  find a set of current rates~\(\CS_\lambda\), such that 
\begin{align}
\label{eq:currentTCLdef}
\Ir = q_\lambda\Tr [\CS_\lambda(0,\eta) \rhopb],
\end{align}
where we add the reservoir index~\(\lambda\) to the generator; we will show below that the rates~\(\CS_\lambda\) can be obtained directly from the {\TCL} rates~\(\CS\) by \textit{filtering/weighting} different processes.
\subsubsection{The current generator~\(\CS_\lambda\)} 
We follow a similar program as in the derivation of the {\TCL} generator~\(\CS\) in section~\ref{sec:background:TCL}.
First, we construct the full density matrix at time~\(t\) from the projected one at time~\(t_0\) using the evolution~\(\CU\) from Eq.~\eqref{eq:superU}.
We then use the inverse projected evolution~\eqref{eq:superUPExpansion} to obtain the full density matrix at time~\(t\) from the projected one at the same time
\begin{align}
\label{eq:reconstruct}
\rho(t) = \CU \rhop (t_0) =\CU \CUP^{-1} \rhop(t).
\end{align} 
Here,~\(\CUP^{-1} = \CU_0^{-1} (\mathcal{I}+\CG)^{-1}\) propagates the projected density matrix~\(\rhop(t)\) back in time to~\(t_0\), where~\(\rhop(t_0) = \rho(t_0)\), cf.~Eqs.~\eqref{eq:superUP} and~\eqref{eq:superUPExpansion}.
The full density matrix~\(\rho(t)\) is then forward propagated using the time evolution~\(\CU\).
Note that, unlike in Sec.~\ref{sec:background}, we do not immediately apply the projector again once we arrive at time~\(t\).
We substitute~\eqref{eq:reconstruct} into Eq.~\eqref{eq:currentExp} to obtain a time-local expression for the currents in terms of the projected density matrix
\begin{align}
\label{eq:currentComplicated}
\Ir = \pt\Tr
[\projector \Qr \CU \CUP^{-1}\rhop],
\end{align}
where we used~\(\Tr\projector=\Tr\), see Eq.~\eqref{eq:superTrace}, to insert a projector before the trace and thus define the projected space superoperators~\(\projector \Qr \CU\CUP^{-1}\).
At this point, it seems that it were sufficient to compute the matrix elements~\((\projector \Qr \CU\CUP^{-1})^{ijfg}\), order by order in~\(V\), to reconstruct the current~\(\Ir\).
Unfortunately, these matrix elements diverge in the limit~\(\eta\to 0\) (irrespective of the derivative~\(\pt\)), even though the trace over them, and therefore the current, is finite
.
We overcome this problem by a subtraction of superoperators whose matrix elements diverge identically to~\(\projector \Qr \CU\CUP^{-1}\) but that vanish when traced.
We will thus be left with a set of convergent rates~\(\CS_\lambda^{ijfg}\).
Furthermore, we will show that the resulting rates~\(\CS_\lambda^{ijfg}\) evaluated for quadratic environments are obtained by adding a set of Kronecker deltas to the diagrams used to compute the {\TCL} rates~\(\CS^{ijfg}\).
This \textit{filter} in the calculation of diagrams implies that the currents~\(I_\lambda\) can be obtained with little additional effort as compared with the determination of the distribution~\(\rhop\).
\textit{The charge transfer propagator}.
As the current is the time derivative of the expectation value of the charge, we can subtract the (vanishing) time derivative of the charge at time~\(t_0\) 
\begin{align}
\label{eq:vanish1}
\pt\expval{\Qr}(t_0) =  \pt\Tr  [\CU \Qr \CUP^{-1}\rhop]=0,
\end{align}
where, different from~\eqref{eq:currentComplicated}, the charge operator~\(\Qr\) is now applied prior to the forward propagation in time~\(\CU\).
Subtracting~\eqref{eq:vanish1} from~\eqref{eq:currentComplicated} we obtain 
\begin{align}
\label{eq:currentG}
\Ir =q_\lambda \pt\Tr[\CG_\lambda (\mathcal{I}+\CG)^{-1} \rhop],
\end{align}
where we have used~\(\Tr\projector=\Tr\) to insert the projector~\(\projector\), made use of the definition~\eqref{eq:superUPExpansion} of~\(\CUP^{-1}\), used~\([\Qr,\CU_0]=0\), and introduced the charge-difference propagator 
\begin{align}
\label{eq:superGlambda}
\CG_\lambda = \projector [\Nr,\CU\CU_0^{-1}]\projector.
\end{align}
Physically, Eq.~\eqref{eq:currentG} can be understood as propagating the projected density matrix~\(\rhop\) backward in time to the distant past, before evolving it forward again.
On the way forward, the charge transfer propagator in Eq.~\eqref{eq:superGlambda} tracks the number of particles in reservoir~\(r\) that have been created minus the ones that have been annihilated.
This physical interpretation of the propagator~\(\CG_\lambda\) is commonly used in T-matrix calculations of currents~\cite{Bruus2004ManyBodyQuantumTheory}.
\textit{Applying the derivative}.
We apply the time derivative~\(\pt\) in Eq.~\eqref{eq:currentG}, and use the simplification~\eqref{eq:LVU1} (which is valid at~\(t_0\to -\infty\)).
The expressions then all become time independent in the limit~\(\eta\to 0\) and we thus use the steady-state projected density matrix to obtain
\begin{align}
\label{eq:currentDiv2}
I_\lambda = q_\lambda\Tr [(\CGH_\lambda+\CGL_\lambda) (\mathcal{I}+\CG)^{-1} \rhopb].
\end{align}
Here, we have introduced the time-differentiated counterparts~\(\CGH_\lambda\) and~\(\CGL_\lambda\) of the charge transfer propagator~\(\CG_\lambda\).
These two terms are obtained via the expansion~\(\CG_\lambda = \sum_\alpha \CG_{\lambda\alpha}\) in combination with~\(\CGH_{\lambda\alpha} = \alpha\eta \CG_{\lambda\alpha}/\hbar\) and~\(\CGL_\lambda = -i[\CG_\lambda, \liouvillian_0]/\hbar\), in full analogy to the steps in Eqs.~\eqref{eq:superUdot1}--\eqref{eq:superGdot} for the propagator~\(\CG\) in section~\ref{sec:formal}.
Notice that Eq.~\eqref{eq:currentDiv2} and Eq.~\eqref{eq:currentTCLdef} are very similar, but that the matrix elements 
\begin{align}
\label{eq:currentComplicatedMatrixElements}
\lim_{\eta\to 0}[(\CGH_\lambda+\CGL_\lambda) (\mathcal{I}+\CG)^{-1} ]^{ijfg},
\end{align}
of the projected space superoperator in Eq.~\eqref{eq:currentDiv2} are not well defined in the limit~\(\eta\to 0\) (even though the current is), just as was the case for~\eqref{eq:currentComplicated}.
The easiest way to verify that~\eqref{eq:currentComplicatedMatrixElements} is ill defined in the limit~\(\eta\to 0\) is to evaluate the matrix elements explicitly at fourth-order, which can be done using the methods developed in Sec.~\ref{sec:quadratic} and Appendix~\ref{app:rates}.
\textit{Adding zero}.
In order to obtain convergent rates, we add a compensating term to Eq.~\eqref{eq:currentDiv2}, that vanishes when traced over, but contains divergent matrix elements that cancel the diverging part of Eq.~\eqref{eq:currentComplicatedMatrixElements}.
In Eq.~\eqref{eq:currentDiv2} the projected space superoperators that propagate backwards~\((I+\CG)^{-1}\) do not contain any occurrences of~\(\Nr\), whereas those that propagate forward with~\(\CGH_\lambda+\CGL_\lambda\) do.
This asymmetry between the backward and forward propagation is the root causing  the diverging matrix elements~\eqref{eq:currentComplicatedMatrixElements}.
We therefore add the vanishing term 
\begin{align}
\label{eq:vanishSuper}
q_\lambda\Tr\{(\CGH+\CGL)[(\mathcal{I}+\CG)^{-1}]_\lambda \rhop\} = 0,
\end{align}
to the current~\eqref{eq:currentComplicated}, where we have introduced the backward charge transfer propagator
\begin{align}
[(\mathcal{I}+\CG)^{-1}]_\lambda \equiv  -\CG_\lambda +\CG_\lambda\CG +\CG\CG_\lambda -
\dots \,,
\end{align}
It will become clear later, when adapting our diagrammatic approach to the currents, how the term~\eqref{eq:vanishSuper} regularises the matrix elements~\eqref{eq:currentComplicatedMatrixElements}.
To verify that Eq.~\eqref{eq:vanishSuper} is indeed vanishing, we rewrite~\(\CGH\) and~\(\CGL\) in terms of~\(\CG\) using Eqs.~\eqref{eq:superGprime} and~\eqref{eq:superGdot}, and expand all occurrences of~\(\CG\) in powers of~\(V\).
All of the resulting terms take on the form 
\begin{align}
\label{eq:vanish2}
\Tr [\CG_\alpha\CA \rhop] = 0,
\end{align}
where~\(\CA\) is composed of a product of charge transfer propagators~\(\CG_{\lambda \beta}\) and regular propagators~\(\CG_\beta\).
To verify that Eq.~\eqref{eq:vanish2} vanishes, we rewrite 
\begin{align}
\Tr [\CG_\alpha \CA \rhop] = \Tr [\CU_\alpha \CU_0^{-1}\projector \CA \rhop],
\end{align} 
where we used~\(\Tr\projector = \Tr\), see Eq.~\eqref{eq:superTrace}.
We then use the invariance of the trace of an operator under unitary transformation to write~
\begin{align}
\label{eq:unitaryTrace}
\Tr[\CU A]=\Tr A =\Tr[\CU_0 A],
\end{align} with~\(A\) an arbitrary operator and~\(\Tr[\CU A]=\Tr[UAU^\dagger]\) and similarly for~\(\CU_0\).
Expanding~\(\CU\) in powers of~\(V\) in Eq.~\eqref{eq:unitaryTrace} we then immediately obtain~\(\Tr[\CU_\alpha A]=0\) for~\(\alpha\neq0\).
Defining the operator~\(A = \CU_0^{-1}\projector \CA \rhop\) then tells us that Eq.~\eqref{eq:vanish2} holds true.
We sum Eqs.~\eqref{eq:currentDiv2} and~\eqref{eq:vanishSuper}, expand the result in~\(V\), and compare it order by order with~\eqref{eq:currentTCLdef}.
We thus arrive at the series expansion for the  {\TCL} current generator 
\begin{align}
\label{eq:currentSalpha}
S_{\lambda\alpha} \equiv \, \,&\CGH_{\lambda\alpha}+ \CGL_{\lambda\alpha}
\\&
\nonumber
-\sum_{\beta=1}^{\alpha-1}
\left[
\CS_{\lambda\beta}\CG_{\alpha-\beta}
+
(\CGH_{\beta}+\CGL_{\beta})\CG_{\lambda\alpha-\beta}
\right],
\end{align}
which can be used to reconstruct the {\TCL} current generator 
\begin{align}
\label{eq:currentS}
\CS_\lambda = \sum_\alpha \CS_{\lambda\alpha}.
\end{align}
The expansion of the current generator~\eqref{eq:currentSalpha} is very similar to the expansion~\eqref{eq:simpleSuperS} for the {\TCL} generator~\(\CS\) but with an additional regularising term in the sum.
Constructing the formally exact convergent current rates~\eqref{eq:currentSalpha} is one of the main results of this work.

\begin{figure}
	\includegraphics[width=8.6cm]{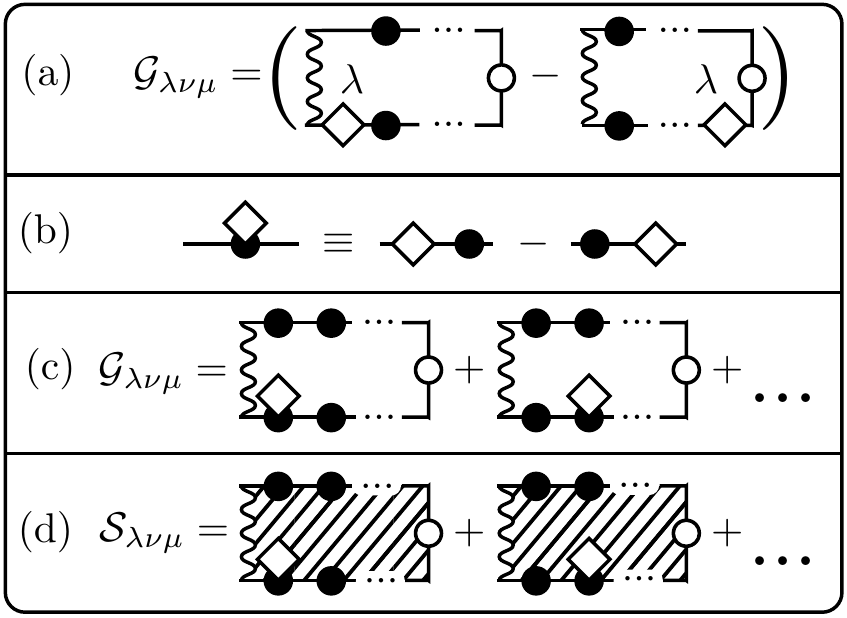}
	\caption{
		\label{Fig:currentGeneric}
		Current generator~\(\CS_{\lambda}\).
		(a) The diagram for the charge transfer propagator~\(\CG_{\lambda\nu\mu}\)~\eqref{eq:diagrammaticGlambda} involves the commutator of the number operator~\(\Nr\) (open diamond) with the propagator~\(U_\mu U_0^\dagger\).
		As a convention, we choose to place the commutator on the lower branch;
		choosing the upper branch would lead to the same result.
		(b) Diagramatic representation (black dot with open diamond) of the commutator~\([\Nr,V]\).
		(c) The~\(\mu\) diagrams that each differ by one commutator~\([\Nr,V]\) from~\(\CG_{\nu\mu}\).
		When summed, these~\(\mu\) diagrams give rise to~\(\CG_{\lambda\nu\mu}\).
		(d) The~\(\CS_{\lambda\nu\mu}\) diagrams, obtained by producing~\(\mu\) copies of~\(\CS_{\nu\mu}\), with each of the scattering events on the lower branch replaced by a commutator~\([\Nr,V]\).
	}
\end{figure}
\subsubsection{Diagrammatics} 
\begin{figure}
	\includegraphics[width=8.6cm]{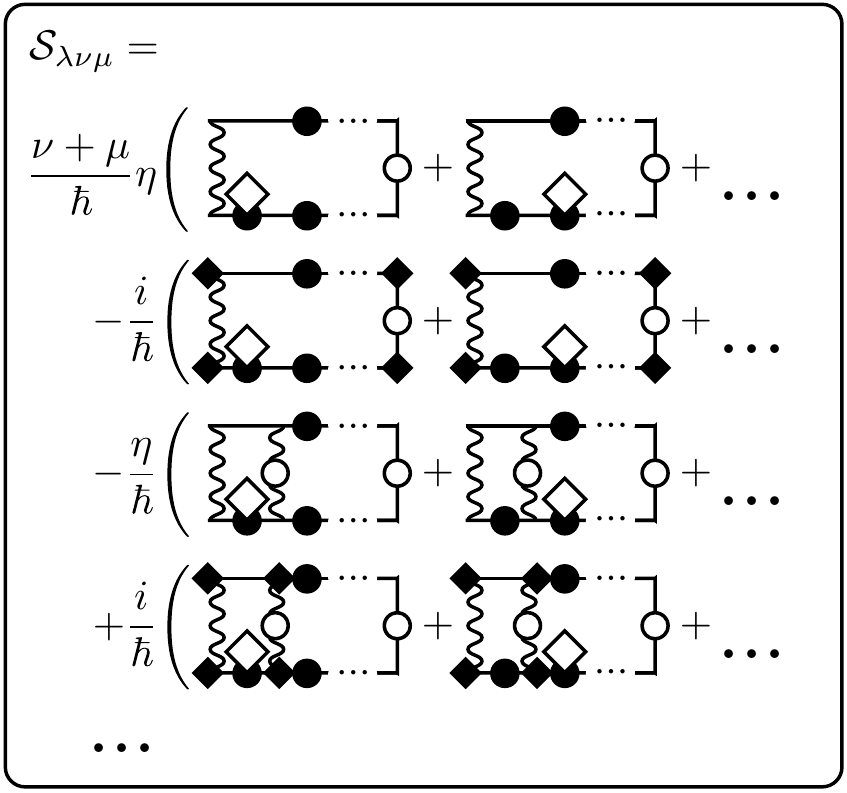}
	\caption{
		\label{Fig:currentGenericP2}
		Diagrams for the current generator~\(\CS_{\lambda\nu\mu}\).
		These are obtained by replacing the dashed area in Fig.~\ref{Fig:currentGeneric} by diagrams with different sets of cuts and prefactors according to the rules in Fig.~\ref{Fig:genericSDiagram}.
		Notice that the terms with a commutator (black dot and open diamond) to the left of the leftmost \textit{cut} arise from Eq.~\eqref{eq:currentG}.
		On the other hand, any diagram with a commutator to the right of the leftmost cut arises from the term~\eqref{eq:vanishSuper}.
		These latter terms ensure that the matrix elements of~\(\CS_{\lambda\nu\mu}\) are convergent.
		Specifically they are a regularising contribution to the corresponding diagram with no cuts.
		For example, the right diagram in the third row of this figure is a regularising contribution to the top right diagram, and would not exist without the addition of the term in Eq.~\eqref{eq:vanishSuper}.
	}
\end{figure}

We now develop a diagrammatic approach to compute the (divergent) matrix elements of~\(\CG_{\lambda\nu\mu}\) and thereof the (convergent) matrix elements of~\(\CS_{\lambda\nu\mu}\).
Here, as for~\(\CG_{\nu\mu}\), the~\(\nu\mu\) indices indicate the number of scattering events on the upper and lower branches respectively, and thus 
\begin{align}
\label{eq:superGlambdaConstruction}
\CG_{\lambda} = \sum_\alpha \CG_{\lambda\alpha} 
=\sum_{\alpha}\sum_{\nu+\mu=\alpha}\CG_{\lambda\nu\mu},
\end{align}
cf.
Eq.
~\eqref{eq:superGConstruction}.
We insert the expansion~\eqref{eq:superUaUnumu} of~\(\CU\) in terms of the expansion~\eqref{eq:Uexpansion} of the unitary evolution~\(U\) into Eq.~\eqref{eq:superGlambda} and obtain 
\begin{align}
\CG_{\lambda\nu\mu}^{ijfg} = &
\Tr_\env\left[ 
\Nr 
\bra{f} U_\nu^\pdagger U_0^\dagger\ket{i}
\rho_\env^0
\bra{j}U_0^\pdagger U_\mu^\dagger\ket{g}\right]
\\\nonumber
&-\Tr_\env \left[
\bra{f} U_\nu^\pdagger U_0^\dagger\ket{i}
\Nr 
\rho_\env^0
\bra{j}U_0^\pdagger U_\mu^\dagger\ket{g}\right].
\end{align}
Next, we use the cyclic nature of the trace and the fact that the charge operator commutes with the unperturbed environment distribution~\([\Nr,\rho_\env^0]=0\), to obtain 
\begin{align}
\label{eq:diagrammaticGlambda}
\!\!\!\CG^{ijfg}_{ \lambda \nu \mu} \!= \!\Tr_\env
\!
\left[
\bra{g}  [\Nr , U^\pdagger_\mu U_0^\dagger]\ket{j}^{\!\dagger}
\!
\bra{f} U_\nu^\pdagger U_0^\dagger \ket{i} 
\rho_\env^0\right],
\end{align}
which is identical to~\(\CG_{\nu\mu}^{ijfg}\) in Eq.~\eqref{eq:diagramaticG} up to the commutator with~\(\Nr\).
Diagrammatically, the commutator in~\(\CG_{\lambda\nu\mu}\) is obtained by taking the difference of two terms, where we insert~\(\Nr\) to the far left on one of the Keldysh branches, and the second with~\(\Nr\) inserted to the far right on the same Keldysh branch, cf.~Figs.~\ref{Fig:specificDiagram}(a) and~\ref{Fig:currentGeneric}(a).
On the other hand, mathematically, we could deal with the commutator~\([N_\lambda,U_\mu U_0^{-1}]\) by separately evaluating the two terms~\(N_\lambda U_\mu U_0^{-1}\) and~\(U_\mu U_0^{-1} N_\lambda\) before subtracting them.
Here, we make use of the structure of~\(U_\mu\) to simplify the commutator before performing any explicit calculation.
We insert the expansion~\eqref{eq:Uexpansion} into the commutator from~\eqref{eq:diagrammaticGlambda} and obtain an expression of the form 
\begin{align}
\label{eq:commutatorNrUmu}
[N_\lambda,U_\mu U_0^{-1}] = [N_\lambda, \Pi_0 V \Pi_0 V ... \Pi_0 V],
\end{align}
where~\(\Pi_0\) are the free propagators, see Eq.~\eqref{eq:freePropagator}.
The number operator~\(\Nr\) commutes with the unperturbed Hamiltonian~\(H_0\) and therefore also with the free propagator~\([\Nr, \Pi_0]=0\), allowing us to rewrite Eq.~\eqref{eq:commutatorNrUmu} as a sum of terms 
\begin{align}
\label{eq:commutatorNrUmu2}
&[N_\lambda,U_\mu U_0^{-1}] = \Pi_0 [N_\lambda,V ]\Pi_0 V ... \Pi_0 V
\\\nonumber
&+\Pi_0 V\Pi_0 [N_\lambda,V ] ... \Pi_0 V+...+\Pi_0 V \Pi_0 V ... \Pi_0 [N_\lambda,V ],
\end{align}
each with one occurrence of the perturbation~\(V\) replaced by the commutator~\([N_\lambda,V]\), see Fig.~\ref{Fig:currentGeneric}(b).
Diagrammatically, Eq.~\eqref{eq:commutatorNrUmu2} means that we go from~\(\CG_{\nu\mu}\) to~\(\CG_{\lambda\nu\mu}\) by drawing~\(\mu\) copies of~\(\CG_{\nu\mu}\) and replacing one of the system--environment scatterings (\(V\)) on the lower Keldysh branch by~\([\Nr,V]\), see Fig.~\ref{Fig:currentGeneric}(c).
We could equivalently choose the upper Keldysh branch and create~\(\nu\) copies, leading to the same final result.
\textit{Convergent matrix elements of~\(\CS_{\lambda\nu\mu}\)}.
Thanks to the addition of the vanishing terms~\eqref{eq:vanishSuper}, the relationship between~\(\CS_{\lambda\nu\mu}\) and~\(\CS_{\nu\mu}\) is identical to the relation between~\(\CG_{\lambda\nu\mu}\) and~\(\CG_{\nu\mu}\), cf.
Figs.
~\ref{Fig:currentGeneric}(c) and (d).
Transforming from the usual superoperator to the current (\(\lambda\)) superoperator, we again draw~\(\mu\) copies of the diagram and  replace scattering events~\(V\) on the lower branch with commutators~\([\Nr,V]\) (without the vanishing terms~\eqref{eq:vanishSuper}, commutators~\([\Nr,V]\) would only replace scattering events to the left of the leftmost \textit{cut}, see Fig.~\ref{Fig:currentGenericP2}).
As the dependence on the slow switch-on parameter~\(\eta\) lies entirely within the unperturbed propagator~\(\Pi_0\), the exact form of the scattering events does not influence the convergence in the limit~\(\eta\to 0\).
Thus if the matrix elements~\(\CS_{\nu\mu}^{ijfg}\) are finite, the same will be true for the matrix elements~\(\CS_{\lambda\nu\mu}^{ijfg}\) of the current generator.
\subsubsection{Quadratic environment and Wick's theorem} \label{sec:currents:derivation:Wick} 
\begin{figure}
	\includegraphics[width=8.6cm]{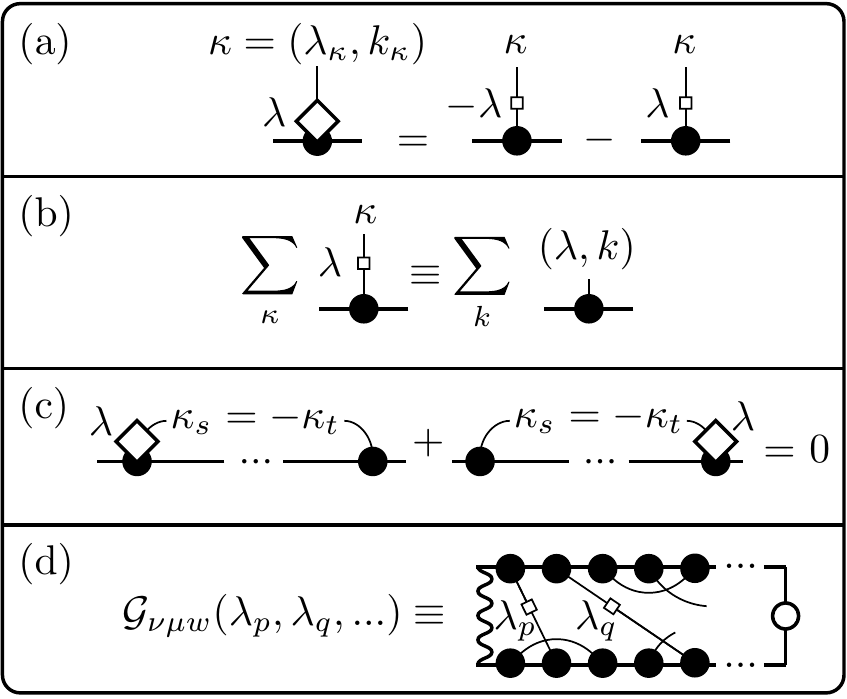}
	\caption{
		\label{Fig:currentWick}
		Charge difference propagator diagrammatics, for a quadratic environment.
		(a) Diagrammatic representation of the commutator~\([\Nr,V]\).
		Each commutator gives rise to two contributions, where the discrete degree of freedom~\(\lambda_\kappa\) is associated with Kronecker deltas (small open squares) to the particle reservoir~\(\lambda\) or anti-particle reservoir~\(-\lambda\), see Eqs.~\eqref{eq:commutatorRule} and~\eqref{eq:bookie}.
		(b) Definition of the constraining symbol (small open square).
		A constrained contraction line only contains the specific discrete degree of freedom~\(\lambda\), i.e., the symbol acts as a Kronecker delta.
		(c) The commutators of two terms contracted on the same Keldysh branch (\(s,t\leq\mu\)) vanish, see Eq.~\eqref{eq:currentRule2}.
		As a result, only contractions that connect the upper and lower Keldysh branches contribute to the charge difference.
		(d) Diagrams for the constrained propagator~\(\CG_{\nu\mu w}(\lambda_p,\lambda_q,...)\).
		All sums over environment degrees of freedom (discrete and continuous) are implicit except for the sums over the discrete part of contractions that connect the upper and lower Keldysh branches (\(\lambda_p,\lambda_q,...\)).
		Not performing the latter allows us to obtain both~\(\CG_{\lambda\nu\mu w}\) and~\(\CG_{\nu\mu w}\) from~\(\CG_{\nu\mu w}(\lambda_p,\lambda_q,...)\), see Fig.~\ref{Fig:currentWickP2}.
	}
\end{figure}
\begin{figure}
	\includegraphics[width=8.6cm]{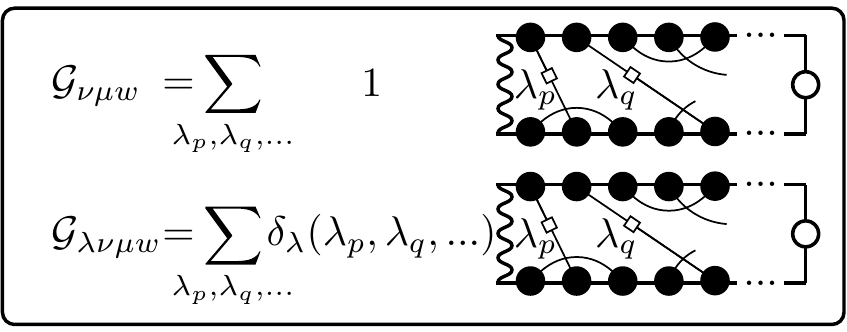}
	\caption{
		\label{Fig:currentWickP2}
		The filtering scheme that produces~\(\CG_{\nu\mu w}\) and~\(\CG_{\lambda\nu\mu w}\) from the constrained propagator~\(\CG_{\nu\mu w}(\lambda_p,\lambda_q,...)\).
		Of all the sums over~\(\kappa_1,\kappa_2,...\kappa_{\nu+\mu}\), we make explicit the sums over the discrete part~\(\lambda_p,\lambda_q,...\) of contractions that connect upper and lower Keldysh branches, while all other sums (over discrete and continuous degrees of freedom) are implicit.
	}
\end{figure}

We now consider the specific case of quadratic environments with linear coupling, see Sec.~\ref{sec:quadratic}.
Here, the commutator~\([\Nr,V]\) assumes a particularly simple form 
\begin{align}
\label{eq:commutatorRule}
[\Nr,c_\kappa] 
= (\delta_{\lambda\lambda_\kappa} -\delta_{\lambda-\lambda_\kappa}) c_{\kappa},
\end{align}
which can be thought of as a \textit{filter} associated with the environment operators~\(c_\kappa\), see Figs.~\ref{Fig:currentWick}(a).
Each contribution to the filter is a Kronecker delta which constrains the discrete degree of freedom~\(\lambda_\kappa\) associated with a scattering event to either~\(\lambda\) or~\(-\lambda\).
In the diagram for~\(\CG_\lambda\), we have to insert this commutator subsequently for every scattering event on the lower branch of~\(\CG_{\lambda\nu\mu}\), see Fig.~\ref{Fig:currentGeneric}(c).
As a result, we obtain a simple diagrammatic rule that takes us from~\(\CG\) to~\(\CG_\lambda\) via the substitution  
\begin{align}
\label{eq:currentRule}
\!\!\!\!\expval{c_{\kappa_1}\dots c_{\kappa_{\nu+\mu}}} \to 
\delta_\lambda(\lambda_1,\lambda_2,\dots, \lambda_\mu) 
\expval{c_{\kappa_1}\dots c_{\kappa_{\nu+\mu}}}\!,
\end{align} 
in the unperturbed environment correlator~\eqref{eq:envCorrelator}, with the filtering function  
\begin{align}
\label{eq:bookie}
\delta_\lambda(\lambda_1,\dots,\lambda_\mu) \equiv 
\sum_{s=1}^\mu [\delta_{\lambda-\lambda_{s}}-\delta_{\lambda\lambda_{s}}],
\end{align}
and the shorthand notation~\(\lambda_s \equiv \lambda_{\kappa_s}\).
The additional minus sign when compared to Eq.~\eqref{eq:commutatorRule} arises from the Hermitian conjugation on the lower branch, see Fig.~\ref{Fig:genericDiagram} in Sec.~\ref{sec:formal}.
\textit{Wick's theorem.} The environment correlator~\eqref{eq:currentRule} for the charge transfer propagator~\(\CG_{\lambda\nu\mu}\) is identical to the one for~\(\CG_{\nu\mu}\), Eq.~\eqref{eq:envCorrelator}, up to the prefactor $\delta_\lambda$ composed of a set of Kronecker deltas for the discrete degrees of freedom $\lambda_s$ associated with the scatterings on the lower Keldysh branch.
Hence, we can apply Wick's theorem in the same way as before and write~\(\CG_{\lambda\nu\mu}\) in terms of a sum over contractions with Wick index~\(w\), 
\begin{align}
\CG_{\lambda \nu\mu} = \sum_{w} \CG_{\lambda \nu \mu w}.
\end{align}
As a further simplification, we note that the filters associated with two contracted scattering events~\(\kappa_s\) and~\(\kappa_t\) on the lower Keldysh branch, such that~\(\kappa_s = -\kappa_t\), vanish
\begin{align}
\label{eq:currentRule2}
( \delta_{\lambda-\lambda_s} \!-\! \delta_{\lambda\lambda_s}
+
\delta_{\lambda-\lambda_t} \!-\! \delta_{\lambda\lambda_t}) 
\expval{c_{\kappa_s}c_{\kappa_t}}\expval{\dots} = 0,
\end{align}
see Fig.~\ref{Fig:currentWick}(c).
Thus, the only filters that contribute to~\(\CG_{\lambda\nu\mu}\) are those associated with contractions that connect the upper and lower branch.
To unify the calculation of both~\(\CG_{\lambda \nu\mu w}\) and~\(\CG_{\nu\mu w}\), we introduce the constrained propagator~\(\CG_{\nu\mu w}(\lambda_p,\lambda_q,...)\), see Fig.~\ref{Fig:currentWick}(d).
This constrained propagator is identical to the propagator~\(\CG_{\nu\mu w}\) except for the fact that the sum over discrete degrees of freedom~\(\lambda_p,\lambda_q,...\) associated to contractions that connect upper and lower Keldysh branches are not performed.
Once~\(\CG_{\nu\mu w}(\lambda_p,\lambda_q,...)\) has been computed one can immediately obtain either the charge difference propagator or the usual propagator
\begin{subequations}
	\label{eq:currentRules}
\begin{align}
\CG_{\lambda\nu\mu w} &= \sum_{\lambda_p,\lambda_q,...} \delta_\lambda(\lambda_p,\lambda_q,...) \CG_{\nu\mu w}(\lambda_p,\lambda_q,...),
\\
\CG_{\nu\mu w} &= \sum_{\lambda_p,\lambda_q,...} \CG_{\nu\mu w}(\lambda_p,\lambda_q,...),
\end{align}
\end{subequations}
see Fig.~\ref{Fig:currentWickP2}.
The relationships between the rates~\(\CS_{\nu\mu w}(\lambda_p,\lambda_q,...)\),~\(\CS_{\nu\mu w}\) and~\(\CS_{\lambda\nu\mu w}\) are identical to the relationships between the corresponding propagators~\(\CG\) (and similarly for Pauli equivalents). This conclusion follows from equivalence of the rules for rates and propagators shown in Figs.~\ref{Fig:currentGeneric}(c) and~(d).

\begin{figure}
	\includegraphics[width=8.6cm]{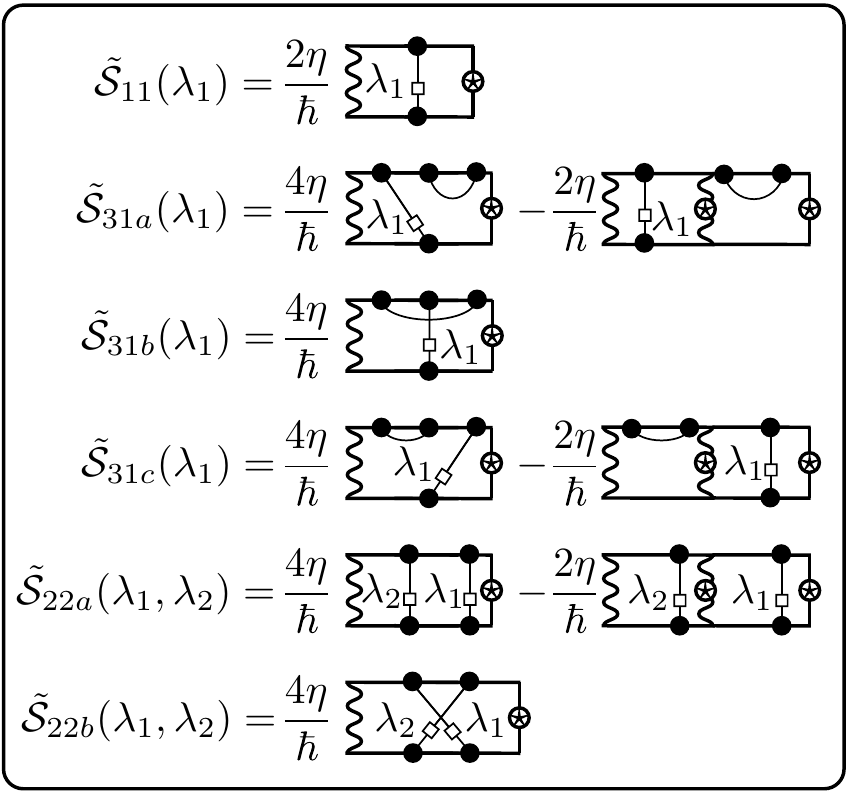}
	\caption{
		\label{Fig:superSWickCotCurrent}
		Constrained Pauli diagrams, up to fourth-order, that contribute to the current for quadratic environments and linear coupling.
		\(\tilde{\CS}_{11}(\lambda_1)\):  Sequential tunnelling rates involving an environment particle in reservoir~\(\lambda_1\).
		The probability conserving rates~\(\tilde{\CS}_{20}\) and~\(\tilde{\CS}_{02}\) do not contribute to the current as the two vertices lie on the same Keldysh branch and thus the initial and final environment states are the same, see Fig.~\ref{Fig:currentWick}(c).
		\(\tilde{\CS}_{31}(\lambda_1)\): The virtually-assisted sequential tunnelling diagrams that contribute to the current.
		These rates only have one leg connecting the upper and lower branches and therefore only one constrained environment index~\(\lambda_1\).
		\(\tilde{\CS}_{22}(\lambda_1,\lambda_2)\): The co- and pair-tunnelling diagrams that contribute to the current.
		Note that each leg that connects the upper and lower branches counts once, leading to two contributions per diagram.
	}
\end{figure}
The \textit{filtering} rules~\eqref{eq:currentRule},~\eqref{eq:currentRule2}, and~\eqref{eq:currentRules} that distinguish the Pauli rates~\(\tilde{\CS}^{if}\) from the current rates~\(\tilde{\CS}^{if}_\lambda\) are the same ones that are commonly employed in T-matrix calculations~\cite{Bruus2004ManyBodyQuantumTheory, Koch2006QuantumTransportSinglemolecule} or the RT approach~\cite{Konig1995ResonantTunnelingCoulomb} and provide an intuitive physical description of the processes that contribute to currents:
one simply counts the change in the environment's charge during a process to identify its contribution to transport
.
\subsubsection{Steady-state currents between reservoirs} 
We proceed with the calculation of the steady-state current~\(I_\lambda\) flowing out of the reservoir~\(\lambda\) using the current rates~\(\tilde{\CS}_{\lambda\alpha}^{nm}\) and the steady-state probability distribution~\(P_n\) as obtained from the rates~\(\tilde{\CS}_{\alpha}^{nm}\) in section~\ref{sec:quadratic}.
We remind the reader that we work with the index~\(\lambda>0\) instead of the reservoir index~\(r\) in order to keep the notation consistent.
The Pauli equivalent to~\eqref{eq:currentTCLdef}, written in matrix element form, reads 
\begin{align}
\label{eq:currentPauliTCLdef}
\Ir=q_\lambda\sum_{nm}\tilde{S}^{nm}_\lambda P_n.
\end{align}
We expand this expression order by order and. obtain 
\begin{subequations}
	\label{eq:currentExpansion}
\begin{align}
\label{eq:currentExpansion2}
\Ir^{(2)} &= q_\lambda\sum_{nm}\tilde{\CS}^{nm}_{\lambda 2} P_n^{(0)},
\\
\label{eq:currentExpansion4}
\Ir^{(4)} &= q_\lambda\sum_{nm}\tilde{\CS}^{nm}_{\lambda 2} P_n^{(2)}
+q_\lambda\sum_{nm}\tilde{\CS}^{nm}_{\lambda 4} P_n^{(0)},
\end{align}
\end{subequations}
which allows us, along with~\(P_n^{(0)}\) and~\(P_n^{(2)}\) from Eq.~\eqref{eq:probabilityExpansion02}, to compute the current from a reservoir~\(\lambda\) in any setup with a quadratic environment up to fourth-order.
\subsection{Current rates}
We now apply the diagrammatic rules for the current generators, see Figs.~\ref{Fig:currentGeneric}--\ref{Fig:superSWickCotCurrent}, and obtain explicit expressions for the associated current rates; the latter are closely related to the Pauli {\TCL} rates we computed in Sec.~\ref{sec:quadratic}.
\textit{Fermi's golden rule.
} At second order, there are three diagrams~\(\tilde{\CS}_{11}\),~\(\tilde{\CS}_{20}\), and~\(\tilde{\CS}_{02}\) for the {\TCL} rates.
The first of these three is Fermi's golden rule, while the latter two contributions ensure conservation of probability~\eqref{eq:conserved}.
Summing up these three terms gives rise to the second-order {\TCL} contribution~\(\tilde{\CS_2}\).
The current rates~\(\tilde{\CS}_{\lambda}^{if}\) only contain contributions from constrained diagrams that connect the upper and lower Keldysh branches.
At second order, there is only one such diagram~\(\tilde{\CS}_{2}(\lambda_1) =\tilde{\CS}_{11}(\lambda_1)\), see Fig~\ref{Fig:superSWickCotCurrent}(a).
We compute the constrained diagram for Fermi's golden rule rates by not performing the discrete environment sum in Eq.~\eqref{eq:FGR} and obtain 
\begin{align}
\label{eq:FGRfixed}
\tilde{\CS}_{11}^{if}(\lambda_1) = \frac{2\pi}{\hbar}
|V_{if\lambda_1}|^2 C_{\lambda_1}(\chi_f-\chi_i).
\end{align}
We then use the filtering scheme~\eqref{eq:currentRules} to obtain the  second order current rates 
\begin{align}
\label{eq:FGRcurrents}
\tilde{\CS}^{if}_{\lambda 2} 
= \frac{2 \pi}{\hbar}
|V_{if\lambda}|^2C_{\lambda}(\delta\chi_{fi})-\frac{2\pi}{\hbar}|V_{fi\lambda}|^2C_{-\lambda}(\delta\chi_{fi}),
\end{align}
which describes sequential tunnelling where a particle of type~\(\lambda\) tunnels out of (first term) or into (second term) the environment.
\textit{Fourth-order.} At fourth-order there are three types of constrained contributions,~\(\tilde{\CS}_{31}(\lambda_1)\),~\(\tilde{\CS}_{13}(\lambda_1)\), and~\(\tilde{\CS}_{22}(\lambda_1,\lambda_2)\). 
The first two contain exactly one contraction that connects the upper and lower branch, irrespective of the Wick index~\(w\), see Fig.~\ref{Fig:superSWickCotCurrent}.
These three diagrams are  identical to the ones for~\(\tilde{\CS}_{31w}\) or~\(\tilde{\CS}_{13w}\), except for the fact that we do not sum over the discrete degrees of freedom that connect upper and lower Keldysh branches, as in Eq.~\eqref{eq:FGRfixed}.
We determine these constrained rates in Appendix~\ref{app:rates}, along with the corresponding {\TCL} rates and can use them to reconstruct the current rates
\begin{align}
\tilde{\CS}_{\lambda31}^{if}=\sum_{w,\lambda_1}\delta_\lambda(\lambda_1)\tilde{\CS}^{if}_{31 w}(\lambda_1),
\end{align}
and similarly for~\(\tilde{\CS}_{\lambda13}^{if}\).

The co- and pair-tunnelling current rates~\(\tilde{\CS}_{\lambda22}^{if}\) arise only from the~\(a\) and~\(b\) contractions, see Fig.~\ref{Fig:superSWickCotCurrent}.
These diagrams have two contractions connecting the upper and lower Keldysh branches, in contrast to the~\(c\) contraction which has none, see Fig.~\ref{Fig:mesoDiagrams22}.
We constrain the contractions (\(\lambda_1\) and~\(\lambda_2\)) that connect the upper and lower branches
\begin{align}
\label{eq:currentCot}
\tilde{\CS}_{22w}^{if}&: \sum_{\lambda_1 \lambda_2, ...}
\to \, \tilde{\CS}_{22w}^{if}(\lambda_1,\lambda_2):\sum_{ \xcancel{\lambda_1 \lambda_2}\,,... }
,
\end{align}
in Eqs.~\eqref{eq:S22a} and~\eqref{eq:S22b}, for~\(w=a,b\).
Here the crossed out sum indices~\(\lambda_1,\lambda_2\) indicate that we do not sum over any discrete environment degree of freedoms.
We then again use the filtering scheme~\eqref{eq:currentRules} to obtain the co- and pair-tunnelling current rates
\begin{align}
\!\!\!\tilde{\CS}_{\lambda22}^{if} =\!\! \sum_{\lambda_1,\lambda_2}
\!\delta_\lambda(\lambda_1,\lambda_2)\left[
\tilde{\CS}_{22a}^{if}(\lambda_1,\lambda_2)
+
\tilde{\CS}_{22b}^{if}(\lambda_1,\lambda_2)
\right]\!.\!\!
\end{align}
Summing the contributions from~\(\tilde{\CS}_{31}\),~\(\tilde{\CS}_{13}\), and~\(\tilde{\CS}_{22}\), we obtain the total fourth-order current rate 
\begin{align}
\label{eq:currentRates4}
\!\!\tilde{\CS}_{\lambda4}^{if} = 
\tilde{\CS}_{\lambda22}^{if}
+2\Re
\tilde{\CS}_{\lambda31}^{if}
.
\end{align}
We are now in a position to calculate both the steady-state system probability distribution and the environment currents to next-to-leading order.
The former is calculated as detailed in Sec.~\ref{sec:quadratic} and then used along with the current rates~\eqref{eq:FGRcurrents} and~\eqref{eq:currentRates4} in the expression~\eqref{eq:currentExpansion} for the currents up to fourth-order.

\subsection{Non-interacting model}
\label{sec:currents:nonInteracting}
\begin{figure}
	\includegraphics[width=8.6cm]{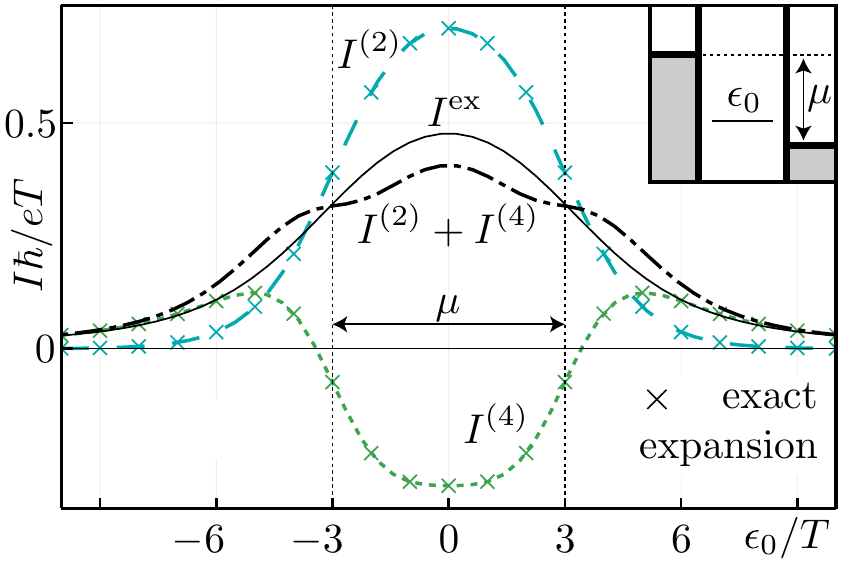}
	\caption{
		\label{Fig:nonInteractingCurrent}
		The non-interacting resonant-level out-of-equilibrium, for~\(\Gamma_0/T = \pi\) and~\(\mu/T =6\).
		The inset provides a sketch of the configuration.
		The chemical potentials in the two leads are offset by~\(\mu\).
		The level moves with~\(\epsilon_0\), and passes through the \textit{bias window}, when it lies between the two chemical potentials.
		The main figure shows the current flowing through the level when the system is driven out-of-equilibrium, lowest order (blue dashed), first correction (green dotted), their sum (black dotted-dashed) and the exact result (black full).
		The corresponding (blue and green) crosses are obtained from an expansion of the exact result and have to be matched by any formally exact perturbative method.
	}
\end{figure}

As a first application and test of the formalism, we focus on the non-interacting resonant-level~\eqref{eq:NINEsys}--\eqref{eq:NINEV} and take the setup out-of-equilibrium (\(\mu\neq0\)), implying that a steady-state current will flow across the device, see Fig.~\ref{Fig:nonInteractingCurrent}.
As for the (equilibrium) probability distribution~\eqref{eq:probabilityExact}, it is possible to compute the resulting (out-of-equilibrium) current exactly.
This can be done using a scattering matrix or Green's function approach~\cite{Bruus2004ManyBodyQuantumTheory,Meir1992LandauerFormulaCurrent} and produces a current  
\begin{align}
\label{eq:currentExact}
I_{\rsR} = \frac{e\Gamma_0}{4\pi \hbar}\bigg[
\Im& \psi\left(\frac{1}{2}+\frac{\Gamma_0-2i\epsilon_0+i\mu}{4\pi T}\right)
\\\nonumber
&-
\Im \psi\left(\frac{1}{2}+\frac{\Gamma_0-2i\epsilon_0-i\mu}{4\pi T}\right)
\bigg],
\end{align}
into the right (\(R\)) reservoir with~\(e\) the unit charge, see Appendix~\ref{app:exact} for a brief sketch of the calculation.
The setup is fully described by the level width~\(\Gamma_0 = 2\pi|\hop|^2\DOS\), the single level energy~\(\epsilon_0\), the chemical potential difference~\(\mu\) between the leads, and the temperature~\(T\).
The expression~\eqref{eq:currentExact} can be expanded in powers of~\(V\) (or equivalently~\(\Gamma_0\)), leading to 
\begin{align}
\label{eq:currentNI2}
I_{\rsR}^{(2)} = \frac{e\Gamma_0}{4\hbar}\left[
n_{\rs F} \left(\epsilon_0-\mu/2\right) 
- 
n_{\rs F} \left(\epsilon_0+\mu/2\right)
\right],
\end{align}
at lowest order.
This same result can be obtained from the {\TCL}.
First we compute the steady-state probability~\(P_1^{(0)}\) (\(P_0^{(0)}\)) of finding the level occupied (empty) as in section~\ref{sec:quadratic:nonInteracting}, but under finite bias conditions.
We then use these probabilities and the sequential current rates~\eqref{eq:FGRcurrents} in the expressions~\eqref{eq:currentExpansion2} for the lowest-order current.
As for the probabilities in Sec.~\ref{sec:quadratic:nonInteracting}, this lowest-order result~\eqref{eq:currentNI2} is exact in the case of infinitely sharp levels.
At higher orders, see Refs.~\cite{Bruus2004ManyBodyQuantumTheory,Koch2006QuantumTransportSinglemolecule}, the broadening of the level due to its coupling to the environment manifests.
A straightforward but lengthy calculation using the fourth-order {\TCL} rates provides the current~\(I_{\rsR}^{(4)}\) and recovers the exact expansion, see Fig.~\ref{Fig:nonInteractingCurrent}.
%

%

\section{Conclusion and outlook}
\label{sec:conclusion}
In this work, we developed an operator-based diagrammatic approach to the steady-state time-convolutionless master equation.
We greatly reduced the number of diagrams to be computed at any given order~\(\alpha\) when compared to a superoperator formulation, from~\(2^\alpha\) to~\(\alpha+1\).
We thus kept the complexity of STCL calculations as low as in the T-matrix approach, while remaining formally exact as in the real-time master equation.
Going beyond the analysis of a steady-state system distribution, we extended the {\TCL} to perturbatively evaluate the steady-state current through the system.
We then applied our diagramatic approach to setups with non-interacting environments and a bilinear system--environment coupling for both steady-state distributions and currents.
As an example, we verified our methodology on a non-interacting setup, a single-level coupled to leads, where we demonstrate perfect agreement between our expansion and the expansion of the exact result.
These results show that the {{\TCL}} is a versatile tool which can be used for practical perturbative calculations.

We identify a number of future research topics  based on the work presented here, with three key examples: extensions to higher orders, resummation schemes, and dynamics.
A sixth order analysis is well within reach as there are only 45 new diagrams to be computed. 
Specifically, these are~\(\tilde{\CG}_{33w}\),~\(\tilde{\CG}_{42w}\), and~\(\tilde{\CG}_{51w}\) for the fifteen Wick contraction contractions~\(w\) at sixth order. 
The sixth order Pauli STCL rates~\(\tilde{\CS}_6\), where Kondo signatures are expected to become apparent~\cite{Bruus2004ManyBodyQuantumTheory}, can then be constructed from these (and lower-order) terms.
This is also the order at which certain backaction effects are expected to appear~\cite{Zilberberg2014MeasuringCotunnelingIts}.
Furthermore, the {\TCL} provides a formally exact expansion for the steady-state observables, and can therefore be used as a base for resummation schemes.
While existing resummation schemes for the SRT method~\cite{Schoeller1994ResonantTunnelingCharge,Konig1995ResonantTunnelingCoulomb,Konig1996ResonantTunnelingUltrasmall} can be directly applied to the superoperator formulation of the {\TCL}, our operator-based approach enables the resummation of different sets of diagrams.
Finally, while this work was exclusively concerned with steady-state properties, our operator-based simplifications can readily be applied to the TCL description of dynamics after a quench~\cite{Breuer2001TimeConvolutionlessProjectionOperator, FergusonThesis}.
We conclude that the time-convolutionless master equation still holds many surprises and a large potential for future applications.
\acknowledgments{
	We thank 
	T.~M.~R.~Wolf, 
	C.~Müller, 
	J.~L.~Lado, 
	F.~Gaggioli, 
	M.~Soriente, 
	A.~\v{S}trkalj, 
	D.~Sutter, 
	C.~Gold, 
	A.~Khedri, 
	S.~Gurvitz, and 
	G.~M.~Graf 
	for illuminating discussions and acknowledge financial support from the Swiss National Science Foundation through Division~2, grants PP00P2 1163818 and PP00P2 190078, and the National Center of Competence in Research on Quantum Science and Technology (QSIT).
}

\appendix

\section{Order on the Keldysh contour}
\label{app:timeorder}

\begin{figure}
	\includegraphics[width=8.6cm]{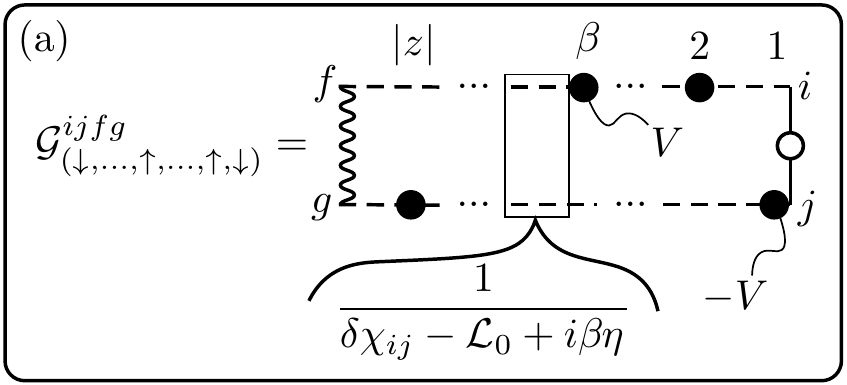}
	\caption{
		\label{Fig:timeOrdered}
		Diagram for~\(\CG_z\), to be contrasted with the diagrams for~\(\CG_{\nu\mu}\), see Fig.~\ref{Fig:genericDiagram}. The upper and lower branch are now linked by the free propagator of the density matrix (upper and lower dashed lines combined, example highlighted by the rectangle).
		The counter for the prefactor of~\(\eta\) is shared between the upper and lower branches and runs from~\(1\) to~\(|z|\).
		Here the specific index is~\(z=(\downarrow,...,\uparrow,...\uparrow,\downarrow)\).
		Note the lack of Hermitian conjugation on the lower branch (cf. Fig.~\ref{Fig:genericDiagram}), the minus sign on the lower branch scatterings, and the free propagator to the left of the leftmost scattering.
	} 
\end{figure}

In this appendix, we compare two approaches to diagrams for the propagator~\(\CG\).
Our diagrams from Sec.~\ref{sec:formal} are based on a decomposition of the density matrix evolution~\(\CU_\alpha\) into unitary evolutions (\(U_\nu^\pdagger\) and~\(U_\mu^\dagger\)) on each Keldysh branch.
They therefore do not rely on ordering (inherited from time ordering) between the upper and lower branches and we term them \textit{unordered}.
In contrast, the common diagrammatic formulation of the RT method~\cite{Schoeller1994MesoscopicQuantumTransport,Koller2010DensityoperatorApproachesTransport}, which can also be applied to the {\TCL} or the propagator~\(\CG\), relies directly on superoperators~\(\CU_\alpha\).
In these latter diagrams there is an ordering relation between events on different Keldysh branches inherited from the superoperator time-ordering, such that we term this formulation \textit{ordered}.
This \textit{ordered} description of the {\TCL} (SRT) contains significantly more diagrams (\(2^\alpha\) vs~\(\alpha+1\)) than the \textit{unordered} version.
The \textit{ordered} formulation, however, has other strengths, for example, in the proof of the convergence of the expansion of~\(\CS\) (\(\CZ\))~\cite{Timm2011TimeconvolutionlessMasterEquation,Koller2010DensityoperatorApproachesTransport}, or in the resummation of certain diagrams~\cite{ Schoeller1994MesoscopicQuantumTransport, Schoeller1994ResonantTunnelingCharge,Konig1995ResonantTunnelingCoulomb, Konig1997CotunnelingResonanceSingleElectron}.
In essence, in the ordered formulation,~\(\CG_{\nu\mu}\) is replaced by related propagators~\(\CG_z\), where the discrete index~\(z\) tracks the upper/lower Keldysh branch which the liouvillian~\(\liouvillian_V\) acts on, see Fig.~\ref{Fig:timeOrdered} and the discussion below for details.
As a safety check, we have computed the fourth-order rates using both of these diagrammatic formulations and find proper agreement.

\begin{figure*}
	\includegraphics[width=17.8cm]{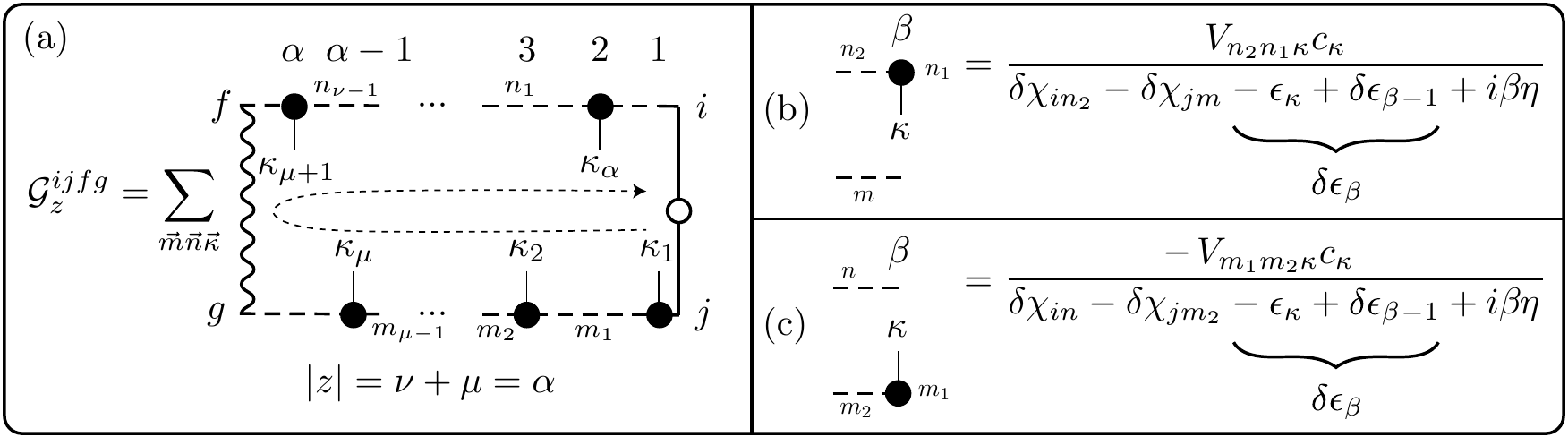}
	\caption{
		\label{Fig:timeOrderedWick}
		Ordered diagram for~\(\CG_{z}\), to be contrasted with the diagrams for~\(\CG_{\nu\mu }\), see Fig.~\ref{Fig:specificDiagram}. 
		As in Sec.~\ref{sec:quadratic}, the Wick index 	\(w\) is added by creating all pairings of the different environment operators.
		(a) The full diagram, which is related to Fig.~\ref{Fig:timeOrdered} in the same way as Fig.~\ref{Fig:specificDiagram}(a) is related to Fig.~\ref{Fig:genericDiagram}(c).
		The scattering events~\(V\) (black dots) pick up a contraction line (thin vertical lines) which are indexed clockwise from the bottom right. 
		The system indices along the free propagation (dashed lines on upper and lower branch) are identical to those in Fig.~\ref{Fig:specificDiagram}(a), while the counter from~\(1\) to~\(\alpha\) is shared between the upper and lower branch.
		(b)-(c)
		Factors that contribute to the diagram in (a) with a scattering event on the upper (b) or lower (c) Keldysh branch.
		Unlike in the unordered case of Fig.~\ref{Fig:specificDiagram} the environment energy counter~\(\delta\epsilon_\beta\) is shared between the branches.
		Note the negative sign when the scattering event lies on the lower branch (c).
	} %
\end{figure*}

\subsection{Ordered diagrams}

The ordered diagrams are constructed by inserting the expansion~\eqref{eq:superUexpansion} of the evolution~\(\CU\) into the definition~\eqref{eq:superS} of the {\TCL} generator~\(\CS\).
First, we construct~\(\CG_\alpha\) in this way.
We substitute the expansion~\eqref{eq:superUexpansion} for~\(\CU\) into the expression~\eqref{eq:superGa} for~\(\CG_\alpha\), and obtain
	\begin{align}
	\label{eq:superGzAbstract}
	\CG_\alpha^{ijfg} &= e^{\alpha\eta t/\hbar}(\delta\chi_{ij}-\delta\chi_{fg}+\alpha i\eta)^{-1}
	\\\nonumber
	&
	\times
	\Tr_\env
	\bra{f}
	[V,(\delta\chi_{ij}-\liouvillian_0+i(\alpha-1)\eta)^{-1}[V,...
	\\\nonumber
	&\,\,\,\quad... [V,(\delta\chi_{ij}-\liouvillian_0+i\eta)^{-1}[V,\ketbra{i}{j}\otimes\rho_\env^0]]]]\ket{g},
	\end{align}
	where the commutator~\([V,\rho]\) is always evaluated before the unperturbed Liouvillian~\(\liouvillian_0\) is applied.
	For each order~\(\alpha\), there are then~\(2^\alpha\) diagrams for~\(\CG_\alpha\), due to the~\(\alpha\) commutators with~\(V\) arising from~\(\alpha\) occurences of~\(\liouvillian_V\).
	We therefore introduce a new multi-index~\(z\), which keeps track of the specific sequence of the commutators and thus replaces~\(\alpha\) in a similar way that the indices~\(\nu\mu\) replaced~\(\alpha\) in section~\ref{sec:formal}.
	This is best illustrated with an example, such as~\(z = (\uparrow,\downarrow)\) that gives rise to the contribution
	\begin{align}
	\label{eq:superGsuperUexample}
	\CG^{ijfg}_{(\uparrow,\downarrow)} &= -e^{\alpha\eta t/\hbar}(\delta\chi_{ij}-\delta\chi_{fg}+\alpha i\eta)^{-1}
	\\\nonumber
	&\times\Tr_\env\!
	\bra{f}\![V(\delta\chi_{ij}\!-\!\liouvillian_0+i\eta)^{-1}(\ketbra{i}{j}\otimes\rho_\env^0V)]\!\ket{g},
	\end{align}
where we consider the positive part of the leftmost (latest in time) commutator, and the negative part of the rightmost (earliest in time) commutator.
Note the negative sign in Eq.~\eqref{eq:superGsuperUexample}, which arises from the negative sign in the rightmost commutator.
We use~\(\uparrow\) and~\(\downarrow\), for the individual commutators, because the positive (negative) contributions act on the upper (lower) Keldysh branch.

To obtain one full order of the propagator we sum over the index~\(z\) with the condition~\(|z|=\alpha\), i.e., the total number of entries in~\(z\) is~\(\alpha\).
Similarly, we can obtain the~\(\nu\mu\) contribution to the propagator~\(\CG\) by summing over the~\(\CG_z\) terms that fulfil the constraint~\(|z_\uparrow|=\nu\) (\(|z_\downarrow|=\mu\)), i.e., the total number of scattering events on the upper (lower) Keldysh branch is~\(\nu\) (\(\mu\)).
These two rules are summarised mathematically by
\begin{align}
\CG_\alpha = \sum_{|z|=\alpha} \CG_z , \quad\textrm{and},\quad\CG_{\nu\mu} =\sum_{|z_\uparrow|=\nu,|z_\downarrow|=\mu} \CG_{z},
\end{align}
which complement the existing relationships~\eqref{eq:superGConstruction} between indices for~\(\CG\).

By inspecting~\eqref{eq:superGzAbstract} and comparing it to the expression~\eqref{eq:diagramaticG} for~\(\CG_{\nu\mu}\), we notice a few key differences, which we encode diagrammatically, cf. Figs.~\ref{Fig:timeOrdered} and~\ref{Fig:genericDiagram}.
\begin{enumerate}
	\item Replace the single branch free propagator by a double branch free propagator.
	\item Keep track of the order between upper and lower scattering events.
	\item Introduce an additional prefactor~\(-1\) to each scattering event on the lower branch, thus accounting for the sign of the commutators.
	\item Remove the Hermitian conjugate operation on the lower branch.
\end{enumerate}
These diagrammatic considerations carry over directly to the specific example of quadratic environments which we considered in section~\ref{sec:quadratic}, see Fig.~\ref{Fig:timeOrderedWick}.
In contrast to our unordered diagrams, the diagrams in Fig.~\ref{Fig:timeOrderedWick} contain a single environment energy counter~\(\delta\epsilon\), which is incremented by scattering events both on the upper and lower branches.
We can further apply Wick's theorem to the diagrams for~\(\CG_z\) to obtain the diagrams~\(\CG_{zw}\)  in exactly the same way as for the~\(\CG_{\nu\mu}\).
The {\TCL} generator terms~\(\CS_{zw}\), are generated from the diagrams~\(\CG_{zw}\) in exactly the same ways as~\(\CS_{\nu\mu w}\) are generated from~\(\CG_{\nu\mu}\).
Due to the ordering relation between upper and lower branches in the \textit{ordered} diagrams, any \textit{cut} must be vertical.

\begin{widetext}
\section{Fourth-order {\TCL} rates}
\label{app:rates}

In this appendix, we show the steps that lead to the explicit expressions for the fourth-order rates.
We will take care to include the~\(\eta\) dependence and only take the~\(\eta\to0\) limit for convergent expressions.
We start with the~\(\tilde{\CS}_{22a}\) rate, as it includes most of the technical difficulties which arise.
We first write down the diagram from Fig.~\ref{Fig:mesoDiagrams22}(a) as
\begin{align}
\tilde{\CS}_{22a}^{if}&=4\eta\tilde{\CG}_{22a}^{if}
-
2\eta\sum_m 
\tilde{\CG}_{11}^{im}
\tilde{\CG}_{11}^{mf}
\\
\nonumber
&=
\sum_{mn\vec{\kappa}}
\frac{V_{in\kappa_1}V_{nf\kappa_2}V_{fm\kappa_3}V_{mi\kappa_4} \expval{c_{\kappa_1}c_{\kappa_4}}\expval{c_{\kappa_2}c_{\kappa_3}}}
{(\delta\chi_{in}+\epsilon_{\kappa_1}-i\eta)
	(\delta\chi_{im}+\epsilon_{\kappa_1}\!+i\eta)}
\left[\frac{4\eta}
{(\delta\chi_{if}+\epsilon_{\kappa_1}+\epsilon_{\kappa_2})^2+4\eta^2}
-
\frac{2\eta\delta_{nm}}{(\delta\chi_{mf}+\epsilon_{\kappa_2})^2+\eta^2}\right],
\end{align}
where the second term in the bracket (\(\propto\delta_{nm}\)) arises from the diagram with a cut.
As a next step, we change the environment sums into sums over the discrete parts of the environment and integrals over the continuous parts. We then perform a change of variables such that both terms in the bracket have the same denominator
\begin{align}
\tilde{\CS}^{if}_{22a}&=\sum_{nm\lambda_1\lambda_2}
\int
\frac{2\eta V_{in\lambda_1}V_{nf\lambda_2}V_{fm-\lambda_2}V_{mi-\lambda_1} C_{\lambda_1}(\epsilon_1)}
{
	(\delta\chi_{in}+\epsilon_1-i\eta) (\delta\chi_{im}+\epsilon_1+i\eta)}
\left[\frac{C_{\lambda_2}(2\epsilon_2-\epsilon_1+\delta\chi_{mi}+\delta\chi_{mf})-\delta_{nm}
	C_{\lambda_2}(\epsilon_2)}{
	(\epsilon_2+\chi_m-\chi_f)^2+\eta^2}\right]
\dd{\epsilon_1}\dd{\epsilon_2}.
\end{align}
We then split this expression into two, evaluate the Kronecker delta, and perform some manipulations on the (convergent)~\(n\neq m\) sum
\begin{align}
\tilde{\CS}^{if}_{22a}
=
2\pi
&\sum_{\substack{n\neq m\\\lambda_1\lambda_2}}
\frac{V_{in\lambda_1}V_{nf\lambda_2}V_{fm-\lambda_2}V_{mi-\lambda_1}}{\chi_n-\chi_m}
\int C_{\lambda_1}(\epsilon)\left[
\frac{C_{\lambda_2}(\chi_f-\chi_i-\epsilon)}{\chi_i-\chi_n+\epsilon-i\eta}
-
\frac{C_{\lambda_2}(\chi_f-\chi_i-\epsilon)}{\chi_i-\chi_m+\epsilon+i\eta}
\right]
\dd{\epsilon}
\\
\label{eq:S22aml1l2}
+&{2}\sum_{m\lambda_1\lambda_2}
|V_{im\lambda_1}|^2|V_{mf\lambda_2}|^2
\int C_{\lambda_1}(\epsilon_1)
\eta\frac{ C_{\lambda_2}(2\epsilon_2-\epsilon_1+2\chi_m-\chi_f-\chi_i) - C_{\lambda_2}(\epsilon_2)}
{[(\chi_i-\chi_m+\epsilon_1)^2+\eta^2][(\chi_m-\chi_f+\epsilon_2)^2+\eta^2]}
\dd{\epsilon_1}\dd{\epsilon_2}.
\end{align}
In the~\(n\neq m\) sum, we have performed the~\(\epsilon_2\) integral in the~\(\eta\to 0\) limit and taken a partial fraction decomposition of the remaining denominator.
The remaining integral is common in higher-order rate equation calculations.
In Appendix~\ref{app:integrals}, we outline one way of performing it analytically which leads to the expression in Eq.~\eqref{eq:S22anneqm}.
The second line~\eqref{eq:S22aml1l2} with~\(m=n\) is also convergent, which we will now show, before computing it explicitly.
We rewrite Eq.~\eqref{eq:S22aml1l2} in the functional form
\begin{align}
\label{eq:functional}
F[g,\eta] = \frac{1}{\eta}\int L(\epsilon_1,\eta)L(\epsilon_2,\eta)g(\epsilon_1,\epsilon_2)\dd{\epsilon_1}\dd{\epsilon_2},
\qquad
\textrm{ with, }
\qquad
L(\epsilon,\eta) = \frac{\eta}{\epsilon^2+\eta^2},
\end{align}
where we have extracted a~\(1/\eta\) factor to guarantee that the integral is convergent, irrespective of the function~\(g\).
The latter is given by 
\begin{align}
\label{eq:dummyFunction}
g(\epsilon_1,&\epsilon_2) = 
2 \sum_{m\lambda_1\lambda_2}
|V_{im\lambda_1}|^2|V_{mf\lambda_2}|^2
C_{\lambda_1}(\epsilon_1-\delta\chi_{im})
[
C_{\lambda_2}(2\epsilon_2-\epsilon_1-\delta\chi_{mf}) - C_{\lambda_2}(\epsilon_2-\delta\chi_{mf})
].
\end{align}
We find the convergence condition on~\(F\) by multiplying~\eqref{eq:functional} by~\(\eta\) and use~\(L(\epsilon,\eta) = \pi\delta(\epsilon)+\bigO(\eta)\) to obtain
\begin{align}
\eta F[g,\eta] = \pi^2g(0,0) +\bigO(\eta) = \bigO(\eta),
\end{align}
which confirms that Eq.~\eqref{eq:S22aml1l2} is convergent in the~\(\eta\to 0\) limit.
As a consequence of L'Hôpital's rule and the fact that~\(F\) is convergent in the limit~\(\eta\to 0\), we can write
\begin{align}
\lim_{\eta\to 0} F[g,\eta] &= \lim_{\eta\to 0} 
\partial_\eta
\left\{ \eta F[g,\eta] \right\}
=\pi\lim_{\eta\to 0}\partial_\eta\int L(\epsilon,\eta)[g(\epsilon,0)+g(0,\epsilon)] \dd{\epsilon},
\end{align}
where we have further made use of the product rule of differentiation and~\(L(\epsilon,\eta) = \pi\delta(\epsilon)+\bigO(\eta)\).
We substitute the expression~\eqref{eq:dummyFunction} for~\(g\) back into this latest expression and obtain
\begin{align}
\eqref{eq:S22aml1l2} = 
2\pi\sum_{m\lambda_1\lambda_2}
|V_{im\lambda_1}|^2|V_{mf\lambda_2}|^2
\partial_\eta 
\int\eta
\left[\frac{ C_{\lambda_1}(\epsilon)C_{\lambda_2}(\delta\chi_{fi}-\epsilon)}
{(\delta\chi_{im}+\epsilon)^2+\eta^2}
+
\frac{C_{\lambda_1}(\delta\chi_{mi}) C_{\lambda_2}(\epsilon)}
{(\delta\chi_{mf}+\epsilon)^2+\eta^2}
-
\frac{C_{\lambda_1}(\epsilon)C_{\lambda_2}(\delta\chi_{fm})}
{(\delta\chi_{im}+\epsilon)^2+\eta^2}
\right]
\dd{\epsilon},
\end{align}
which is valid in the~\(\eta\to 0\) limit.
The first integrand in the last expression is the same contribution as the regularised T-matrix integral, see Ref~\cite{FergusonThesis}, whereas the next two are corrections.
In Appendix~\ref{app:integrals}, we show how these can be cast into the expressions in Eqs.~\eqref{eq:S22ameqnMat} and~\eqref{eq:S22ameqnTCLCor}.

The~\(\tilde{\CS}^{if}_{22b} = 4\eta\tilde{\CG}^{if}_{22b} \) diagram is much simpler
	\begin{align}
	\tilde{\CS}^{if}_{22b}
	&=
	\pm
	\sum_{nm \vec{\kappa}} 
	\frac{4\eta V_{in\kappa_1}V_{nf\kappa_2}V_{fm\kappa_3}V_{mi\kappa_4}}
	{(\delta\chi_{if}+\epsilon_{\kappa_1}+\epsilon_{\kappa_2})^2+4\eta^2}
	\frac{ \expval{c_{\kappa_1}c_{\kappa_3}}\expval{c_{\kappa_2}c_{\kappa_4}}}
	{(\delta\chi_{in}+\epsilon_{\kappa_1}-i\eta)(\delta\chi_{im}+\epsilon_{\kappa_2}+i\eta)}
	\\
	&=\mp 2\pi\sum_{mn\lambda_1\lambda_2}
	V_{in\lambda_1}V_{nf\lambda_2}V_{fm-\lambda_1}V_{mi-\lambda_2}
	\int\frac{C_{\lambda_1}(\epsilon)C_{\lambda_2}(\delta\chi_{fi}-\epsilon)}
	{(\delta\chi_{in}+\epsilon-i\eta)(\delta\chi_{mf}+\epsilon-i\eta)} \dd{\epsilon},
	\end{align}
where in the second line, we performed the~\(\epsilon_2\) integral, a partial fraction decomposition, and relabelled~\(\epsilon_1\to\epsilon\).
The~\(\pm\) sign is~\(+\) for bosons and~\(-\) for fermions.
The remaining integrals, after a partial fraction decomposition, are again standard and an analytic expression for them can be found in Appendix~\ref{app:integrals}.
We thus obtain the expression in Eq.~\eqref{eq:S22b}.
The last~\(\tilde{\CS}_{22} \) contribution arises from the~\(c\) contraction
\begin{align}
\tilde{\CS}_{22c}^{if} = 4\eta\tilde{\CG}_{22c}^{if}
-
4\eta\delta_{if}\tilde{\CG}^{ii}_{20}\tilde{\CG}^{ii}_{02}
=
(1-\delta_{if})
\sum_{nm \vec{\kappa}} 
\frac{4\eta V_{in\kappa_1}V_{nf\kappa_2}V_{fm\kappa_3}V_{mi\kappa_4}
	\expval{c_{\kappa_1}c_{\kappa_2}}\expval{c_{\kappa_3}c_{\kappa_4}}}
{(\delta\chi_{in}+\epsilon_{\kappa_1}-i\eta)(\delta\chi_{im}+\epsilon_{\kappa_3}+i\eta)[(\delta\chi_{if})^2+4\eta^2]}
= 0 + \bigO(\eta),
\end{align}
and vanishes in the~\(\eta\to 0\) limit.

We can now repeat the same steps as above but for the~\(\tilde{\CS}_{31}\) and~\(\tilde{\CS}_{13}\) diagrams.
For the first contraction~\(a\) we obtain
\begin{subequations}
	\label{eq:S31a}
\begin{align}
\nonumber
\tilde{\CS}^{if}_{31a}=&
\sum_{nm\vec{\kappa}} 
\frac{V_{if\kappa_1}V_{fn\kappa_2}V_{nm\kappa_3}V_{mi\kappa_4}\expval{c_{\kappa_1}c_{\kappa_4}}\expval{c_{\kappa_2}c_{\kappa_3}}}
{(\delta\chi_{if}+\epsilon_{\kappa_1}-i\eta)(\delta\chi_{im}+\epsilon_{\kappa_1}+i\eta)}
\left[\frac{ 4\eta}
{(\delta\chi_{in}+\epsilon_{\kappa_1}+\epsilon_{\kappa_2}+2i\eta)(\delta\chi_{if}+\epsilon_{\kappa_1}+3i\eta)}
+\frac{i\delta_{mf}}{(\delta\chi_{fn}+\epsilon_{\kappa_2}+i\eta)}\right]
\\
\label{eq:S31aP1}
=&
\,2\pi\sum_{\lambda_1}|V_{if\lambda_1}|^2\sum_{\substack{n\lambda_2\\m\neq f}} 
\frac{ V_{fn\lambda_2} V_{nm-\lambda_2} V_{mi-\lambda_1} }{(\chi_f-\chi_m) V_{fi-\lambda_1}}
C_{\lambda_1}(\delta\chi_{fi})
J_{\lambda_2}^+(\delta\chi_{fn})
\\
\label{eq:S31aP2}
&-
2\pi\sum_{n\lambda_1\lambda_2} 
|V_{if\lambda_1}|^2 |V_{fn\lambda_2}|^2 
C_{\lambda_1}(\delta\chi_{fi})
\partial_{\chi_n}
J_{\lambda_2}^+(\delta\chi_{fn})
-
i\sum_{n\lambda_1\lambda_2} 
|V_{if\lambda_1}|^2 |V_{fn\lambda_2}|^2 
\partial_{\chi_i}
J_{\lambda_1}^+(\delta\chi_{if})
J_{\lambda_2}^+(\delta\chi_{fn}),
\end{align}
\end{subequations}
where we have again used L'Hôpital's rule and the integrals from Appendix~\ref{app:integrals}.
The~\(b\) contraction is again simpler, leading to
\begin{align}
\nonumber
\tilde{\CS}_{if}^{31b}&=
\pm\sum_{nm\vec{\kappa}} 
\frac{4\eta V_{if\kappa_1}V_{fn\kappa_2}V_{nm\kappa_3}V_{mi\kappa_4}}
{(\delta\chi_{if}+\epsilon_{\kappa_1}-i\eta)(\delta\chi_{if}+\epsilon_{\kappa_1}+3i\eta)}
\frac{ \expval{c_{\kappa_1}c_{\kappa_3}}\expval{c_{\kappa_2}c_{\kappa_4}}}
{(\delta\chi_{in}+\epsilon_{\kappa_1}+\epsilon_{\kappa_2}+2i\eta)(\delta\chi_{im}+\epsilon_{\kappa_2}+i\eta)}
\\
\label{eq:S31b}
&=
\pm 2\pi
\sum_{nm\lambda_1\lambda_2}{V_{if\lambda_1}V_{fn\lambda_2}V_{nm-\lambda_1}V_{mi-\lambda_2}}
C_{\lambda_1}(\delta\chi_{fi})
\frac{J_{\lambda_2}^+(\chi_{fn})-J_{\lambda_2}^+(\chi_{im})}{\chi_i-\chi_f+\chi_n-\chi_m},
\end{align}
where the~\(\pm\) differentiates bosons (\(+\)) and fermions (\(-\)).
The last~\(\tilde{\CS}_{31}\) contribution arises from the~\(c\) contraction.
It requires the use of L'Hôpital's rule and the integrals from Appendix~\ref{app:integrals} to obtain
\begin{align}
\label{eq:S31c}
\tilde{\CS}^{if}_{31c}&=
\sum_{nm\kappa} 
\frac{V_{if\kappa_1}V_{fn\kappa_2}V_{nm\kappa_3}V_{mi\kappa_4}\expval{c_{\kappa_1}c_{\kappa_2}}\expval{c_{\kappa_3}c_{\kappa_4}}}
{(\delta\chi_{in}+2i\eta)(\delta\chi_{im}+\epsilon_{\kappa_3}+i\eta)(\delta\chi_{if}+\epsilon_{\kappa_1}-i\eta)}
\Bigg[
\frac{4\eta }
{(\delta\chi_{if}+\epsilon_{\kappa_1}+3i\eta)}
-
\frac{\delta_{in}2\eta}{(\delta\chi_{if}+\epsilon_{\kappa_1}+i\eta)}
\Bigg]
\\\nonumber
\\\nonumber
&=2\pi\sum_{\substack{m\lambda_1\lambda_2\\n\neq i}} 
V_{if\lambda_1}V_{fn-\lambda_1}V_{nm\lambda_2}V_{mi-\lambda_2}
C_{\lambda_1}(\delta\chi_{fi})
\frac{J_{\lambda_2}^+(\delta\chi_{im})}{\delta\chi_{in}}
-i\sum_{m\lambda_1\lambda_2}|V_{if\lambda_1}|^2|V_{im\lambda_2}|^2
\partial_{\chi_f}
J_{\lambda_1}^+(\delta\chi_{if})
J_{\lambda_2}^+(\delta\chi_{im}).
\end{align}
\end{widetext}
The~\(\tilde{\CS}_{13} = \tilde{\CS}_{31}^* \) terms are obtained by complex conjugation. We do not compute the~\(\tilde{\CS}_{40}\) or~\(\tilde{\CS}_{04}\) terms as their sum can be obtained easily from the conservation of probability, see Eq.~\eqref{eq:conserved}.

The last terms in each of~\(\tilde{\CS}^{if}_{31a}\) and~\(\tilde{\CS}^{if}_{31c}\) contain a~\(J\) integral which is not differentiated.
They thus include ultraviolet divergences~\(\Lambda\), which we will show cancel in physical setups in the wide-band limit.
We sum the two contributions
\begin{align}
\label{eq:constraint}
&\tilde{\CS}^{if}_{31a}+\tilde{\CS}^{if}_{31c} 
 = \order{\Lambda^0}
\\ \nonumber
&-i
\sum_{n\lambda_1\lambda_2} 
|V_{if\lambda_1}|^2
\left[ 
\partial_{\chi_i} J_{\lambda_1}^+(\delta\chi_{if})
\right]
 (|V_{fn\lambda_2}|^2 - |V_{in\lambda_2}|^2)
\Lambda_{\lambda_2},
\end{align}
and focus on the terms that are linear in~\(\Lambda\) to obtain the condition
\begin{align}
\sum_{n\lambda} |V_{in\lambda}|^2 \Lambda_\lambda= K,\quad  \forall\,\, i,
\end{align}
where~\(K\) is a constant.
If we assume that the environment is approximately particle--hole symmetric, the two ultraviolet cutoffs~\(\Lambda_\lambda \approx \Lambda_{-\lambda}\) are related.
We can then further simplify the constraint~\eqref{eq:constraint} to
\begin{align}
\label{eq:constraintStronger}
\sum_n \left(|V_{in\lambda}|^2+|V_{ni\lambda}|^2\right) = |V_\lambda|^2,\quad  \forall\,\, i,\lambda,
\end{align}
where~\(V_\lambda\) are reservoir dependent constants.
This is the case in any setup constructed purely from second quantised operators, such as the non-interacting level considered in this work or Anderson's impurity model.
It is, however, possible to make the ultraviolet cutoff relevant.
One method to do so is to introduce a large (of order unity compared to the bandwidth) particle--hole asymmetry, which breaks the assumption allowing us to go from Eq.~\eqref{eq:constraint} to Eq.~\eqref{eq:constraintStronger}.
Alternatively, we can discard system states that are considered to be at very high energies.
In the Kondo model, for example, high energy intermediate system states are not included, as their presence is purely encoded in an effective coupling.
This leads to a bandwidth-dependent Kondo-temperature~\cite{Bruus2004ManyBodyQuantumTheory}.
On the other hand, a Kondo temperature computed directly from the Anderson impurity, where the empty and doubly-occupied states are included does not depend on the bandwidth~\cite{Haldane1978ScalingTheoryAsymmetric}.

\subsection{Current rates}

Finally we find the current rates for the~\(\tilde{\CS}_{13}\) contributions.
We use the diagrammatic rules from Fig.~\ref{Fig:currentWick} and thus immediately conclude that we must replace the sums over the reservoir index that connects the two Keldysh contours
\begin{align}
\tilde{\CS}_{31}: \sum_{\lambda_1 \lambda_2, ...} 
\to \tilde{\CS}_{31}(\lambda_1): \sum_{\xcancel{\lambda_1}\lambda_2,...}.
\end{align}

\section{Useful integrals and properties}
\label{app:integrals}

In this appendix, we provide a collection of useful integrals, many of which can be found in similar form in Refs.~\cite{Koch2004ThermopowerSinglemoleculeDevices,Koch2006QuantumTransportSinglemolecule}.
These are used extensively when computing cotunnelling rates of electronic setups.
The integrals~\(I_{\lambda_1 \lambda_2}^\pm(\delta_1,\delta_2)\) and~\(J_\lambda^\pm(\delta)\), from Sec.~\ref{sec:rates:cot} and Appendix~\ref{app:rates}, can be evaluated analytically in terms of digamma functions~\(\psi\).
The latter is closely related to the Fermi-Dirac distribution~\(n_{\rs F}\) and Bose-Einstein distributions~\(n_{\rs B}\) as both distributions can be written in terms of digamma functions
\begin{align}
\!\!
2\pi n_{\rs F}(z) &=
\pi + 
i\psi\left(\frac{1}{2}+\frac{iz}{2\pi T}\right) 
-
i\psi\left(\frac{1}{2}-\frac{iz}{2\pi T}\right)\!,
\!\!
\\ \nonumber
\!\!
2\pi n_{\rs B}(z) &= 
-\pi -\frac{2\pi T}{z}
-i\psi\left(\frac{iz}{2\pi T}\right) 
+ 
i\psi\left(-\frac{iz}{2\pi T}\right)\!.
\end{align}
Here, we work with fermionic environments and a constant temperature~\(T\) across the entire environment.
To avoid notational clutter, we introduce the integral
\begin{align}
\label{eq:Integral0}
I_0^-(\mu_1,\mu_2,\gamma) &= \int_{-\infty}^\infty d\epsilon \frac{n_{\rs F}(\epsilon-\mu_1) n_{\rs F}(\mu_2-\epsilon)}{\epsilon-i\gamma},
\end{align}
and its complex conjugate~\(I^{+}_0 = I^{-*}_0\), where~\(\gamma>0\).
This integral is related to the ones in Sec.~\ref{sec:rates:cot} by
\begin{align}
\label{eq:IntegralI}
I_{\lambda_1 \lambda_2}^\pm(\delta_1,\delta_2) &= \lim_{\eta\to 0} I_0^\pm(\delta_1+\mu_{\lambda_1},\delta_2-\mu_{\lambda_2},\eta),
\\
\label{eq:IntegralJ}
J_\lambda^\pm(\delta)&= \lim_{\eta\to 0}I_0^\pm(\delta+\mu_\lambda,-\Lambda_\lambda,\eta),
\end{align}
where we have introduced the reservoir dependent cutoff~\(\Lambda_\lambda \) for the continuous degree of freedom.
In a physical system this cutoff is finite due to the bandwidth of the electronic environment and can thus be properly accounted for.
Here, however, we restrict ourselves to the wide band limit where~\(\Lambda_\lambda\to \infty\), which is justified as long as~\(\Lambda\) is much larger than all other energy scales in the setup and all observables remain finite in the limit.
We use the substitution
\begin{align}
\frac{1}{K} = \int_0^\infty e^{-Kt}\dd{t},
\end{align} 
to replace the denominator in  Eq.~\eqref{eq:Integral0} and thus obtain
\begin{align}
\nonumber
\eqref{eq:Integral0}
&=
i\int_{-\infty}^\infty d\epsilon \int_0^\infty dt \,
n_{\rs F}(\epsilon-\mu_1)  n_{\rs F}(\mu_2-\epsilon)e^{-i\epsilon t-\gamma t}.
\end{align}
We then perform the Fourier transform of the product of Fermi distributions using contour integration and the residue theorem to obtain
\begin{align}
\eqref{eq:Integral0}=n_{\rs B}\left(\mu_2-\mu_1\right)\nonumber
\int_0^\infty \!\!\!\dd{x}
e^{-(1/2 +\tilde{\gamma})x}
\frac{e^{-i\tilde{\mu}_2x}
-	
e^{-i\tilde{\mu}_1x}}{1-e^{-x}},
\end{align}
where we used the notation~\(\tilde{a} = {a}/({2\pi T})\), for~\(a=\gamma,\mu_1,\mu_2\). We can then use the integral representation of the digamma function
\begin{align}
\label{eq:digammaInt}
\psi(z)
=
\int_{0}^{\infty}
\left(
\frac{e^{-t}}{t}-\frac{e^{-zt}}{1-e^{-t}}
\right)
\dd{t},
\end{align}
to conclude that
\begin{align}
I_0^-(\mu_1,&\mu_2,\gamma) = n_{\rs B}(\mu_2-\mu_1)
\\\nonumber
&\times\left[\psi\left(\frac{1}{2}+\frac{\gamma+i\mu_1}{2\pi T}\right)
-
\psi\left(\frac{1}{2}+\frac{\gamma+i\mu_2}{2\pi T}\right)\right].
\end{align}
Inserting this last result into Eq.~\eqref{eq:IntegralI},  we obtain the expression~\eqref{eq:IntegralIpm} for the integral~\(I\) in section~\ref{sec:rates:cot}.
Inserting the same result into Eq.~\eqref{eq:IntegralJ}, taking the large band limit, and using
\begin{align}
\psi(i\Lambda_\lambda) \approx i\frac{\pi}{2}+\ln\Lambda_\lambda,
\end{align}
we obtain an expression~\eqref{eq:IntegralJpm} for the~\(J\) integral in section~\ref{sec:rates:cot}.

\section{Exact methods for the non-interacting setup}
\label{app:exact}

This appendix contains a brief description of the exact results for the non-interacting level that are derived in a broad set of different pedagogical texts~\cite{Bruus2004ManyBodyQuantumTheory,Ryndyk2015TheoryQuantumTransport}.
A key element, both in and out of equilibrium, is the imaginary part of the retarded Green's function (spectral function)
\begin{align}
\label{eq:spectralFunction}
A(\omega) = \frac{\Gamma_0}{(\omega-\epsilon_0)^2+\Gamma_0^2/4},
\end{align}
of the non-interacting level~\cite{Bruus2004ManyBodyQuantumTheory}, recall~\(\Gamma_0=2\pi|\hop|^2\DOS\).
It is obtained in a straightforward manner using an equation-of-motion for Green's functions approach, see for example chapter 9.2 of Ref.~\cite{Bruus2004ManyBodyQuantumTheory}.

\subsection{Equilibrium occupation}

The probability~\(P_1=\langle d_0^\dagger d_0^\pdagger \rangle\) of finding the level occupied is the expectation value of the associated number operator~\(d_0^\dagger d_0^\pdagger\), which in equilibrium (\(\mu=0\)) is simply
\begin{align}
\label{eq:exactExpectation}
\expval{d_0^\dagger d_0^\pdagger} 
=
\int_{-\infty}^{\infty}\frac{\dd{\omega}}{2\pi}A(\omega)n_{\rs F}(\omega),
\end{align} 
the convolution of the spectral weight~\(A\) and the occupation probability~\(n_{\rs F}\), see for example Eq.~(8.62) of Ref.~\cite{Bruus2004ManyBodyQuantumTheory}.
Using the integral from Eq.~\eqref{eq:IntegralJ} in the limit~\(\Lambda_\lambda \to \infty\) and a partial fraction decomposition on~\(A\) we arrive at the exact expression~\eqref{eq:probabilityExact} for the occupation probability~\(P_1\).

\subsection{Out-of-equilibrium current}

The out-of-equilibrium(~\(\mu\neq 0\)) current across a level is an archetypal observable quantity.
For a non-interacting system it can be obtained~\cite{Bruus2004ManyBodyQuantumTheory,Ryndyk2015TheoryQuantumTransport} using scattering matrices or the Green's function formalism.
Alternatively it can be derived from the more general Meir-Wingreen formula~\cite{Meir1992LandauerFormulaCurrent}, which is valid for both interacting and non-interacting systems and takes the Green's function as an input.
For the non-interacting level, the current is given by the exact expression~\cite{Koch2006QuantumTransportSinglemolecule}
\begin{align}
\!I = \frac{e\Gamma_0}{8\pi \hbar}\int_{-\infty}^{\infty}\!\!\!\dd{\omega} A(\omega)[ n_{\rs F}(\omega-\mu/2)-n_{\rs F}(\omega+\mu/2)],\!
\end{align}
see for example equation (9) of Ref.~\cite{Meir1992LandauerFormulaCurrent}.
Physically, the transport occurs in an energy window of width~\(\mu\), reduced by the temperature~\(T\).
Within this window the tunnelling probability is proportional to the spectral function~\(A\) of the non-interacting level, i.e., how much weight the non-interacting level makes available at a given energy.
Using the integral Eq.~\eqref{eq:IntegralJ} in the limit~\(\Lambda_{\lambda}\to\infty\) and a partial fraction decomposition on~\(A\), we arrive at the exact expression~\eqref{eq:currentExact} for the current across the non-interacting level.

\section{Combining operators and superoperators}
\label{app:mixing}
In Sec.~\ref{sec:currents} we obtained expressions which contain both superoperators and the number operator~\(N_\lambda\).
To maintain an unambiguous order, such expressions require large numbers of brackets.
For example the series~\(\CA A \CB B \rho\) of operators~\(A,B\) and superoperators~\(\CA,\CB\) acting on a density matrix~\(\rho\) can be bracketed in a multitude of ways including but not limited to
\begin{align}
(\CA A) (\CB B) \rho, \textrm{ or }
\{\CA [A (\CB B)]\} \rho, \textrm{ or } \CA \{A [\CB (B\rho)]\},
\end{align} 
which in general may produce different results.
In this Appendix, we explain the notational shorthand we use to avoid this large number of brackets.

Whenever an operator~\(\Nr\) appears to the right of a projector~\(\projector\)~\eqref{eq:superP}, we define
\begin{align}
\projector \Nr \CA \rho \equiv 
\projector [ \Nr (\CA \rho)]=
\projector [ (\CA \rho)\Nr ],
\end{align} 
where~\(\CA\) is an arbitrary superoperator and we have used the fact that~\(\Nr\) is an environment operator and the cyclic property of the environment trace in~\(\projector\).
Similarly, whenever an operator~\(\Nr\) appears to the left of a projector we define
\begin{align}
\CB \Nr\projector \CA \rho \equiv 
\CB[\Nr(\projector \CA \rho)] =
\CB[(\projector \CA \rho)\Nr],
\end{align}
where~\(\CB\) is an arbitrary superoperator and we used both the fact that~\(\Nr\) is an environment operator and the fact that~\([\Nr,H_0]=0\) (which in turn implies~\([\Nr,\rho_\env^0]=0\)).

\end{document}